\documentclass[11pt,a4paper]{article}
\usepackage[T1]{fontenc}
\usepackage[utf8]{inputenc}
\usepackage{mathptmx}
\usepackage[scaled=0.92]{helvet}
\usepackage{textcomp}
\usepackage{amsmath,amssymb}
\usepackage[margin=2.3cm]{geometry}
\usepackage{graphicx}
\usepackage{longtable,booktabs,array,calc}
\usepackage[table]{xcolor}
\usepackage{microtype}
\usepackage{newunicodechar}
\usepackage{needspace}
\usepackage[hidelinks]{hyperref}
\providecommand{\tightlist}{\setlength{\itemsep}{0pt}\setlength{\parskip}{0pt}}

\newunicodechar{₀}{\textsubscript{0}}
\newunicodechar{₁}{\textsubscript{1}}
\newunicodechar{₂}{\textsubscript{2}}
\newunicodechar{₃}{\textsubscript{3}}
\newunicodechar{₄}{\textsubscript{4}}
\newunicodechar{₅}{\textsubscript{5}}
\newunicodechar{₆}{\textsubscript{6}}
\newunicodechar{₇}{\textsubscript{7}}
\newunicodechar{₈}{\textsubscript{8}}
\newunicodechar{₉}{\textsubscript{9}}
\newunicodechar{ₐ}{\textsubscript{a}}
\newunicodechar{ₓ}{\textsubscript{x}}
\newunicodechar{ₗ}{\textsubscript{l}}
\newunicodechar{ₘ}{\textsubscript{m}}
\newunicodechar{ₚ}{\textsubscript{p}}
\newunicodechar{ᵢ}{\textsubscript{i}}
\newunicodechar{ₕ}{\textsubscript{h}}
\newunicodechar{⁰}{\textsuperscript{0}}
\newunicodechar{⁸}{\textsuperscript{8}}
\newunicodechar{⁹}{\textsuperscript{9}}
\newunicodechar{⁴}{\textsuperscript{4}}
\newunicodechar{⁵}{\textsuperscript{5}}
\newunicodechar{⁶}{\textsuperscript{6}}
\newunicodechar{⁷}{\textsuperscript{7}}
\newunicodechar{⁺}{\textsuperscript{+}}
\newunicodechar{⁻}{\textsuperscript{\textminus}}
\newunicodechar{⁽}{\textsuperscript{(}}
\newunicodechar{⁾}{\textsuperscript{)}}
\newunicodechar{ⁿ}{\textsuperscript{\textit{n}}}
\newunicodechar{Γ}{\ensuremath{\Gamma}}
\newunicodechar{Δ}{\ensuremath{\Delta}}
\newunicodechar{Σ}{\ensuremath{\Sigma}}
\newunicodechar{Ω}{\ensuremath{\Omega}}
\newunicodechar{α}{\ensuremath{\alpha}}
\newunicodechar{β}{\ensuremath{\beta}}
\newunicodechar{γ}{\ensuremath{\gamma}}
\newunicodechar{ε}{\ensuremath{\varepsilon}}
\newunicodechar{θ}{\ensuremath{\theta}}
\newunicodechar{κ}{\ensuremath{\kappa}}
\newunicodechar{ν}{\ensuremath{\nu}}
\newunicodechar{π}{\ensuremath{\pi}}
\newunicodechar{σ}{\ensuremath{\sigma}}
\newunicodechar{ω}{\ensuremath{\omega}}
\newunicodechar{→}{\ensuremath{\rightarrow}}
\newunicodechar{−}{\ensuremath{-}}
\newunicodechar{∝}{\ensuremath{\propto}}
\newunicodechar{≈}{\ensuremath{\approx}}
\newunicodechar{≡}{\ensuremath{\equiv}}
\newunicodechar{≤}{\ensuremath{\le}}
\newunicodechar{≥}{\ensuremath{\ge}}
\newunicodechar{≪}{\ensuremath{\ll}}
\newunicodechar{≲}{\ensuremath{\lesssim}}
\newunicodechar{⟨}{\ensuremath{\langle}}
\newunicodechar{⟩}{\ensuremath{\rangle}}
\newunicodechar{ů}{\r{u}}
\newunicodechar{č}{\v{c}}
\newunicodechar{š}{\v{s}}
\newunicodechar{ž}{\v{z}}
\begin{document}

\begin{center}
{\LARGE\bfseries Low-temperature vibrational anomalies and thermal transport in polycyclic aromatic hydrocarbon crystals\par}
\vspace{10pt}
{\large V.~Sokolenko\textsuperscript{1}, A.~I.~Krivchikov\textsuperscript{1,2,*}, A.~Je\.zowski\textsuperscript{1}, V.~Boiko\textsuperscript{1}, O.~A.~Korolyuk\textsuperscript{2}, D.~Szewczyk\textsuperscript{1}\par}
\vspace{6pt}
{\small \textsuperscript{1}Institute of Low Temperature and Structure Research, Polish Academy of Sciences, Ok\'olna 2, 50-422 Wroc\l{}aw, Poland\\\textsuperscript{2}B.~Verkin Institute for Low Temperature Physics and Engineering, National Academy of Sciences of Ukraine, Nauky Ave.~47, 61103 Kharkiv, Ukraine\\\textsuperscript{*}Corresponding author. E-mail: \href{mailto:krivchikov@ilt.kharkov.ua}{krivchikov@ilt.kharkov.ua}\par}
\end{center}

\begin{abstract}
\noindent The boson peak is an excess of low-frequency vibrational states over the Debye prediction and is commonly associated with structural disorder. Similar heat-capacity anomalies also occur in ordered molecular crystals, but their dependence on molecular structure and their relation to heat transport remain incompletely understood. Here we compare six polycyclic aromatic hydrocarbon (PAH) crystals---pyrene, tetracene, perylene, pentacene, coronene and corannulene---using low-temperature heat capacity, low-frequency Raman spectroscopy and thermal conductivity measurements. For compounds with well-resolved maxima, the rising side of the heat-capacity excess is consistent with a raised-cosine scaling function with no adjustable shape parameter, whereas the decay is compound-dependent. The number and widths of resolved lattice-mode Raman bands vary across the series, although their assignment is constrained by spectral background and resolution. Thermal conductivity ranges from crystal-like to glass-like behaviour. An activated contribution is resolved in the five planar compounds but is not required to describe the corannulene data within the sensitivity of the present measurements. These results show that glass-like thermal and vibrational signatures are not unique to amorphous materials and are associated with differences in molecular structure and crystal packing; they do not, by themselves, isolate packing as the sole causal variable.
\end{abstract}

\section*{Introduction}

The boson peak---an excess of vibrational states over the Debye prediction, commonly manifested as a maximum in the reduced heat capacity Cₚ/T\textsuperscript{3} and, in many glasses, as a plateau in the thermal conductivity κ(T)---is a long-standing problem in the physics of disordered solids\textsuperscript{1,2}. Structural disorder can scatter and hybridize acoustic phonons and redistribute vibrational spectral weight\textsuperscript{3,4}, but the microscopic origin of the boson peak remains debated. Proposed explanations include soft anharmonic modes\textsuperscript{5}, heterogeneous elasticity\textsuperscript{6--9}, localized structural defects\textsuperscript{10,11}, non-phononic quasi-localized excitations\textsuperscript{12,13} and deviations from linear acoustic-phonon dispersion\textsuperscript{14}. Studies of ultrastable and aged glasses further show that tunnelling states and boson-peak modes can respond differently to preparation history\textsuperscript{15--17}, arguing against a single universal microscopic explanation\textsuperscript{18}. An interacting two-level-system (ITLS) framework has recently been proposed in which localized soft excitations dominate at the lowest energies, while higher-energy hybridized excitations couple local configurations to elastic vibrations and contribute near and above the boson-peak region\textsuperscript{19}.

Boson-peak-like heat capacity anomalies have also been reported in orientationally ordered and minimally disordered molecular crystals\textsuperscript{20--22}. In periodic lattices, low-lying optical modes and their hybridization with acoustic branches can produce an excess vibrational density of states without requiring structural disorder\textsuperscript{23,24}. A useful way to examine this possibility is to compare a family of molecular crystals with related intermolecular interactions but different molecular shapes and packing arrangements. Such a comparison can identify structure--property correlations, although a causal role for packing requires structural characterization of the actual measured specimens and, ideally, controlled comparisons within the same compound.

Polycyclic aromatic hydrocarbons (PAHs) provide a useful family for such a comparison. Tetracene, pentacene, pyrene, perylene and coronene are established organic semiconductors used in the context of field-effect transistors, light-emitting devices and photovoltaic materials\textsuperscript{25,26}; tetracene and pentacene are also prototypical singlet-fission materials\textsuperscript{27}. In these van der Waals crystals, low-frequency lattice vibrations modulate intermolecular electronic transfer integrals and can contribute to dynamic disorder, which affects charge transport\textsuperscript{28--30}; the anharmonicity of these modes is recognized as a key material property\textsuperscript{31}. The same lattice vibrations contribute to heat transport, while the relatively low and anisotropic thermal conductivity of organic molecular crystals and films is relevant to thermal management in organic electronics and to organic thermoelectric materials\textsuperscript{32--35}. Reliable heat-capacity data are also important for thermodynamic reference functions of PAHs\textsuperscript{36}. PAHs are additionally abundant in the interstellar medium, where their vibrational properties are relevant to the interpretation of infrared emission features\textsuperscript{37}.

Here we investigate six PAH crystals---pyrene, tetracene (benz{[}b{]}anthracene), perylene, pentacene, coronene and corannulene (Fig. 1)---that share π-conjugated molecular frameworks and are predominantly held together by van der Waals interactions, but differ in molecular shape, symmetry and crystal packing. Coronene is a compact, approximately disc-shaped D₆\textsubscript{h} molecule. Pyrene, perylene, tetracene and pentacene have D₂\textsubscript{h} molecular symmetry; tetracene and pentacene form a linear-acene pair of different length. Corannulene is bowl-shaped (C₅\textsubscript{v}) and exhibits temperature-dependent orientational behaviour in the solid state\textsuperscript{38,39}. The universal empirical scaling of the rising side of low-temperature heat capacity anomalies has recently been reported for classical and quantum cryocrystals\textsuperscript{42}. This study extends the comparison to a chemically related series of molecular crystals using raised-cosine scaling. It examines the asymmetry of the heat-capacity anomaly alongside Raman spectra and thermal conductivity. We address three questions: whether the rising side of the heat-capacity excess follows the same normalized function across the series; how the resolved low-frequency Raman response varies among the compounds; and whether the thermal conductivity can be described by a common phenomenological form containing an activated contribution. The results establish correlations across these observables, while the microscopic interpretation and the causal role of packing remain open questions.

\begin{figure}[!htbp]
\centering
\includegraphics[width=0.75\linewidth]{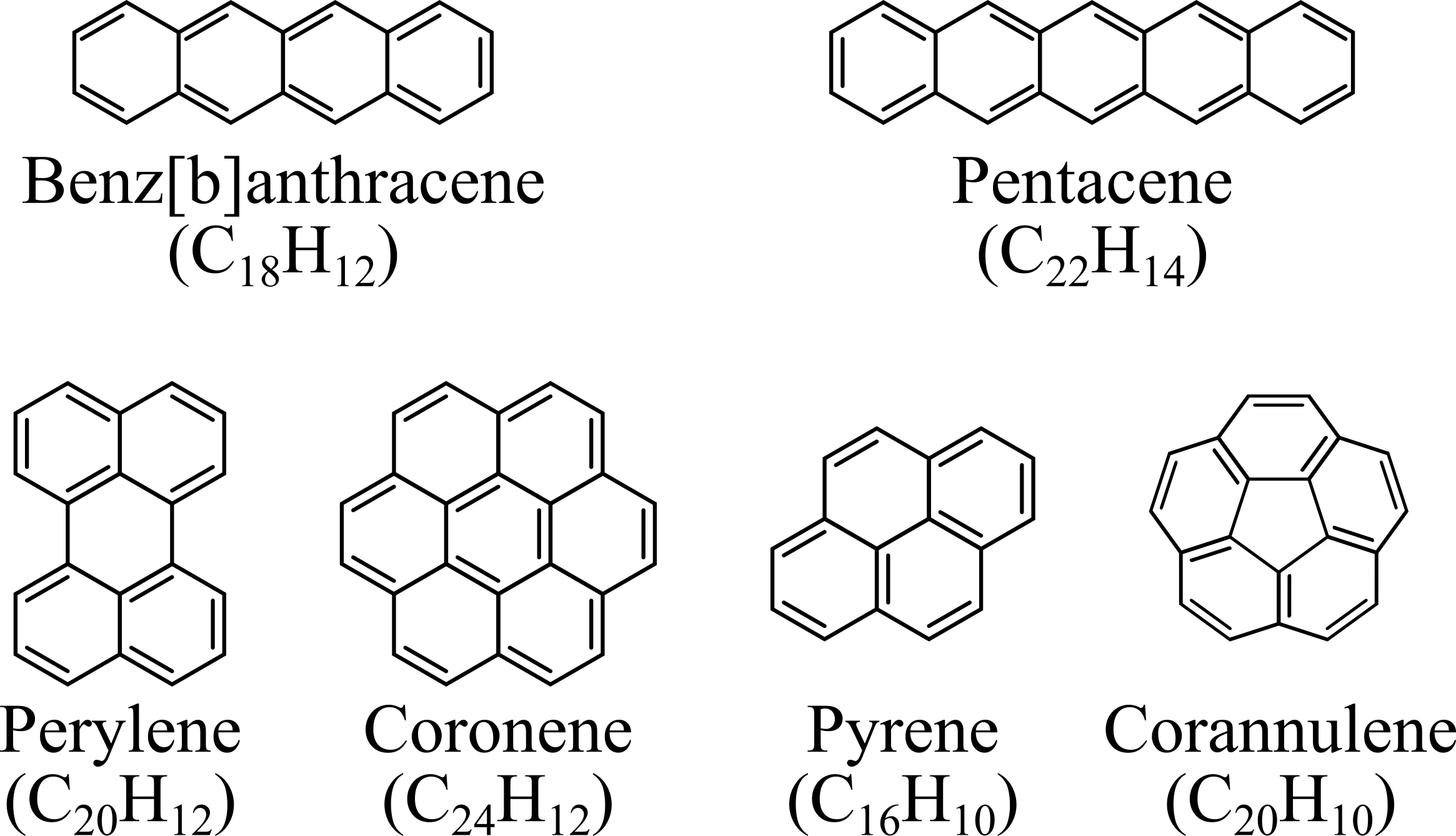}

\medskip
\begin{minipage}{\linewidth}\small \textbf{Fig. 1 \textbar{}} Molecular structures of the six PAHs studied: tetracene (benz{[}b{]}anthracene, C₁₈H₁₂), pentacene (C₂₂H₁₄), perylene (C₂₀H₁₂), coronene (C₂₄H₁₂), pyrene (C₁₆H₁₀) and corannulene (C₂₀H₁₀). Crystal space groups, numbers of molecules per unit cell (Z) and isolated-molecule point groups are summarized in Table 2.\end{minipage}
\end{figure}

\section*{Results}

\subsection*{The General Cosine--Cosine (GCC) model: a two-branch description of the heat-capacity excess}

At temperatures below approximately 5--10 K, the heat capacity of glasses, disordered crystals and some molecular solids is often represented by Cₚ(T) = c₁T + c₃T\textsuperscript{3} + Δ(T), where c₁T is a linear contribution often associated with tunnelling states\textsuperscript{40,41}, c₃T\textsuperscript{3} is the Debye term and Δ(T) denotes the excess contribution. The corresponding reduced quantity is Cₚ/T\textsuperscript{3} = c₁/T\textsuperscript{2} + c₃ + Δ(T)/T\textsuperscript{3}. For this analysis, we define Tₘₐₓ and Hₘₐₓ from the maximum of the background-corrected quantity Cₚ/T\textsuperscript{3} − c₁/T\textsuperscript{2}; thus Hₘₐₓ is the value of this quantity at Tₘₐₓ and includes the Debye baseline c₃. This definition differs slightly from the maximum of the uncorrected Cₚ/T\textsuperscript{3} curve. We normalize the excess as Δ*(x) = {[}Cₚ/T\textsuperscript{3} − c₁/T\textsuperscript{2} − c₃{]}/(Hₘₐₓ − c₃), with x = T/Tₘₐₓ. A raised-cosine function, ½{[}1 − cos(πx){]}, has been used to describe the rising side of related anomalies\textsuperscript{42},\textsuperscript{43}. Because experimental peaks are not generally symmetric, we use separate cosine expressions for the two sides of the maximum:

\begin{quote}
\begin{equation*}\frac{C_p}{T^3}=\frac{c_1}{T^2}+c_3+\left(H_{\max}-c_3\right)\tfrac12\left[1-\cos\!\left(\frac{\pi T}{T_{\max}}\right)\right],\qquad T\le T_{\max}.\tag{1}\end{equation*}

\begin{equation*}\frac{C_p}{T^3}=\frac{c_1}{T^2}+c_3+\left(H_{\max}-c_3\right)\tfrac12\left[1+\cos\!\left(d\left(\frac{T}{T_{\max}}-1\right)\right)\right],\qquad T_{\max}\le T\le T_2.\tag{2}\end{equation*}
\end{quote}

The parameters c₁, c₃, Hₘₐₓ and Tₘₐₓ are determined from the measured data before the cosine functions are applied (Methods). The left branch therefore contains no adjustable shape parameter after the independent scaling parameters have been established. The right branch introduces one dimensionless shape parameter, d; T₂ denotes the upper limit over which that branch is fitted. The two branches join continuously at Tₘₐₓ and have zero slope there. Expanding Eq. (1) for T ≪ Tₘₐₓ gives Cₚ(T) = c₁T + c₃T\textsuperscript{3} + {[}(Hₘₐₓ − c₃)π\textsuperscript{2}/(4Tₘₐₓ\textsuperscript{2}){]}T\textsuperscript{5} + O(T\textsuperscript{7}). Thus, the GCC form introduces a T\textsuperscript{5} correction to the Debye heat capacity. A corresponding ω\textsuperscript{4} term in the low-frequency vibrational density of states can arise from quasi-localized excitations, but also from corrections to acoustic-phonon dispersion; the expansion alone does not identify its microscopic origin. The model reduces to c₁T + c₃T\textsuperscript{3} as T approaches zero. Near the maximum, the two branches have different curvatures, proportional to π\textsuperscript{2} and d\textsuperscript{2}, respectively; d = π gives a symmetric peak, while d \textless{} π gives a more gradual decay on the high-temperature side. A single Einstein oscillator also produces an asymmetric reduced-heat-capacity feature\textsuperscript{44,45}, so it should not be rejected on symmetry grounds. Its suitability must instead be assessed by comparing its fit quality with that of the GCC expression using the same data points, weighting and model-selection criteria. The full comparison is provided in Supplementary Note 1.

Figures 2 and 3 show Cₚ/T\textsuperscript{3} and (Cₚ − c₁T)/T\textsuperscript{3}, respectively; the parameters obtained from the GCC analysis are listed in Table 1. The fitted maxima occur at Tₘₐₓ ≈ 2.0--7.9 K, with Hₘₐₓ ≈ 3.3--7.1 mJ K⁻\textsuperscript{4} mol⁻\textsuperscript{1}. The excess amplitude relative to the Debye baseline, (Hₘₐₓ − c₃)/c₃, varies substantially: approximately 60\% for corannulene, 20--30\% for perylene, coronene and pyrene, and 7--10\% for tetracene and pentacene. Pentacene has the lowest fitted Tₘₐₓ, approximately 2.0 K (Fig. 3a). After subtracting both the linear and Debye contributions, the excess was normalized using the independently determined Tₘₐₓ and Hₘₐₓ values. For coronene, corannulene, perylene and pyrene, the normalized rising branches are consistent with Δ*(x) = ½{[}1 − cos(πx){]} for x ≤ 1, without readjusting the normalization. This comparison tests the shape of the rising branch; the amplitude and temperature scale are determined separately for each compound. For tetracene and pentacene, the excess is weaker and the normalized data are less discriminating; an independent automated refit also fails to recover the adopted maxima (Supplementary Table 1). Their peak positions should therefore be regarded as weakly resolved, and they do not provide an equally strong test of the scaling. The fitted Tₘₐₓ values span a factor of approximately four, while Hₘₐₓ varies by about a factor of two. The high-temperature decay is compound dependent. For the adopted fits, d ranges from 1.9 to 3.1, below π, indicating a more gradual decay than rise within this parameterization. Several mechanisms could produce such asymmetry, including changes in the vibrational density of states in soft-potential or heterogeneous-elasticity descriptions and the smoothing of van Hove features by acoustic--optical or acoustic--librational hybridization. These possibilities are not distinguished by the present heat-capacity data and should be regarded as candidate interpretations rather than established assignments.

\needspace{8\baselineskip}
\textbf{Table 1 \textbar{} Parameters of the General Cosine--Cosine description of the low-temperature heat capacity.}\par\nopagebreak
\begin{longtable}[]{@{}
  >{\raggedright\arraybackslash}p{(\columnwidth - 12\tabcolsep) * \real{0.2120}}
  >{\raggedright\arraybackslash}p{(\columnwidth - 12\tabcolsep) * \real{0.2001}}
  >{\raggedright\arraybackslash}p{(\columnwidth - 12\tabcolsep) * \real{0.1804}}
  >{\raggedright\arraybackslash}p{(\columnwidth - 12\tabcolsep) * \real{0.1804}}
  >{\raggedright\arraybackslash}p{(\columnwidth - 12\tabcolsep) * \real{0.0907}}
  >{\raggedright\arraybackslash}p{(\columnwidth - 12\tabcolsep) * \real{0.0606}}
  >{\raggedright\arraybackslash}p{(\columnwidth - 12\tabcolsep) * \real{0.0759}}@{}}
\toprule\noalign{}
\endhead
\bottomrule\noalign{}
\endlastfoot
Compound & c₁

(mJ K⁻\textsuperscript{2} mol⁻\textsuperscript{1}) & c₃

(mJ K⁻\textsuperscript{4} mol⁻\textsuperscript{1}) & H\textsubscript{max}

(mJ K⁻\textsuperscript{4} mol⁻\textsuperscript{1}) & T\textsubscript{max}

(K) & d & T₂

(K) \\
Tetracene (benz{[}b{]}anthracene) & 0.15 & 3.10 & 3.33 & 7.9 & 2.7 & 10.0 \\
Corannulene & 0.29 & 4.39 & 6.93 & 6.62 & 2.5 & 14.7 \\
Coronene & 0.39 & 4.19 & 5.45 & 7.14 & 3.1 & 13.7 \\
Pentacene & 0.34 & 5.0 & 5.49 & 2.0 & 1.9 & 5.0 \\
Perylene & 0.23 & 5.48 & 6.55 & 4.81 & 2.9 & 9.0 \\
Pyrene & 0.5 & 5.45 & 7.05 & 5.5 & 2.3 & 12.0 \\
\end{longtable}

The parameter uncertainties are estimated to be approximately ±10\% of the stated values; the uncertainty and stability of individual parameters are not uniform across compounds. The parameter d is dimensionless, with d = π corresponding to a symmetric peak; T₂ is the upper limit of the right-branch fit.

\begin{figure}[!htbp]
\centering
\begin{minipage}[t]{0.49\linewidth}\raggedright\textbf{a}\\\centering\includegraphics[width=0.95\linewidth]{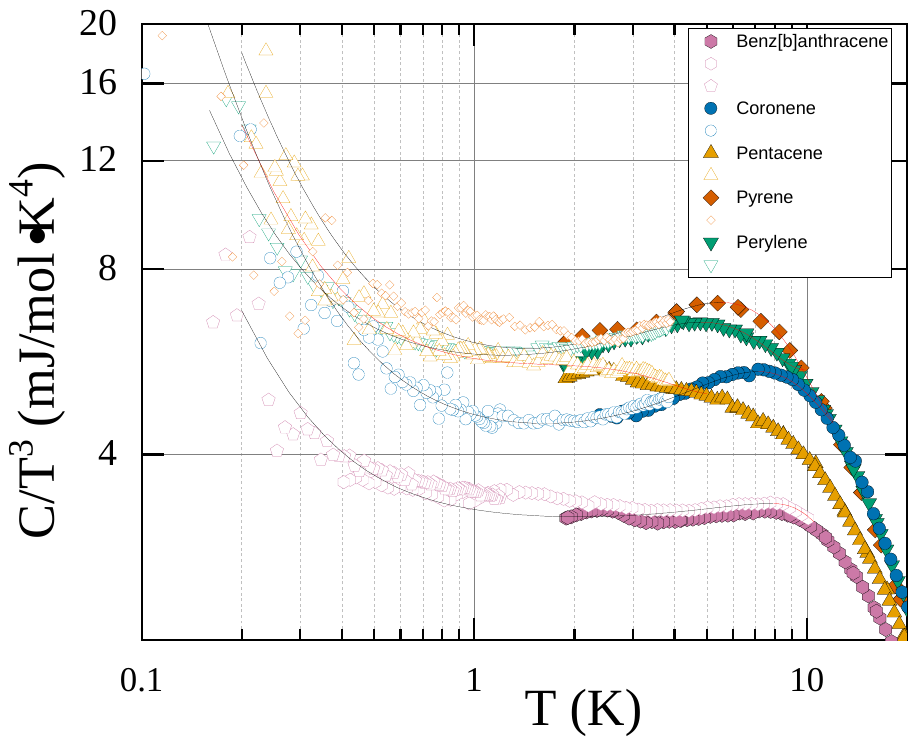}\end{minipage}\hfill\begin{minipage}[t]{0.49\linewidth}\raggedright\textbf{b}\\\centering\includegraphics[width=0.95\linewidth]{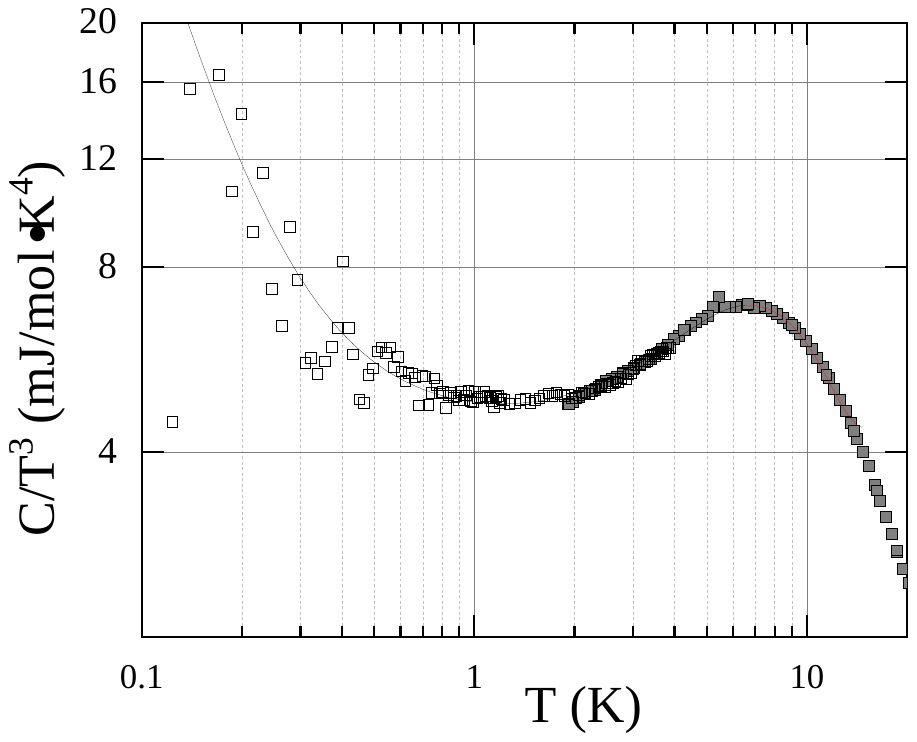}\end{minipage}

\medskip
\begin{minipage}{\linewidth}\small \textbf{Fig. 2 \textbar{}} Reduced heat capacity of the six PAH crystals. a, Cₚ/T\textsuperscript{3} versus temperature for pyrene, perylene, coronene, pentacene and tetracene. b, Corannulene, plotted separately for clarity. Open symbols denote measurements with the dilution insert below approximately 2 K; filled symbols denote PPMS measurements. Black curves show the left branch of Eq. (1), and red curves show the right branch of Eq. (2). The low-temperature upturn is associated with the fitted linear contribution c₁/T\textsuperscript{2}.\end{minipage}
\end{figure}

\begin{figure}[!htbp]
\centering
\begin{minipage}[t]{0.49\linewidth}\raggedright\textbf{a}\\\centering\includegraphics[width=0.95\linewidth]{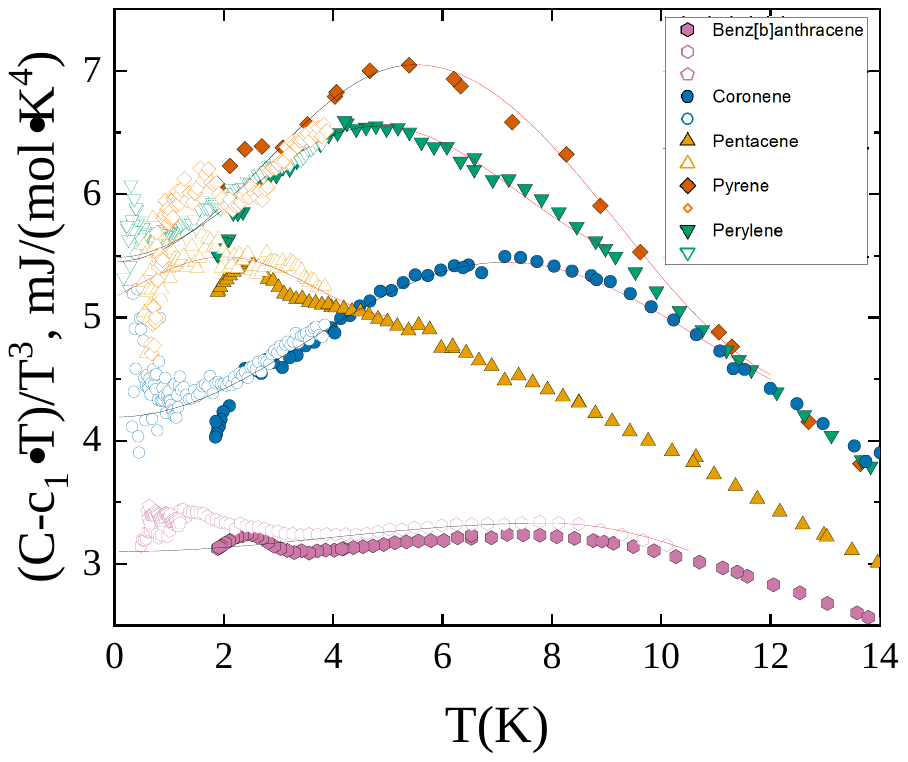}\end{minipage}\hfill\begin{minipage}[t]{0.49\linewidth}\raggedright\textbf{b}\\\centering\includegraphics[width=0.95\linewidth]{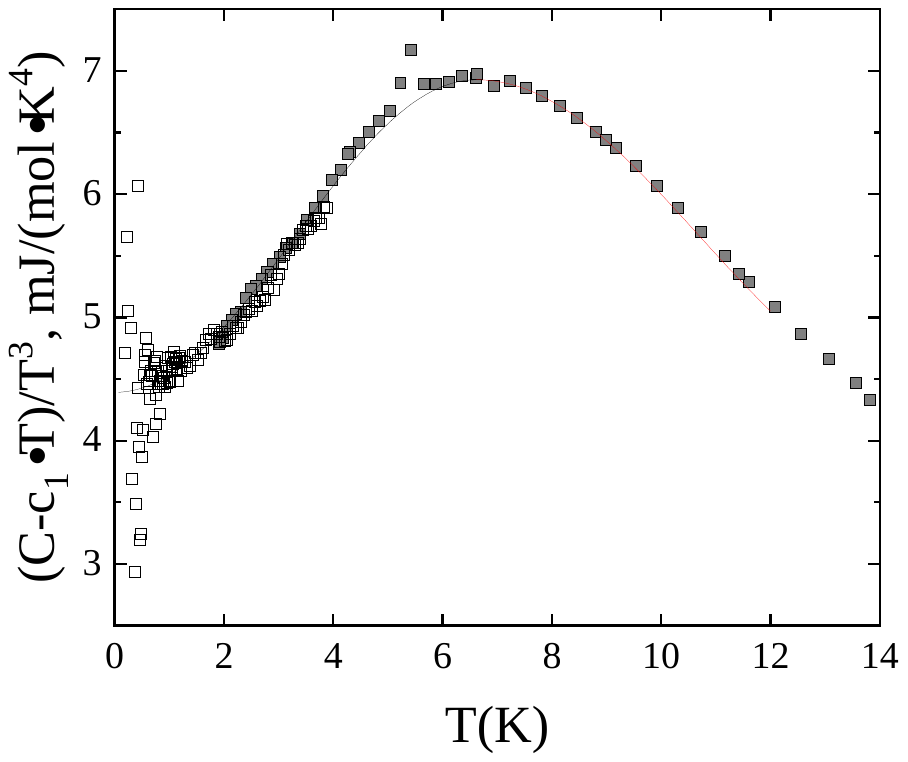}\end{minipage}

\medskip
\begin{minipage}{\linewidth}\small \textbf{Fig. 3 \textbar{}} Reduced heat capacity after subtraction of the linear term. a, (Cₚ − c₁T)/T\textsuperscript{3} versus temperature for pyrene, perylene, coronene, pentacene and tetracene. b, Corannulene. The plotted quantity retains the Debye contribution c₃; the excess above c₃ is Δ(T)/T\textsuperscript{3}. Symbols and fitted branches are as in Fig. 2.\end{minipage}
\end{figure}

\begin{figure}[!htbp]
\centering
\begin{minipage}[t]{0.49\linewidth}\raggedright\textbf{a}\\\centering\includegraphics[width=0.95\linewidth]{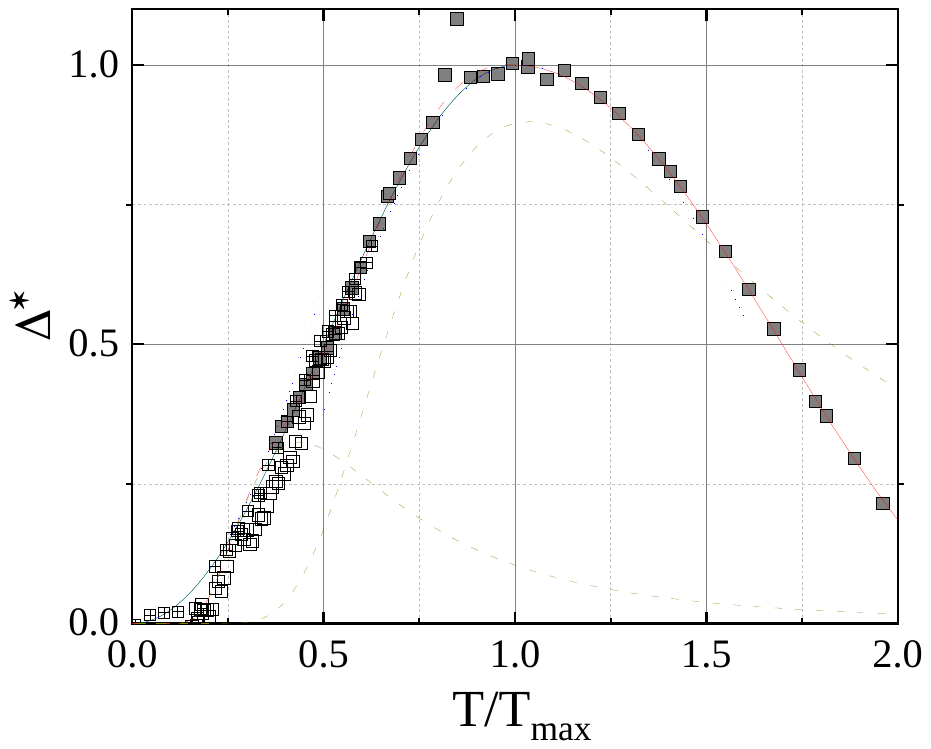}\end{minipage}\hfill\begin{minipage}[t]{0.49\linewidth}\raggedright\textbf{b}\\\centering\includegraphics[width=0.95\linewidth]{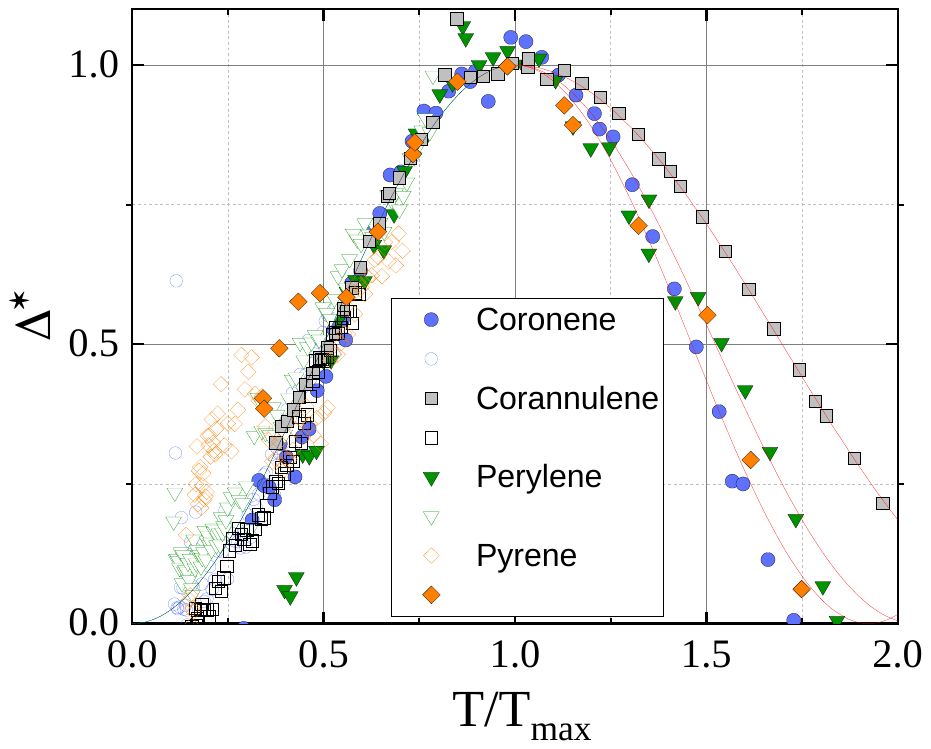}\end{minipage}

\medskip
\begin{minipage}{\linewidth}\small \textbf{Fig. 4 \textbar{}} Normalized heat-capacity excess. Δ*(x) versus x = T/Tₘₐₓ. a, Corannulene; black and red curves show the left and right GCC branches, respectively. Red dashed curves indicate the parabolic approximations near x = 1; dash-dotted curves show the two components of the two-oscillator Einstein fit. b, Coronene, corannulene, perylene and pyrene. The black curve is the raised-cosine function ½{[}1 − cos(πx){]}; red curves are the compound-specific right branches from Table 1. The rising branches are consistent with a common function within the resolution of the data, whereas the decays differ among compounds.\end{minipage}
\end{figure}

\subsection*{Lattice-phonon Raman spectra vary with symmetry, Z and rigidity}

Low-frequency Raman scattering probes rigid-body translations and librations of molecules in the crystal lattice, typically below about 200 cm⁻\textsuperscript{1}. It therefore provides complementary information on low-energy lattice vibrations, although the measured Raman response is restricted by selection rules, instrumental sensitivity and the finite frequency window. The Bose-corrected spectra are shown in Fig. 5, and the resolved band positions are summarized in Table 2. Coronene (D\textsubscript{6h}) exhibits two prominent bands in the measured 10--200 cm⁻\textsuperscript{1} range, at approximately 59 and 86 cm⁻\textsuperscript{1}, whereas tetracene and pentacene, which also have Z = 2, show six and five resolved bands, respectively. This difference does not imply that only two Raman-active lattice modes exist in coronene; other modes may be weak, overlapping or outside the measured window. The two Z = 4 compounds, perylene and pyrene, exhibit comparatively dense spectra, with seven resolved bands for pyrene. However, Z alone does not determine the number of observed Raman bands: Raman activity in a crystal depends on the factor group and molecular site symmetry, as well as on mode intensity and spectral resolution. Tetracene, coronene and pyrene show relatively narrow resolved features (FWHM approximately 4--13 cm⁻\textsuperscript{1}) on comparatively flat backgrounds. Pentacene and perylene also show narrow features but have stronger backgrounds that rise after Bose correction. Corannulene exhibits broader features (approximately 16--18 cm⁻\textsuperscript{1}) superimposed on a broad response. Its nearly frequency-independent raw intensity means that a rising reduced spectrum can result from the Bose correction of a constant background; it is therefore not, by itself, evidence for relaxational dynamics. Broadening may also reflect overlapping modes, inhomogeneity or the instrumental response rather than overdamping.

The observed spectra are broadly consistent with previous reports. For pentacene (P-1, Z = 2, with molecules on inversion centres), the crystal symmetry restricts the Raman-active external modes; five modes below 200 cm⁻\textsuperscript{1} have been reported for single crystals\textsuperscript{46,47}, and their frequencies depend on polymorph and measurement conditions\textsuperscript{48}. Lattice-dynamics calculations for β- and γ-coronene predict two dominant external modes in ranges that bracket the bands observed here\textsuperscript{49}. Dynamic bowl reorientation in corannulene has been discussed on the basis of crystallographic studies\textsuperscript{38,39}, but the room-temperature Raman spectrum measured here does not independently establish the orientational dynamics. The rising reduced background observed for pentacene could be related to anharmonic lattice dynamics\textsuperscript{31}; however, a frequency-independent fluorescence or instrumental background produces a similar trend after Bose correction. The present spectra therefore do not distinguish these explanations.

\begin{figure}[!htbp]
\centering
\includegraphics[width=\linewidth]{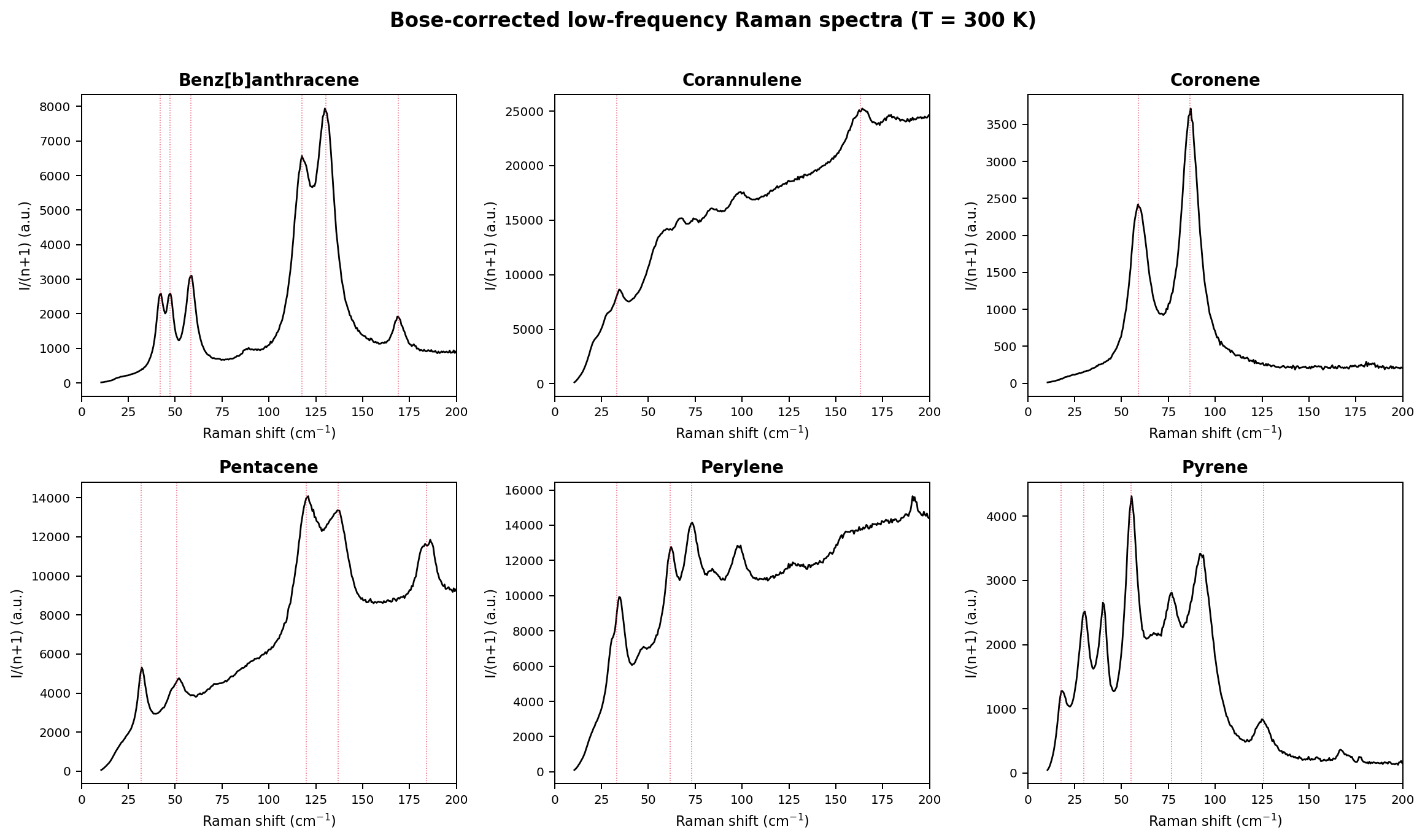}

\medskip
\begin{minipage}{\linewidth}\small \textbf{Fig. 5 \textbar{}} Low-frequency Raman spectra of the six PAH crystals at a nominal temperature of 300 K, after Bose correction. Dotted vertical lines indicate fitted band positions; positions for corannulene are approximate. Different vertical scales are used. No background was subtracted before Bose correction; consequently, a constant background in the raw spectrum appears as a frequency-dependent contribution in the reduced spectrum.\end{minipage}
\end{figure}

\needspace{8\baselineskip}
\textbf{Table 2 \textbar{} Crystal structure, molecular symmetry and external-mode Raman bands.}\par\nopagebreak
\begin{longtable}[]{@{}
  >{\raggedright\arraybackslash}p{(\columnwidth - 8\tabcolsep) * \real{0.2001}}
  >{\raggedright\arraybackslash}p{(\columnwidth - 8\tabcolsep) * \real{0.2000}}
  >{\raggedright\arraybackslash}p{(\columnwidth - 8\tabcolsep) * \real{0.2000}}
  >{\raggedright\arraybackslash}p{(\columnwidth - 8\tabcolsep) * \real{0.2000}}
  >{\raggedright\arraybackslash}p{(\columnwidth - 8\tabcolsep) * \real{0.2000}}@{}}
\toprule\noalign{}
\begin{minipage}[b]{\linewidth}\raggedright
Compound
\end{minipage} & \begin{minipage}[b]{\linewidth}\raggedright
Space group; Z (reference)
\end{minipage} & \begin{minipage}[b]{\linewidth}\raggedright
Molecular point group
\end{minipage} & \begin{minipage}[b]{\linewidth}\raggedright
Resolved Raman bands (cm⁻\textsuperscript{1})
\end{minipage} & \begin{minipage}[b]{\linewidth}\raggedright
Observed κ(T) behaviour
\end{minipage} \\
\midrule\noalign{}
\endhead
\bottomrule\noalign{}
\endlastfoot
Tetracene & P-1; Z = 2 (Ref. 50) & D\textsubscript{2h} & 42, 47, 58, 118, 130, 169 & Crystal-like \\
Pentacene & P-1; Z = 2 (Ref. 51) & D\textsubscript{2h} & 32, 51, 120, 137, 184 & Glass-like \\
Perylene & P2₁/a; Z = 4 (Ref. 52) & D₂\textsubscript{h} & 33, 62, 73 (weak: 99, 191) & Crystal-like \\
Pyrene & P2₁/a; Z = 4 (Ref. 53) & D₂\textsubscript{h} & 18, 30, 40, 55, 77, 93, 126 & Glass-like \\
Coronene & P2₁/a; Z = 2 (Ref. 54) & D₆\textsubscript{h} & 59, 86 & Crystal-like \\
Corannulene & P2₁/c; Z = 8 (Refs. 38, 39) & C₅\textsubscript{v} & \textasciitilde32, \textasciitilde163 (broad) & Glass-like \\
\end{longtable}

Space groups and Z values refer to published room-temperature crystal structures; they have not been independently verified for the pressed specimens used here. Molecular point groups refer to isolated molecules. Raman activity in the crystal is governed by the site symmetry and factor group, not by the isolated-molecule point group alone. The number of resolved bands is not necessarily the total number of Raman-active external modes.

\subsection*{Thermal conductivity from crystal-like to glass-like}

In well-ordered crystals, thermal conductivity commonly exhibits a pronounced phonon maximum that can be described within the Peierls--Boltzmann framework\textsuperscript{55}. Glasses often show a low-temperature power law, a plateau and a renewed rise at higher temperatures\textsuperscript{1,56}. Models of glassy transport invoke resonant scattering by tunnelling states and soft quasi-localized vibrations\textsuperscript{40,41,57--59}, as well as diffuson-mediated transport\textsuperscript{60}. The Wigner formulation provides a framework in which particle-like (Peierls) and wave-like coherence contributions can both enter the thermal conductivity\textsuperscript{61--63}. Following our previous phenomenological analyses\textsuperscript{64--68}, we use the expression

\begin{equation*}\kappa(T)=\left[\left(\beta T^{n}\right)^{-1}+\left(A/T+\kappa_{\mathrm{pl}}\right)^{-1}\right]^{-1}+\kappa_0\exp(-E/T).\tag{3}\end{equation*}

The first term in Eq. (3) is a phenomenological Peierls-like contribution in which the low-temperature power-law channel, βTⁿ, is combined with the higher-temperature term A/T + κₚₗ through an inverse-additivity (Matthiessen-type) construction\textsuperscript{69}. The second term is an activated contribution with characteristic temperature E (so E is expressed in kelvin in this parameterization). Following Refs. 64--68, we discuss it as a possible analogue of the wave-like coherence contribution in the Wigner formulation. This assignment is tentative: a successful fit to κ(T) does not, on its own, establish the microscopic origin of the activated term.

Figure 6 shows κ(T) measured between 4.5 and 320 K. Tetracene, perylene and coronene exhibit pronounced maxima and are described here as crystal-like; pyrene, pentacene and corannulene show lower conductivities and broad plateaus and are described as glass-like. These labels refer to the observed shapes of κ(T), not to a claim that the latter compounds are amorphous. Equation (3) describes all six data sets over the measured range (Supplementary Table 5). Below 8 K, the fitted power-law exponents are n = 1.0--1.73. For the five planar PAHs, the fitted activated term contributes approximately 12--46\% of the total thermal conductivity at 300 K: 12\% for pentacene, 18\% for tetracene, 36\% for perylene, 41\% for coronene and 46\% for pyrene. Thus, the activated term is appreciable at high temperature but does not exceed the Peierls-like term at 300 K for any of these compounds. The pyrene plateau occurs at a higher temperature and is broader than the plateau reported for the amorphous polymers considered in Ref. 66.

Corannulene, the only compound in this series for which orientational disorder has been reported, exhibits a weak low-temperature power law (n ≈ 1.0) followed by a plateau near 0.3 W m⁻\textsuperscript{1} K⁻\textsuperscript{1} above approximately 40 K. The plateau persists to room temperature, and no separate activated contribution is resolved in the present data. The reported upper bound κ₀ \textless{} 0.01 W m⁻\textsuperscript{1} K⁻\textsuperscript{1} should be understood as conditional on the assumed fitting model and parameter range, rather than as a model-independent detection limit. The measured data are described by the first term of Eq. (3), with A = 9.1 W m⁻\textsuperscript{1} and κₚₗ = 0.31 W m⁻\textsuperscript{1} K⁻\textsuperscript{1} (Supplementary Table 5).

The fitted parameters κ₀ and E for the five compounds with a resolved activated term show an apparent Meyer--Neldel-type compensation relation: ln κ₀ varies approximately linearly with E (Supplementary Fig. 12)\textsuperscript{70}. Because both parameters are obtained from the same nonlinear fit and may be strongly correlated, this relation could partly reflect parameter covariance and the finite temperature range. The fitted κ\textsubscript{C}(T) curves pass near 0.01 W m⁻\textsuperscript{1} K⁻\textsuperscript{1} at a characteristic temperature T\textsubscript{MN} ≈ 144 K. A related compensation relation has been reported for amorphous polymers\textsuperscript{66}. The similarity of the convergence thermal conductivity is noteworthy, but the difference in characteristic temperature and the small number of compounds preclude a strong universality claim. A joint uncertainty analysis of (E, ln κ₀) is needed to establish the robustness of this trend.

\begin{figure}[!htbp]
\centering
\includegraphics[width=\linewidth]{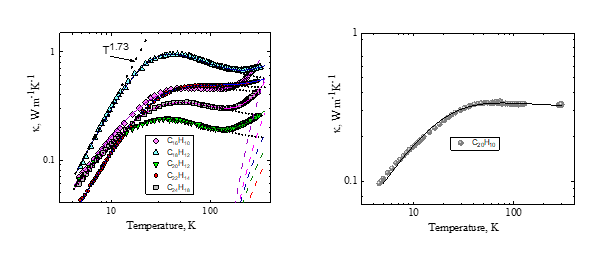}

\medskip
\begin{minipage}{\linewidth}\small \textbf{Fig. 6 \textbar{}} Thermal conductivity of the six PAH crystals. a, κ(T) for pyrene, tetracene, perylene, pentacene and coronene. The black straight line indicates the low-temperature power law κ ∝ T\textsuperscript{1}·\textsuperscript{73} for tetracene. Black dotted curves show the first (Peierls-like) term of Eq. (3), coloured dashed curves show the activated term, and coloured solid curves show the total fits. b, κ(T) for corannulene; the solid curve shows the total thermal conductivity. The estimated measurement uncertainty is below 15\%.\end{minipage}
\end{figure}

\subsection*{Activation energy and the Raman lattice-mode centroid: a preliminary observation}

For the five compounds with a fitted activated contribution, we examined whether the characteristic temperature E correlates with descriptors of the low-frequency Raman spectrum. The highest-frequency band, strongest band, number of resolved bands and unweighted mean band position showed no clear correlation (\textbar r\textbar{} \textless{} 0.5). The intensity-weighted centroid of the Bose-corrected spectrum, ⟨\ensuremath{\tilde{\omega}}⟩ = ΣᵢRᵢ\ensuremath{\tilde{\omega}}ᵢ/ΣᵢRᵢ, calculated without subtracting a background, showed a strong negative Pearson correlation with E (r = −0.99; Fig. 7). However, the result is not robust to the background treatment: subtracting a constant background estimated from the raw spectra reduces the correlation to approximately r = −0.63 (p ≈ 0.25). Changing the upper frequency limit also changes the correlation, and repeat scans show centroid variations of up to approximately 10 cm⁻\textsuperscript{1} for some compounds (Supplementary Note 5). Moreover, Bose correction transforms a frequency-independent raw background B into a frequency-dependent term B{[}1 − exp(−hc\ensuremath{\tilde{\omega}}/k\textsubscript{B}T){]}. On the stated frequency grid at 300 K, this background alone has a centroid of approximately 129 cm⁻\textsuperscript{1}, close to the reported pentacene value. The apparent correlation is therefore sensitive to a processing choice and cannot presently be interpreted as evidence that E is determined by the lattice-mode spectrum. We retain it only as a diagnostic observation and do not use it to support the physical interpretation.

\begin{figure}[!htbp]
\centering
\includegraphics[width=0.6\linewidth]{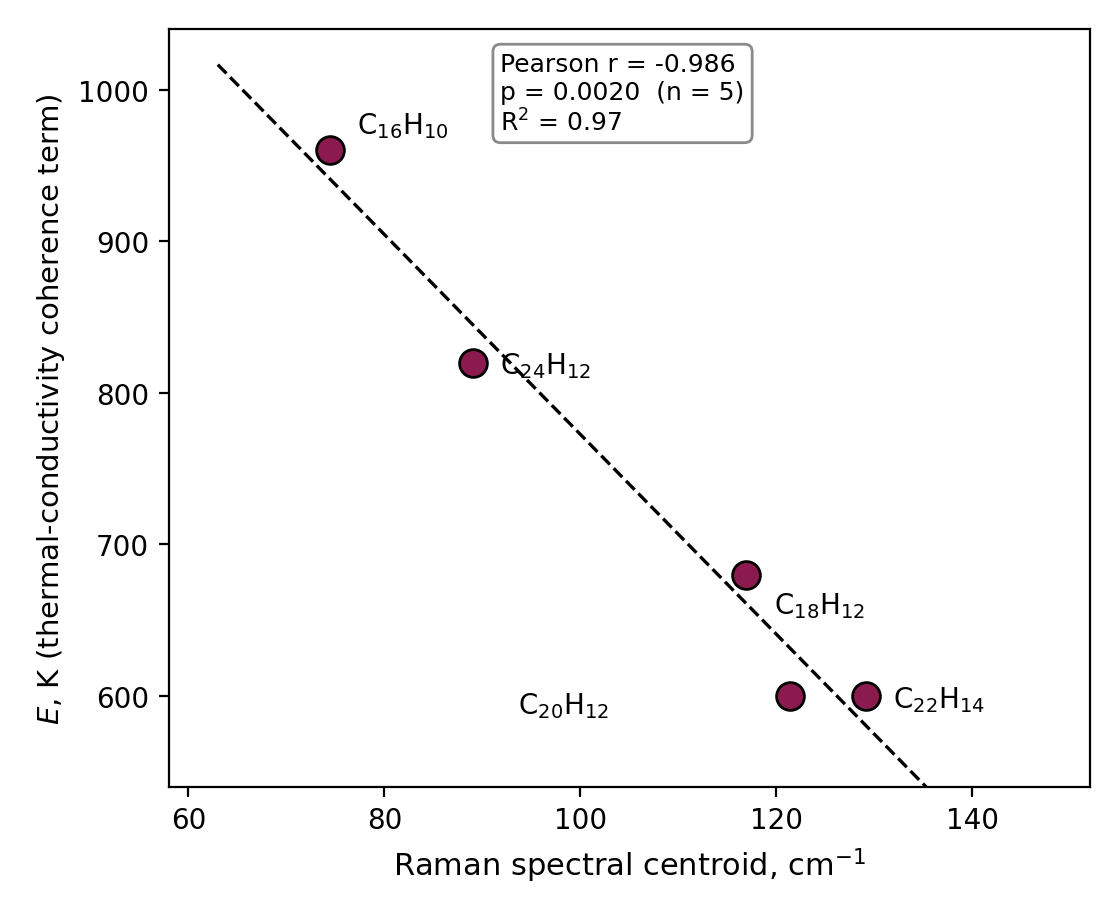}

\medskip
\begin{minipage}{\linewidth}\small \textbf{Fig. 7 \textbar{}} Preliminary comparison of the fitted activation temperature E with the intensity-weighted centroid of the Bose-corrected Raman spectrum, calculated without background subtraction, for the five compounds with a resolved activated contribution. The dashed line is a linear regression (R\textsuperscript{2} = 0.97). Because the centroid is sensitive to the spectral background, this correlation is descriptive and should not be interpreted as an established physical relation.\end{minipage}
\end{figure}

\section*{Discussion}

The measurements reveal related but not identical trends in the heat capacity, Raman response and thermal conductivity of the six PAH crystals. The reduced heat capacity exhibits a low-temperature maximum in each compound, although the maxima of tetracene and pentacene are weakly resolved. For the four compounds with clearer maxima, the normalized rising branch is consistent with the same raised-cosine function, while the decay varies among compounds. The Raman spectra contain different numbers and widths of resolved low-frequency bands, and the thermal conductivity ranges from pronounced phonon maxima to broad glass-like plateaus. Taken together, these observations extend previous reports of glass-like thermal anomalies in ordered or minimally disordered molecular crystals\textsuperscript{20--22} to a comparative series of PAHs. They establish correlations among molecular structure, low-energy vibrational response and thermal transport, but do not isolate a single structural variable as the cause.

No single molecular descriptor separates the two transport classes. Coronene (D\textsubscript{6h}) and tetracene and perylene (D\textsubscript{2h}) show crystal-like thermal conductivity maxima, whereas pyrene and pentacene (D\textsubscript{2h}) and corannulene (C\textsubscript{5v}) show glass-like plateaus. Molecular symmetry alone is therefore insufficient. Aspect ratio is also insufficient: tetracene and pentacene, which belong to the linear-acene family, exhibit markedly different thermal conductivity curves. The observed Raman systematics likewise depend on more than one factor, including molecular symmetry, crystal packing, site symmetry, the number of molecules per unit cell and the relative intensities and widths of the modes. The data are consistent with a multi-parameter structure--property relationship, as anticipated by heterogeneous-elasticity and quasi-localized-mode descriptions\textsuperscript{7},\textsuperscript{9},\textsuperscript{13}, but further structural characterization is needed to identify the relevant microscopic variables.

The magnitude of the heat-capacity excess does not uniquely predict the thermal-transport class. Pyrene and pentacene both show glass-like thermal conductivity, although the excess relative to c₃ is larger for pyrene and the pentacene maximum occurs at the lowest temperature. Among the crystal-like conductors, coronene and perylene have excess amplitudes comparable to pyrene, whereas tetracene---the best conductor in this series---has the weakest excess. Thus, the heat-capacity maximum and the thermal conductivity plateau are related empirical features, not equivalent observables. Corannulene combines the largest relative heat-capacity excess, a broad Raman response and glass-like thermal conductivity, and it is the only compound for which no activated term is resolved. Because corannulene also differs in molecular curvature and orientational behaviour, the present comparison cannot attribute these differences uniquely to orientational disorder.

The GCC parameters may be discussed in relation to the ITLS framework\textsuperscript{19}, but this connection is presently qualitative. In that framework, localized soft excitations contribute at lower energies, while hybridized excitations couple local configurations to elastic vibrations at higher energies. One possible interpretation is that the common rising branch reflects a similar low-energy onset across the compounds, whereas the compound-dependent parameter d captures differences in the high-temperature decay. However, the GCC parameters have not been derived from the ITLS Hamiltonian, and the present PAH data have not been fitted directly with the ITLS model. Numerical comparisons reported for another molecular system do not establish a quantitative mapping for these PAHs. The ITLS discussion should therefore be treated as a hypothesis for future testing, not as a microscopic derivation of the GCC parameters.

The GCC expression provides a compact empirical description of the resolved heat-capacity excess, but its universality is limited by the weakly resolved maxima in tetracene and pentacene and by the instability of some parameters under independent refitting. The asymmetry parameter d may be relevant to soft-potential, heterogeneous-elasticity or avoided-crossing scenarios\textsuperscript{5},\textsuperscript{8},\textsuperscript{24}, but none has been tested quantitatively against the present data. Several limitations should be emphasized. First, first-principles phonon densities of states and lattice-dynamics calculations are not available for the measured specimens. Second, the Raman measurements were performed at a nominal room temperature and probe zone-centre modes only; the spectral background was not independently resolved, and the sample temperature under laser illumination was not measured. Third, the correlation between the fitted activation temperature and the Raman centroid is based on five compounds and is strongly background dependent. Fourth, the specimens were pressed at approximately 1 GPa, but sample-specific PXRD, texture and porosity data are not reported here; the literature crystal structures therefore do not establish the structure of the measured pellets. Future work should combine sample-specific diffraction and density measurements with lattice-dynamics calculations, low-temperature terahertz or inelastic-neutron spectroscopy, and controlled comparisons of polymorphs or annealed specimens of the same compound.

The results support the broader view that boson-peak-like anomalies are not unique signatures of amorphousness and may have different microscopic realizations in disordered and ordered solids\textsuperscript{2,3,10,11}. For PAH-based organic semiconductors, low-frequency lattice vibrations can influence both heat transport and intermolecular electronic coupling\textsuperscript{28,30}. The present structural correlations motivate further investigation of molecular design as a route to controlling vibrational and thermal properties, but a direct link to device performance remains to be established.

\subsection*{Conclusion}

We have investigated the low-temperature vibrational and thermal-transport properties of six polycyclic aromatic hydrocarbon crystals---tetracene, pentacene, pyrene, perylene, coronene and corannulene---using heat-capacity measurements, low-frequency Raman spectroscopy and thermal conductivity measurements. Although these compounds share predominantly van der Waals intermolecular interactions, their molecular shapes, symmetries and crystal packings differ substantially. This series therefore provides an opportunity to examine how molecular structure influences boson-peak-like anomalies in translationally ordered molecular solids.

All six compounds exhibit a maximum in the reduced heat capacity, C\textsubscript{p}/T\textsuperscript{3}, in the temperature range of approximately 2--8 K. The General Cosine--Cosine (GCC) model provides a compact phenomenological description of these anomalies. For the four compounds with the most clearly resolved excess---coronene, corannulene, perylene and pyrene---the rising branches are consistent with a common raised-cosine function after independent normalization by their characteristic temperatures and excess amplitudes. In contrast, the high-temperature decay is compound-dependent and is described by the asymmetry parameter \emph{d}. The weak excesses in tetracene and pentacene do not permit an equally stringent test of the proposed scaling. Thus, the results support a common empirical description of the rising side of the anomaly in several members of the series, but do not establish a universal microscopic mechanism.

Low-frequency Raman measurements reveal substantial differences in the number, positions and widths of the resolved external-mode bands. These differences are consistent with the combined influence of molecular symmetry, the number of molecules per unit cell and molecular rigidity. The Raman spectra provide complementary information on the lattice vibrations, although their interpretation is limited by the absence of an independently resolved spectral background and by the fact that the measurements were performed at room temperature, well above the temperatures of the heat-capacity maxima.

The thermal conductivity exhibits a pronounced variation across the series, ranging from crystal-like behaviour with a distinct phonon maximum in tetracene, perylene and coronene to glass-like behaviour with a broad plateau in pentacene, pyrene and corannulene. The phenomenological model combining a Peierls-type contribution with an activated term describes the measured temperature dependences. For the five planar PAHs, the activated term contributes approximately 12--46\% of the total thermal conductivity at 300 K, whereas no corresponding contribution is resolved for corannulene within the sensitivity of the measurements. This contrast is noteworthy, particularly because corannulene is also the compound with documented orientational disorder. Nevertheless, the activated contribution should not be regarded as direct experimental proof of a particular microscopic coherence mechanism.

Taken together, the results demonstrate that boson-peak-like heat-capacity anomalies and glass-like thermal transport can occur in molecular crystals without conventional amorphous structural disorder. Their characteristics are not governed by molecular symmetry or shape alone, but appear to depend on several coupled structural factors, including molecular geometry, packing and lattice dynamics. The observed correlations between the heat capacity, Raman and thermal-transport characteristics provide a basis for further investigation, rather than a definitive assignment of the underlying vibrational mechanisms.

A more microscopic understanding will require lattice-dynamics calculations for the actual crystal structures, together with measurements that directly probe the low-energy vibrational density of states. In particular, low-temperature Raman or terahertz spectroscopy, inelastic neutron scattering and sample-specific structural characterization would help distinguish the roles of acoustic--optical mode hybridization, quasi-localized excitations and orientational disorder. Extending the comparison to a broader family of molecular crystals should also establish whether the empirical scaling and transport trends identified here are more general. These results suggest that molecular shape and crystal packing offer useful parameters for understanding---and potentially tailoring---the vibrational and thermal properties of organic molecular solids.

\section*{Methods}

\subsection*{Materials and sample preparation}

Commercial PAHs were used as received: coronene (99.99\%, CAS 191-07-1), perylene (≥99.0\%, CAS 198-55-0), pyrene (99.0\%, CAS 129-00-0), pentacene (99.0\%, CAS 135-48-8), tetracene (benz{[}b{]}anthracene; 99.99\% trace-metals basis, CAS 92-24-0) and corannulene (95.0\%, CAS 5821-51-2). Supplier names and catalogue numbers are provided in Supplementary Table 2. For heat-capacity measurements, powders were pressed at room temperature into cylindrical pellets approximately 3 mm in diameter and 2 mm high, with masses of 5--14 mg, at a nominal pressure of approximately 1 GPa. For thermal conductivity measurements, pellets of the same diameter and up to 10 mm in length were prepared under the same nominal pressure.

\subsection*{Heat capacity}

Heat capacity was measured by the thermal-relaxation method using a Physical Property Measurement System (PPMS, Quantum Design) equipped with the Heat Capacity option. Measurements covered 2--300 K; measurements below 2 K, down to 0.05 K, were performed using a \textsuperscript{3}He--\textsuperscript{4}He dilution insert. Apiezon N grease was used to provide thermal contact between the sample and the measurement platform. The stated measurement uncertainty was below 2\% below 50 K and below 5\% above 100 K. Data were processed using the standard PPMS analysis procedure.

\subsection*{Thermal conductivity}

Thermal conductivity was measured from 4.5 to 320 K using a steady-state axial heat-flow method in a custom cryostat\textsuperscript{71}. The sample was mounted under high vacuum and surrounded by temperature-controlled radiation shields. A miniature resistive heater established the heat flow, and the temperature difference was measured with calibrated constantan--manganin differential thermocouples separated by approximately 10 mm. The total estimated uncertainty was below 15\%, including an approximately 5\% contribution from uncertainty in the thermocouple separation. The scatter between repeated runs was below 2\%. The samples were 15 mm long and 5 mm in diameter. Thermal contact between the thermocouples, the heater and the sample was ensured by special glue (see Ref. 71). Heat loss is suppressed using a set of temperature-controlled thermal shields (see Ref. 71).

\subsection*{Raman spectroscopy}

Room-temperature Raman spectra were recorded using a Renishaw inVia microscope (Leica DM 2500) with 830 nm excitation, a 20× objective and a thermoelectrically cooled CCD detector. Spectra were acquired over 10--1800 cm⁻\textsuperscript{1}; the analysis reported here uses the 10.5--200.3 cm⁻\textsuperscript{1} interval sampled on a common 341-point grid. The laser power at the sample was 10 mW. Each spectrum comprised five 10 s exposures, and several positions were measured on each sample. No visible laser-induced changes were observed; however, the sample temperature under illumination was not measured independently. Three scans per compound were averaged and converted to reduced spectra R(\ensuremath{\tilde{\omega}}) = I(\ensuremath{\tilde{\omega}})/{[}n(\ensuremath{\tilde{\omega}},T) + 1{]}, where n is the Bose--Einstein occupation factor evaluated at the nominal temperature T = 300 K. No background was subtracted before this conversion. One perylene scan with a large approximately frequency-independent fluorescence offset was excluded; the exclusion criterion and the effect of alternative baseline treatments are described in Supplementary Notes 3 and 5. Bands were identified using a prominence-based peak search and fitted with Lorentzian profiles on local linear backgrounds (Supplementary Note 3). Because the baseline is not independently resolved, band positions and especially spectral centroids should be interpreted with this limitation in mind.

\subsection*{Data analysis}

The heat-capacity analysis was performed in stages. First, c₁ and c₃ were estimated by fitting Cₚ/T = c₁ + c₃T\textsuperscript{2} + bT\textsuperscript{4} over the selected lowest-temperature interval. Next, Tₘₐₓ was identified as the maximum of a LOESS-smoothed Cₚ/T\textsuperscript{3} − c₁/T\textsuperscript{2} curve, and Hₘₐₓ was defined as the value of that background-corrected curve at Tₘₐₓ. The normalized excess was then calculated using these fixed parameters and compared with the raised-cosine function without readjusting its amplitude or temperature scale; uncertainties in c₁ and c₃ were propagated into the normalized curves. The right-branch parameter d was fitted separately (Supplementary Note 1). For model comparison, the GCC and Einstein-oscillator models were fitted to the same data points with the same relative-residual weighting. Fit quality was compared using rms deviation, AIC and BIC. Equation (3) was fitted to the thermal conductivity data by nonlinear least squares over the measured temperature range. Correlations between E and Raman descriptors were evaluated using Pearson coefficients, leave-one-out analysis and an exact permutation test over all 5! permutations. Their sensitivity to baseline subtraction, the upper frequency limit and individual repeat scans was also examined (Supplementary Note 5).

\section*{References}\small

\begin{enumerate}
\def\labelenumi{\arabic{enumi}.}
\tightlist
\item
  Zeller, R. C. \& Pohl, R. O. Thermal conductivity and specific heat of noncrystalline solids. \emph{Phys. Rev.~B} \textbf{4}, 2029--2041 (1971).
\item
  Ramos, M. A. (ed.) \emph{Low-Temperature Thermal and Vibrational Properties of Disordered Solids: A Half-Century of Universal ``Anomalies'' of Glasses} (World Scientific, 2022).
\item
  Chumakov, A. I. et al.~Equivalence of the boson peak in glasses to the transverse acoustic van Hove singularity in crystals. \emph{Phys. Rev.~Lett.} \textbf{106}, 225501 (2011).
\item
  González-Jiménez, M. et al.~Understanding the emergence of the boson peak in molecular glasses. \emph{Nat. Commun.} \textbf{14}, 215 (2023).
\item
  Gurevich, V. L., Parshin, D. A. \& Schober, H. R. Anharmonicity, vibrational instability, and the Boson peak in glasses. \emph{Phys. Rev.~B} \textbf{67}, 094203 (2003).
\item
  Schirmacher, W., Ruocco, G. \& Scopigno, T. Acoustic attenuation in glasses and its relation with the boson peak. \emph{Phys. Rev.~Lett.} \textbf{98}, 025501 (2007).
\item
  Marruzzo, A., Schirmacher, W., Fratalocchi, A. \& Ruocco, G. Heterogeneous shear elasticity of glasses: the origin of the boson peak. \emph{Sci. Rep.} \textbf{3}, 1407 (2013).
\item
  Schirmacher, W. \& Ruocco, G. in \emph{Low-Temperature Thermal and Vibrational Properties of Disordered Solids} (ed.~Ramos, M. A.) Ch. 9 (World Scientific, 2022). Preprint at https://arxiv.org/abs/2009.05970.
\item
  Mizuno, H., Mori, T., Baldi, G. \& Minamitani, E. Boson peak in covalent network glasses: isostaticity and marginal stability. \emph{Proc. Natl Acad. Sci. USA} \textbf{123}, e2528998123 (2026).
\item
  Hu, Y.-C. \& Tanaka, H. Origin of the boson peak in amorphous solids. \emph{Nat. Phys.} \textbf{18}, 669--677 (2022).
\item
  Xi, Q., Wang, Y. \& Tanaka, H. String-sliding vibrational modes govern the boson peak and phonon anomalies in amorphous materials. \emph{Nat. Mater.} \textbf{25}, 1219--1229 (2026).
\item
  Lerner, E. \& Bouchbinder, E. Low-energy quasilocalized excitations in structural glasses. \emph{J. Chem. Phys.} \textbf{155}, 200901 (2021).
\item
  Lerner, E. \& Bouchbinder, E. Boson-peak vibrational modes in glasses feature hybridized phononic and quasilocalized excitations. \emph{J. Chem. Phys.} \textbf{158}, 194503 (2023).
\item
  Ding, G. et al.~Unified theory of phonon in solids with phase diagram of non-Debye anomalies. \emph{Nat. Phys.} \textbf{21}, 1911--1919 (2025).
\item
  Pérez-Castañeda, T., Rodríguez-Tinoco, C., Rodríguez-Viejo, J. \& Ramos, M. A. Suppression of tunneling two-level systems in ultrastable glasses of indomethacin. \emph{Proc. Natl Acad. Sci. USA} \textbf{111}, 11275--11280 (2014).
\item
  Pérez-Castañeda, T., Jiménez-Riobóo, R. J. \& Ramos, M. A. Two-level systems and boson peak remain stable in 110-million-year-old amber glass. \emph{Phys. Rev.~Lett.} \textbf{112}, 165901 (2014).
\item
  Festi, I. et al.~Effect of glass stability on the low frequency vibrations of vapor deposited glasses. \emph{Phys. Rev.~X} \textbf{16}, 021021 (2026).
\item
  Leggett, A. J. \& Vural, D. C. ``Tunneling two-level systems'' model of the low-temperature properties of glasses: are ``smoking-gun'' tests possible? \emph{J. Phys. Chem. B} \textbf{117}, 12966--12971 (2013).
\item
  Santamaría-Holek, I. \& Pérez-Madrid, A. Boson peak and scaling of vibrational coherence in glasses from an interacting two-level system framework. \emph{Phys. Rev. B} \textbf{114}, 204204 (2026).
\item
  Gebbia, J. F. et al.~Glassy anomalies in the low-temperature thermal properties of a minimally disordered crystalline solid. \emph{Phys. Rev.~Lett.} \textbf{119}, 215506 (2017).
\item
  Krivchikov, A. I. et al.~Role of optical phonons and anharmonicity in the appearance of the heat capacity boson peak-like anomaly in fully ordered molecular crystals. \emph{J. Phys. Chem. Lett.} \textbf{13}, 5061--5067 (2022).
\item
  Szewczyk, D. et al.~Specific heat at low temperatures in quasiplanar molecular crystals: origin of glassy anomalies in minimally disordered crystals. \emph{Phys. Rev.~B} \textbf{110}, 174204 (2024).
\item
  Baggioli, M. \& Zaccone, A. Universal origin of boson peak vibrational anomalies in ordered crystals and in amorphous materials. \emph{Phys. Rev.~Lett.} \textbf{122}, 145501 (2019).
\item
  Baggioli, M. \& Zaccone, A. Low-energy optical phonons induce glassy-like vibrational and thermal anomalies in ordered crystals. \emph{J. Phys. Mater.} \textbf{3}, 015004 (2020).
\item
  Anthony, J. E. Functionalized acenes and heteroacenes for organic electronics. \emph{Chem. Rev.} \textbf{106}, 5028--5048 (2006).
\item
  Coropceanu, V. et al.~Charge transport in organic semiconductors. \emph{Chem. Rev.} \textbf{107}, 926--952 (2007).
\item
  Smith, M. B. \& Michl, J. Singlet fission. \emph{Chem. Rev.} \textbf{110}, 6891--6936 (2010).
\item
  Troisi, A. \& Orlandi, G. Charge-transport regime of crystalline organic semiconductors: diffusion limited by thermal off-diagonal electronic disorder. \emph{Phys. Rev.~Lett.} \textbf{96}, 086601 (2006).
\item
  Fratini, S., Mayou, D. \& Ciuchi, S. The transient localization scenario for charge transport in crystalline organic materials. \emph{Adv. Funct. Mater.} \textbf{26}, 2292--2315 (2016).
\item
  Schweicher, G. et al.~Chasing the ``killer'' phonon mode for the rational design of low-disorder, high-mobility molecular semiconductors. \emph{Adv. Mater.} \textbf{31}, 1902407 (2019).
\item
  Asher, M. et al.~Anharmonic lattice vibrations in small-molecule organic semiconductors. \emph{Adv. Mater.} \textbf{32}, 1908028 (2020).
\item
  Kim, N. et al.~Thermal transport properties of thin films of small molecule organic semiconductors. \emph{Appl. Phys. Lett.} \textbf{87}, 241908 (2005).
\item
  Wang, D., Tang, L., Long, M. \& Shuai, Z. Anisotropic thermal transport in organic molecular crystals from nonequilibrium molecular dynamics simulations. \emph{J. Phys. Chem. C} \textbf{115}, 5940--5946 (2011).
\item
  Zhang, H. \& Brill, J. W. Interlayer thermal conductivity of rubrene measured by ac-calorimetry. \emph{J. Appl. Phys.} \textbf{114}, 043508 (2013).
\item
  Russ, B., Glaudell, A., Urban, J. J., Chabinyc, M. L. \& Segalman, R. A. Organic thermoelectric materials for energy harvesting and temperature control. \emph{Nat. Rev.~Mater.} \textbf{1}, 16050 (2016).
\item
  Roux, M. V., Temprado, M., Chickos, J. S. \& Nagano, Y. Critically evaluated thermochemical properties of polycyclic aromatic hydrocarbons. \emph{J. Phys. Chem. Ref. Data} \textbf{37}, 1855--1996 (2008).
\item
  Tielens, A. G. G. M. Interstellar polycyclic aromatic hydrocarbon molecules. \emph{Annu. Rev.~Astron. Astrophys.} \textbf{46}, 289--337 (2008).
\item
  Hanson, J. C. \& Nordman, C. E. The crystal and molecular structure of corannulene, C20H10. \emph{Acta Crystallogr. B} \textbf{32}, 1147--1153 (1976).
\item
  Gaboardi, M. et al.~Curvature-induced vitrification and polymorphism in corannulene. \emph{Commun. Chem.} \textbf{9}, 173 (2026).
\item
  Anderson, P. W., Halperin, B. I. \& Varma, C. M. Anomalous low-temperature thermal properties of glasses and spin glasses. \emph{Phil. Mag.} \textbf{25}, 1--9 (1972).
\item
  Phillips, W. A. Tunneling states in amorphous solids. \emph{J. Low Temp. Phys.} \textbf{7}, 351--360 (1972).
\item
  Barabashko, M., Jeżowski, A. \& Krivchikov, A. Universal empirical scaling of low-temperature heat capacity van Hove signature in classical and quantum cryocrystals. \emph{Sci. Rep.} \textbf{16}, 12395 (2026).
\item
  Miyazaki, Y. et al. Low-temperature heat capacity anomalies in ordered and disordered phases of normal and deuterated thiophene. \emph{J. Phys. Chem. Lett.} \textbf{12}, 2112--2117 (2021).
\item
  Zhou, Z. et al.~On the existence of Einstein oscillators and thermal conductivity in bulk metallic glass. \emph{Appl. Phys. Lett.} \textbf{89}, 031924 (2006).
\item
  Gamsjäger, E. \& Wiessner, M. Extended regression analysis for Debye--Einstein models describing low temperature heat capacity data of solids. \emph{Entropy} \textbf{26}, 452 (2024).
\item
  Gnoli, L., Venuti, E., Salzillo, T., Masino, M. \& Graziosi, P. Electrons and phonons in pentacene: coupling patterns reveal the microscopic origin of the phonon-limited mobility. \emph{J. Phys. Chem. C} \textbf{129}, 21738 (2025).
\item
  Lynch, P. G., Baughman, S., Mihm, T. N., Sharifzadeh, S. \& Frontiera, R. R. Identifying driving and spectator phonon modes in pentacene exciton transport. \emph{J. Am. Chem. Soc.} \textbf{147}, 23705--23714 (2025).
\item
  Brillante, A., Bilotti, I., Della Valle, R. G., Venuti, E. \& Girlando, A. Probing polymorphs of organic semiconductors by lattice phonon Raman microscopy. \emph{CrystEngComm} \textbf{10}, 937--946 (2008).
\item
  Bannister, N. et al.~Lattice vibrations of γ- and β-coronene from Raman microscopy and theory. \emph{Phys. Rev.~Mater.} \textbf{3}, 125601 (2019).
\item
  Robertson, J. M., Sinclair, V. C. \& Trotter, J. The crystal and molecular structure of tetracene. \emph{Acta Crystallogr.} \textbf{14}, 697--704 (1961).
\item
  Mattheus, C. C. et al.~Polymorphism in pentacene. \emph{Acta Crystallogr. C} \textbf{57}, 939--941 (2001).
\item
  Donaldson, D. M., Robertson, J. M. \& White, J. G. The crystal and molecular structure of perylene. \emph{Proc. R. Soc. Lond. A} \textbf{220}, 311--321 (1953).
\item
  Camerman, A. \& Trotter, J. The crystal and molecular structure of pyrene. \emph{Acta Crystallogr.} \textbf{18}, 636--643 (1965).
\item
  Potticary, J. et al.~An unforeseen polymorph of coronene by the application of magnetic fields during crystal growth. \emph{Nat. Commun.} \textbf{7}, 11555 (2016).
\item
  Peierls, R. Zur kinetischen Theorie der Wärmeleitung in Kristallen. \emph{Ann. Phys.} \textbf{395}, 1055--1101 (1929).
\item
  Nataj, Z. E. et al.~Cryogenic characteristics of graphene composites---evolution from thermal conductors to thermal insulators. \emph{Nat. Commun.} \textbf{14}, https://doi.org/10.1038/s41467-023-38508-3 (2023).
\item
  Karpov, V. G., Klinger, M. I. \& Ignatiev, P. N. Atomic tunneling states and low-temperature anomalies of thermal properties in amorphous materials. \emph{Solid State Commun.} \textbf{44}, 333--337 (1982).
\item
  Buchenau, U. et al.~Interaction of soft modes and sound waves in glasses. \emph{Phys. Rev.~B} \textbf{46}, 2798--2808 (1992).
\item
  Ramos, M. A. \& Buchenau, U. Low-temperature thermal conductivity of glasses within the soft-potential model. \emph{Phys. Rev.~B} \textbf{55}, 5749--5754 (1997).
\item
  Allen, P. B., Feldman, J. L., Fabian, J. \& Wooten, F. Diffusons, locons and propagons: character of atomic vibrations in amorphous Si. \emph{Phil. Mag. B} \textbf{79}, 1715--1731 (1999).
\item
  Simoncelli, M., Marzari, N. \& Mauri, F. Unified theory of thermal transport in crystals and glasses. \emph{Nat. Phys.} \textbf{15}, 809--813 (2019).
\item
  Simoncelli, M., Marzari, N. \& Mauri, F. Wigner formulation of thermal transport in solids. \emph{Phys. Rev.~X} \textbf{12}, 041011 (2022).
\item
  Simoncelli, M., Mauri, F. \& Marzari, N. Thermal conductivity of glasses: first-principles theory and applications. \emph{npj Comput. Mater.} \textbf{9}, 106 (2023).
\item
  Krivchikov, A., Andersson, O., Korolyuk, O. \& Kryvchikov, O. Thermal conductivity of solid triphenyl phosphite. \emph{Molecules} \textbf{27}, 8399 (2022).
\item
  Krivchikov, A. I. et al.~Exponential approximation of the coherence contribution to the thermal conductivity of complex clathrate-type crystals. \emph{Materialia} \textbf{32}, 101944 (2023).
\item
  Krivchikov, A. I. \& Korolyuk, O. A. Empirical universal approach to describing the thermal conductivity of amorphous polymers: effects of pressure, radiation and the Meyer--Neldel rule. \emph{Low Temp. Phys.} \textbf{50}, 328--335 (2024).
\item
  Korolyuk, O. A., Sagan, V. V., Krivchikov, A. I., Konstantinov, V. A. \& Horbatenko, Yu. V. Multi-channel heat transfer in molecular crystals and their solutions. \emph{Low Temp. Phys.} \textbf{51}, 1048 (2025).
\item
  Korolyuk, O. A. et al.~Bridging glass-like and crystal-like thermal transport: a unified model for disordered and complex solids. \emph{J. Non-Cryst. Solids} \textbf{690}, 124273 (2026).
\item
  Berman, R. \emph{Thermal Conduction in Solids} (Clarendon, 1976).
\item
  Meyer, W. \& Neldel, H. Über die Beziehungen zwischen der Energiekonstanten ε und der Mengenkonstanten a in der Leitwerts-Temperaturformel bei oxydischen Halbleitern. \emph{Z. Tech. Phys.} \textbf{18}, 588--593 (1937).
\item
  Jeżowski, A., Mucha, J. \& Pompe, G. Thermal conductivity of the amorphous alloy Fe40Ni40P14B6 between 80 and 300 K. \emph{J. Phys. D} \textbf{20}, 1500--1506 (1987).
\end{enumerate}

\section*{Acknowledgements}

The work of A.I.K. was supported by the National Research Foundation of Ukraine (Project No. 2023.03/0012). The work of V.S., A.J. and D.S. was supported by the National Science Centre, Poland (grant no. 2022/45/B/ST3/02326).

\section*{Author contributions}

A.I.K. conceived the study; V.S. and D.S. measured heat capacity; A.J. measured thermal conductivity; V.S. and V.B. performed Raman spectroscopy; V.S., A.I.K. and O.A.K. analysed the data; V.S., A.I.K. and O.A.K. wrote the manuscript with input from all authors.

\section*{Competing interests}

The authors declare no competing interests.

\clearpage
\setcounter{page}{1}
\renewcommand{\thepage}{S\arabic{page}}
\begin{center}
{\LARGE\bfseries Supplementary Information\par}
\vspace{6pt}
{\large Low-temperature vibrational anomalies and thermal transport in polycyclic aromatic hydrocarbon crystals\par}
\vspace{6pt}
{\normalsize V.~Sokolenko\textsuperscript{1}, A.~I.~Krivchikov\textsuperscript{1,2,*}, A.~Je\.zowski\textsuperscript{1}, V.~Boiko\textsuperscript{1}, O.~A.~Korolyuk\textsuperscript{2}, D.~Szewczyk\textsuperscript{1}\par}
\end{center}

\textbf{Contents}

\begin{itemize}
\tightlist
\item
  Supplementary Note 1. The General Cosine--Cosine (GCC) model: formulation, fitting protocol and parameter uncertainties
\item
  Supplementary Note 2. Materials
\item
  Supplementary Note 3. Low-frequency Raman spectroscopy: methods and per-compound analysis
\item
  Supplementary Note 4. Thermal conductivity: transport model, single-compound analysis and fit parameters
\item
  Supplementary Note 5. Preliminary correlation between the activation energy and the Raman spectrum
\item
  Supplementary Note 6. Previous thermal studies of PAHs and boson-peak-like anomalies in ordered molecular crystals
\item
  Supplementary Figures 1--12; Supplementary Tables 1--8; Supplementary References
\end{itemize}

Equation numbers prefixed with S refer to this document; equations (1)--(3), figures and tables without the prefix ``Supplementary'' refer to the main text.

\section*{\texorpdfstring{Supplementary Note 1. \textbf{The General Cosine--Cosine (GCC) model: formulation, fitting protocol and parameter uncertainties}}{Supplementary Note 1. The General Cosine--Cosine (GCC) model: formulation, fitting protocol and parameter uncertainties}}

\subsection*{1.1 Formulation}

The General Cosine--Cosine (GCC) model is a phenomenological scaling description of the boson-peak-like excess in the reduced heat capacity of these molecular crystals. It separates the low-temperature background from the excess and represents the excess by two cosine branches joined at the maximum of the background-subtracted reduced heat capacity, Cₚ/T\textsuperscript{3} − c₁/T\textsuperscript{2}. We denote the temperature of this maximum by Tₘₐₓ. The reduced heat capacity is then written as

\emph{Y}(\emph{T}) ≡ \emph{C}\textsubscript{p}(\emph{T})/\emph{T}\textsuperscript{3} = \emph{c}\textsubscript{1}/\emph{T}\textsuperscript{2} + \emph{c}\textsubscript{3} + Δ\textsubscript{BP}(\emph{T}), (S1)

where \emph{c}\textsubscript{1}/\emph{T}\textsuperscript{2} represents the residual linear-in-\emph{T} heat capacity, \emph{c}\textsubscript{3} is the Debye \emph{T}\textsuperscript{3} coefficient and Δ\textsubscript{BP}(\emph{T}) is the boson-peak-like excess. For \emph{T} ≤ \emph{T}\textsubscript{max} the excess is a raised cosine,

Δ\textsubscript{BP}\textsuperscript{(L)}(\emph{T}) = (\emph{H}\textsubscript{max} − \emph{c}\textsubscript{3})·½{[}1 − cos(π\emph{T}/\emph{T}\textsubscript{max}){]}, (S2)

and for \emph{T}\textsubscript{max} ≤ \emph{T} ≤ \emph{T}\textsubscript{2} the high-temperature decay is described by a second cosine branch,

Δ\textsubscript{BP}\textsuperscript{(R)}(\emph{T}) = (\emph{H}\textsubscript{max} − \emph{c}\textsubscript{3})·½{[}1 + cos(\emph{d}(\emph{T}/\emph{T}\textsubscript{max} − 1)){]}. (S3)

Inserting (S2) and (S3) into (S1) gives equations (1) and (2) of the main text. Hₘₐₓ is defined here as the value of Cₚ/T\textsuperscript{3} − c₁/T\textsuperscript{2} at Tₘₐₓ; consequently, the total measured reduced heat capacity at this temperature is Cₚ/T\textsuperscript{3} = c₁/Tₘₐₓ\textsuperscript{2} + Hₘₐₓ. This definition must be distinguished from the maximum of the uncorrected Cₚ/T\textsuperscript{3} curve, which can occur at a slightly lower temperature when c₁ is non-negligible.

The parameters c₁, c₃, Hₘₐₓ and Tₘₐₓ are shared by the two branches. The additional dimensionless parameter d controls the width and asymmetry of the high-temperature decay. If T₂ is the temperature at which the right-branch excess reaches zero, then d(T₂/Tₘₐₓ − 1) = π, and

\emph{d} = π\emph{T}\textsubscript{max}/(\emph{T}\textsubscript{2} − \emph{T}\textsubscript{max}). (S4)

Thus d = π corresponds to equal widths of the two branches in the reduced variable T/Tₘₐₓ; d \textless{} π corresponds to a slower high-temperature decay, and d \textgreater{} π to a faster decay. These statements concern the GCC parametrization and do not, by themselves, identify a microscopic mechanism.

In the fits, T₂ is the upper limit of the temperature interval over which the right branch is retained; it need not coincide with the zero of the cosine excess. If the fit terminates before that zero, d(T₂/Tₘₐₓ − 1) \textless{} π, or d \textless{} πTₘₐₓ/(T₂ − Tₘₐₓ). Equality applies only when the fitted interval extends to the zero. The adopted values for the six compounds are listed in Table 1 of the main text.

\subsection*{\texorpdfstring{1.2 Continuity, asymptotics and the scaling function Δ*(\emph{x})}{1.2 Continuity, asymptotics and the scaling function Δ*(x)}}

Both excess branches have the same value and zero derivative at \emph{T} = \emph{T}\textsubscript{max}:

Δ\textsubscript{BP}\textsuperscript{(L)}(\emph{T}\textsubscript{max}) = Δ\textsubscript{BP}\textsuperscript{(R)}(\emph{T}\textsubscript{max}) = \emph{H}\textsubscript{max} − \emph{c}\textsubscript{3}, dΔ\textsubscript{BP}\textsuperscript{(L)}/d\emph{T}\textbar{}\textsubscript{\emph{T}max} = dΔ\textsubscript{BP}\textsuperscript{(R)}/d\emph{T}\textbar{}\textsubscript{\emph{T}max} = 0. (S5)

The excess is therefore continuously differentiable at the join for any d. The derivative of the complete Cₚ/T\textsuperscript{3} curve at Tₘₐₓ is nevertheless −2c₁/Tₘₐₓ\textsuperscript{3} because the linear-in-T background contributes a temperature-dependent term.

For \emph{T} ≪ \emph{T}\textsubscript{max} the raised cosine expands as

Δ\textsubscript{BP}\textsuperscript{(L)}(\emph{T}) = (\emph{H}\textsubscript{max} − \emph{c}\textsubscript{3}){[}(π\textsuperscript{2}/4)(\emph{T}/\emph{T}\textsubscript{max})\textsuperscript{2} − (π\textsuperscript{4}/48)(\emph{T}/\emph{T}\textsubscript{max})\textsuperscript{4} + \ldots{]}, (S6)

so that

\emph{C}(\emph{T}) = \emph{c}\textsubscript{1}\emph{T} + \emph{c}\textsubscript{3}\emph{T}\textsuperscript{3} + (\emph{H}\textsubscript{max} − \emph{c}\textsubscript{3})(π\textsuperscript{2}/4\emph{T}\textsubscript{max}\textsuperscript{2})\emph{T}\textsuperscript{5} + \emph{O}(\emph{T}\textsuperscript{7}). (S7)

The excess enters at order T\textsuperscript{5}, two powers beyond the Debye term. Accordingly, the model approaches C(T) = c₁T + c₃T\textsuperscript{3} as T → 0. Within a harmonic vibrational description, a T\textsuperscript{5} heat-capacity contribution is consistent with an excess density of states proportional to ω\textsuperscript{4} at low frequency. This correspondence is not unique evidence for quasi-localized modes: corrections to acoustic-phonon dispersion can also generate higher-order terms.

Near the maximum, with \emph{x} = \emph{T}/\emph{T}\textsubscript{max} and the normalized excess

Δ(\emph{T}) ≡ {[}\emph{C}(\emph{T})/\emph{T}\textsuperscript{3} − \emph{c}\textsubscript{1}/\emph{T}\textsuperscript{2} − \emph{c}\textsubscript{3}{]}/(\emph{H}\textsubscript{max} − \emph{c}\textsubscript{3}), (S8)

both branches are parabolic to leading order,

Δ\textsubscript{L}(\emph{x}) ≈ 1 − (π\textsuperscript{2}/4)(\emph{x} − 1)\textsuperscript{2}, Δ\textsubscript{R}(\emph{x}) ≈ 1 − (\emph{d}\textsuperscript{2}/4)(\emph{x} − 1)\textsuperscript{2}. (S9)

The curvature is π\textsuperscript{2}/4\emph{T}\textsubscript{max}\textsuperscript{2} on the low-temperature side and \emph{d}\textsuperscript{2}/4\emph{T}\textsubscript{max}\textsuperscript{2} on the high-temperature side, so \emph{d} quantifies the peak asymmetry in closed form.

For the left branch, equation (S2) gives exactly

Δ*(\emph{x}) = ½(1 − cos π\emph{x}), 0 ≤ \emph{x} ≤ 1. (S10)

The function (S10) contains no material-specific shape parameter. Once the temperature is scaled by the independently determined Tₘₐₓ and the excess is normalized by its own amplitude, the rising edges of the four compounds with the strongest excess (coronene, corannulene, perylene and pyrene) are consistent with this function (Fig. 4b of the main text). The collapse tests the shape of the rising branch; it does not establish a common microscopic origin. For tetracene and pentacene, the excess is only about 7--10\% of the Debye term and the normalized data are too noisy for an equally stringent test. The right branch, ΔR(x) = ½{[}1 + cos(d(x − 1)){]}, has the same functional form but a compound-dependent d.

\subsection*{1.3 Mathematical properties of the raised-cosine function}

\begin{itemize}
\tightlist
\item
  On the physical domain used here, x = T/Tₘₐₓ ≥ 0, the raised cosine has the following properties:
\item
  Δ*(0) = 0, Δ*(0.5) = 0.5 and Δ*(1) = 1.
\item
  It rises quadratically, Δ*(x) ≈ (π\textsuperscript{2}/4)x\textsuperscript{2} as x → 0, and has a stationary maximum at x = 1, where dΔ*/dx = 0 and d\textsuperscript{2}Δ*/dx\textsuperscript{2} = −π\textsuperscript{2}/2.
\end{itemize}

The raised cosine is an analytic, parameter-free representation of the normalized rising branch. Its usefulness is empirical: it compactly describes the measured trend over the fitted interval. The data do not establish that it is a unique function or that its mathematical properties alone identify the underlying lattice dynamics.

\subsection*{1.4 Fitting protocol}

The GCC parameters in Table 1 of the main text were obtained in six steps.

\begin{enumerate}
\def\labelenumi{\arabic{enumi}.}
\tightlist
\item
  Initial estimate of Tₘₐₓ and Hₘₐₓ. The low-temperature upturn was first separated from the boson-peak-like feature by estimating c₁. The candidate maximum was then located in Cₚ/T\textsuperscript{3} − c₁/T\textsuperscript{2}. For weak or shoulder-like features, this estimate was iterated with the low-temperature background fit; the resulting values were checked against the measured curve and the automated refit described below.
\item
  Low-temperature background and left branch. The coefficients c₁ and c₃ were estimated from the low-temperature data using the expansion Cₚ/T = c₁ + c₃T\textsuperscript{2} + bT\textsuperscript{4}, with the fitting interval and weighting stated in the main-text Methods. The left-branch GCC curve was then evaluated using these independently estimated background parameters.
\item
  Right branch (Tₘₐₓ ≤ T ≤ T₂). With c₁, c₃, Hₘₐₓ and Tₘₐₓ fixed, d was adjusted to describe the high-temperature side. T₂ is the upper limit of the interval over which the branch remains within the adopted tolerance of the data. The tolerance was based on the scatter of the left-branch fit (2.5 times its rms relative deviation, constrained to 3--10\%); the additional condition d(T/Tₘₐₓ − 1) ≤ π prevents the cosine excess from becoming negative within the fitted interval.
\item
  \textbf{Normalized excess.} Δ*(\emph{x}) was computed from the data using equation (S8), with \emph{x} = \emph{T}/\emph{T}\textsubscript{max}.
\item
  Refined maximum. Tₘₐₓ and Hₘₐₓ were refined from the maximum of Cₚ/T\textsuperscript{3} − c₁/T\textsuperscript{2} using LOESS smoothing (local quadratic regression with tricube weights in ln T and half-width ln 1.35). The numerical search precision is not an estimate of the physical uncertainty; the parameter uncertainties are substantially larger and are discussed in Section 1.5.
\item
  Rising-branch comparison. The normalized excess Δ*(x) was calculated using the independently determined c₁, c₃, Hₘₐₓ and Tₘₐₓ, and compared with ½(1 − cos πx) for x ≤ 1 without readjusting the normalization parameters. This is the test of the common rising-edge shape reported in the main text.
\end{enumerate}

The resulting fits are shown in Supplementary Figs. 1--3.

\subsection*{1.5 Parameter uncertainties, fit quality and automated refit}

Fit quality was assessed using the relative root-mean-square (rms) deviation over the fitted intervals and the ordinary coefficient of determination R\textsuperscript{2} for each branch. R\textsuperscript{2} can be low or negative when the fitted branch has very little variance, even if its absolute residuals are modest. Thus, the low left-branch R\textsuperscript{2} for pyrene (0.108), alongside a 4.9\% rms deviation, reflects the nearly flat branch and small data variance. The negative left-branch R\textsuperscript{2} and 7.6\% rms deviation for tetracene indicate a less reliable determination for this weak-excess sample. The quoted uncertainties of approximately ±10\% in the main text are estimates, not formal confidence intervals from a full uncertainty-propagation or bootstrap analysis.

As an independent sensitivity check, the fitting procedure was repeated as an automated least-squares calculation, using weighted linear regression for c₁ and c₃ and nonlinear least squares for d. The standard errors in Supplementary Table 1 are the local 1σ estimates obtained from the covariance matrices of these fits. They do not include uncertainty associated with the choice of fitting interval, smoothing procedure or model form.

\needspace{8\baselineskip}
\textbf{Supplementary Table 1 \textbar{}} Independent automated refit of the GCC parameters. Values are local 1σ standard errors from the covariance matrix; they do not include fitting-window or model-selection uncertainty. Units: c₁ in mJ K⁻\textsuperscript{2} mol⁻\textsuperscript{1}; c₃ and Hₘₐₓ in mJ K⁻\textsuperscript{4} mol⁻\textsuperscript{1}. n.d., no stable maximum obtained by the automated procedure.\par\nopagebreak
\begin{longtable}[]{@{}
  >{\raggedright\arraybackslash}p{(\columnwidth - 14\tabcolsep) * \real{0.2075}}
  >{\raggedright\arraybackslash}p{(\columnwidth - 14\tabcolsep) * \real{0.1693}}
  >{\raggedright\arraybackslash}p{(\columnwidth - 14\tabcolsep) * \real{0.1287}}
  >{\raggedright\arraybackslash}p{(\columnwidth - 14\tabcolsep) * \real{0.0370}}
  >{\raggedright\arraybackslash}p{(\columnwidth - 14\tabcolsep) * \real{0.1659}}
  >{\raggedright\arraybackslash}p{(\columnwidth - 14\tabcolsep) * \real{0.1055}}
  >{\raggedright\arraybackslash}p{(\columnwidth - 14\tabcolsep) * \real{0.0904}}
  >{\raggedright\arraybackslash}p{(\columnwidth - 14\tabcolsep) * \real{0.0958}}@{}}
\toprule\noalign{}
\endhead
\bottomrule\noalign{}
\endlastfoot
Compound & \emph{c}\textsubscript{1}

(mJ K⁻\textsuperscript{2} mol⁻\textsuperscript{1}) & \multicolumn{2}{>{\raggedright\arraybackslash}p{(\columnwidth - 14\tabcolsep) * \real{0.1657} + 2\tabcolsep}}{%
\emph{c}\textsubscript{3}

(mJ K⁻\textsuperscript{4} mol⁻\textsuperscript{1})} & \emph{H}\textsubscript{max}

(mJ K⁻\textsuperscript{4} mol⁻\textsuperscript{1}) & \emph{T}\textsubscript{max}

(K) & \emph{d} & \emph{T}\textsubscript{2}

(K) \\
Tetracene (benz{[}b{]}anthracene) & 0.15 & \multicolumn{2}{>{\raggedright\arraybackslash}p{(\columnwidth - 14\tabcolsep) * \real{0.1657} + 2\tabcolsep}}{%
3.10} & 3.33 & 7.9 & 2.7 & 10.0 \\
Corannulene & 0.29 & \multicolumn{2}{>{\raggedright\arraybackslash}p{(\columnwidth - 14\tabcolsep) * \real{0.1657} + 2\tabcolsep}}{%
4.39} & 6.93 & 6.62 & 2.5 & 14.7 \\
Coronene & 0.39 & \multicolumn{2}{>{\raggedright\arraybackslash}p{(\columnwidth - 14\tabcolsep) * \real{0.1657} + 2\tabcolsep}}{%
4.19} & 5.45 & 7.14 & 3.1 & 13.7 \\
Pentacene & 0.34 & \multicolumn{2}{>{\raggedright\arraybackslash}p{(\columnwidth - 14\tabcolsep) * \real{0.1657} + 2\tabcolsep}}{%
5.0} & 5.49 & 2.0 & 1.9 & 5.0 \\
Perylene & 0.23 & \multicolumn{2}{>{\raggedright\arraybackslash}p{(\columnwidth - 14\tabcolsep) * \real{0.1657} + 2\tabcolsep}}{%
5.48} & 6.55 & 4.81 & 2.9 & 9.0 \\
Pyrene & 0.5 & \multicolumn{2}{>{\raggedright\arraybackslash}p{(\columnwidth - 14\tabcolsep) * \real{0.1657} + 2\tabcolsep}}{%
5.45} & 7.05 & 5.5 & 2.3 & 12.0 \\
\end{longtable}

\begin{figure}[!htbp]
\centering
\begin{minipage}[t]{0.49\linewidth}\raggedright\textbf{a}\\\centering\includegraphics[width=0.95\linewidth]{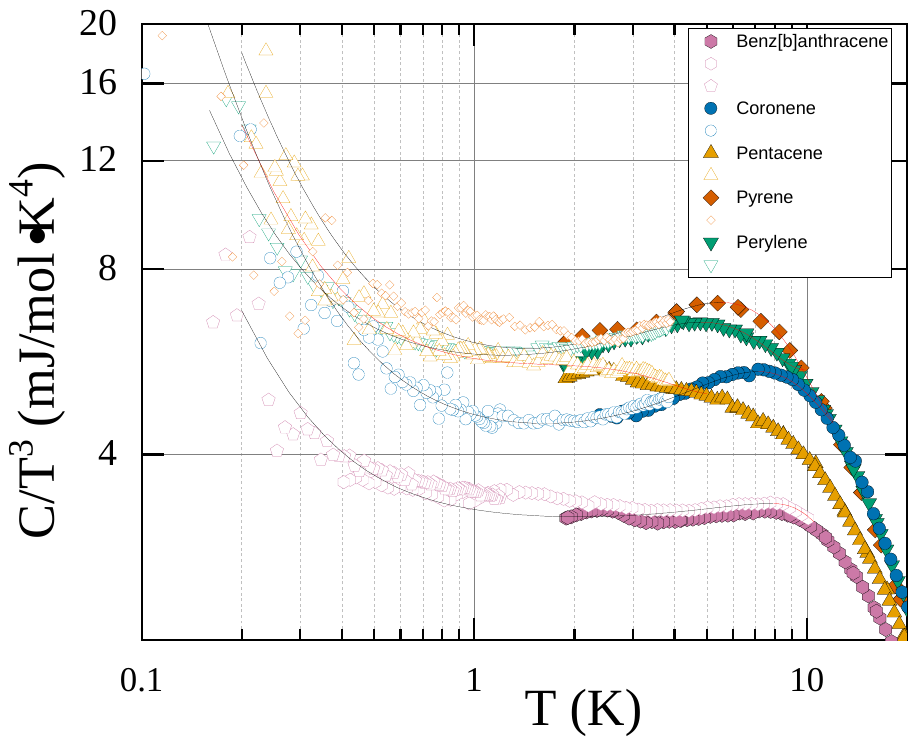}\end{minipage}\hfill\begin{minipage}[t]{0.49\linewidth}\raggedright\textbf{b}\\\centering\includegraphics[width=0.95\linewidth]{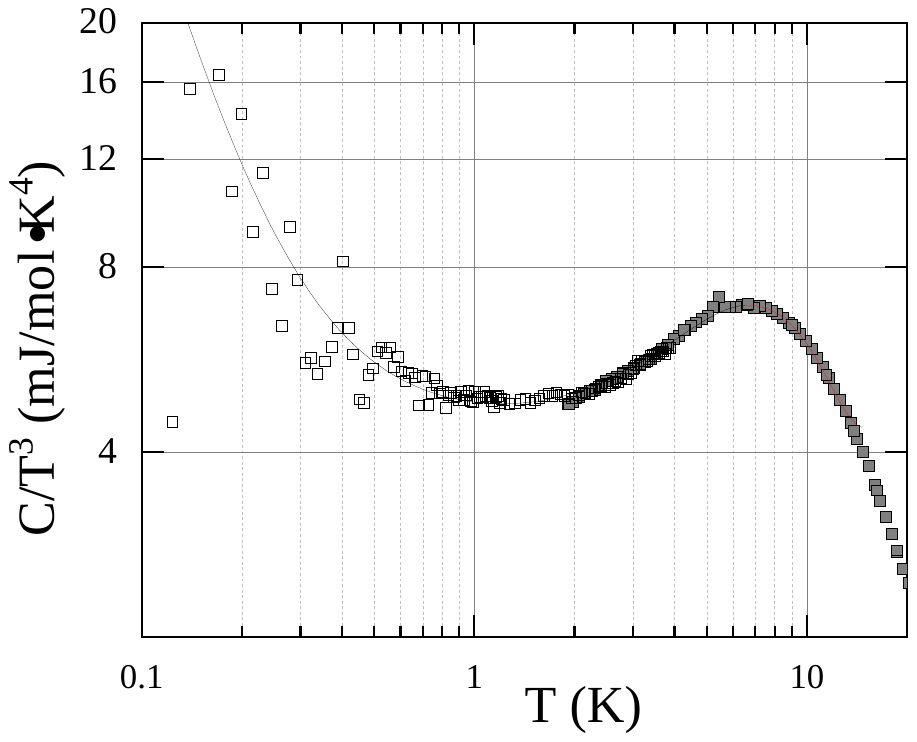}\end{minipage}

\medskip
\begin{minipage}{\linewidth}\small \textbf{Supplementary Figure 1 \textbar{}} Reduced heat capacity of the six PAH crystals. a, Cₚ/T\textsuperscript{3} versus temperature for pyrene, perylene, coronene, pentacene and tetracene. b, Corannulene, plotted separately for clarity. Open symbols denote measurements with the dilution insert below approximately 2 K; filled symbols denote PPMS measurements. Black curves show the left branch of Eq. (1), and red curves show the right branch of Eq. (2). The low-temperature upturn is associated with the fitted linear contribution c₁/T\textsuperscript{2}.\end{minipage}
\end{figure}

\begin{figure}[!htbp]
\centering
\begin{minipage}[t]{0.49\linewidth}\raggedright\textbf{a}\\\centering\includegraphics[width=0.95\linewidth]{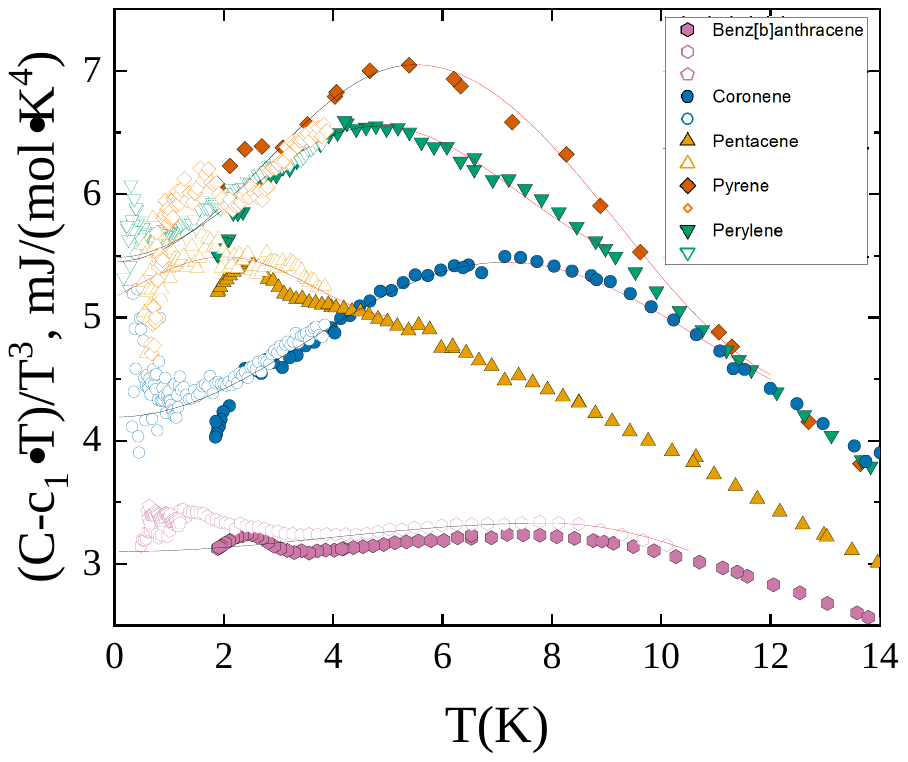}\end{minipage}\hfill\begin{minipage}[t]{0.49\linewidth}\raggedright\textbf{b}\\\centering\includegraphics[width=0.95\linewidth]{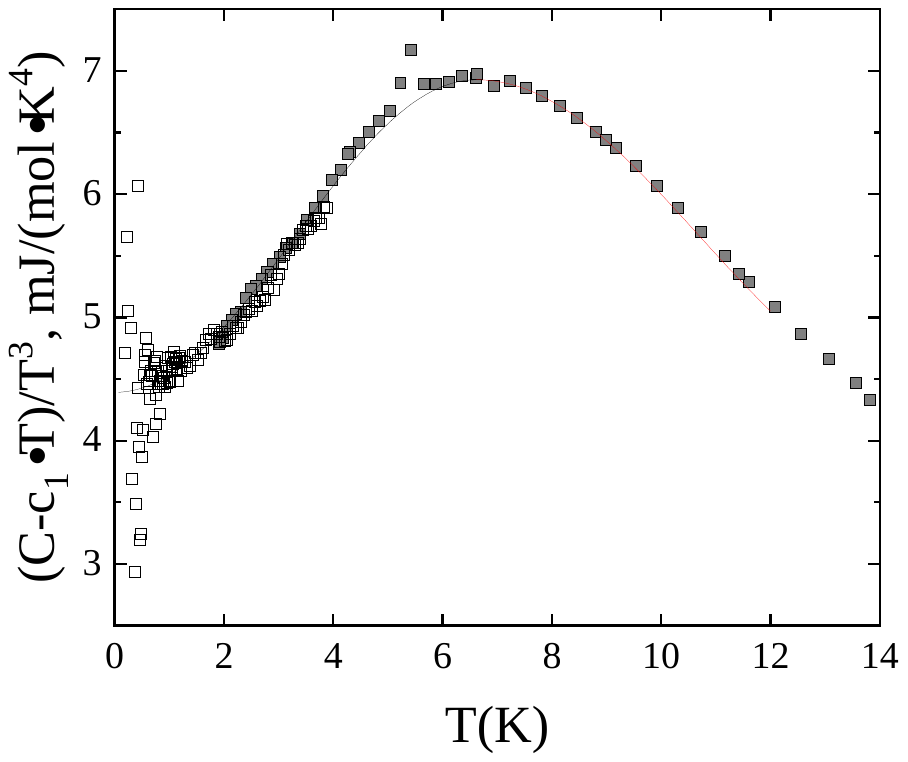}\end{minipage}

\medskip
\begin{minipage}{\linewidth}\small \textbf{Supplementary Figure 2 \textbar{}} Reduced heat capacity after subtraction of the linear term. a, (Cₚ − c₁T)/T\textsuperscript{3} versus temperature for pyrene, perylene, coronene, pentacene and tetracene. b, Corannulene. The plotted quantity retains the Debye contribution c₃; the excess above c₃ is Δ(T)/T\textsuperscript{3}. Symbols and fitted branches are as in Supplementary Fig. 1.\end{minipage}
\end{figure}

\begin{figure}[!htbp]
\centering
\begin{minipage}[t]{0.49\linewidth}\raggedright\textbf{a}\\\centering\includegraphics[width=0.95\linewidth]{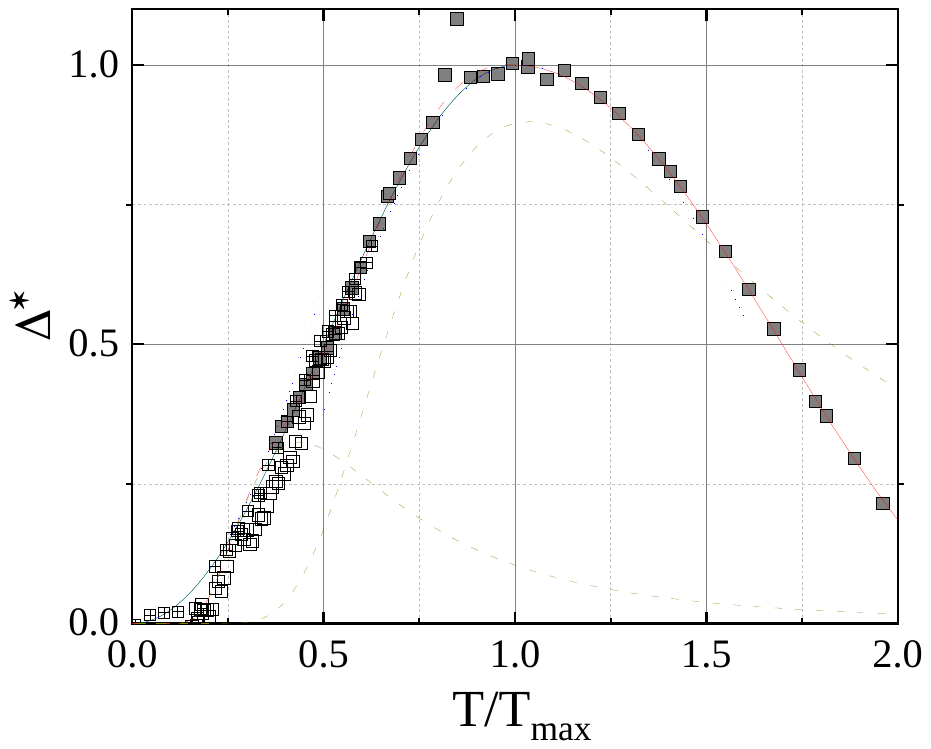}\end{minipage}\hfill\begin{minipage}[t]{0.49\linewidth}\raggedright\textbf{b}\\\centering\includegraphics[width=0.95\linewidth]{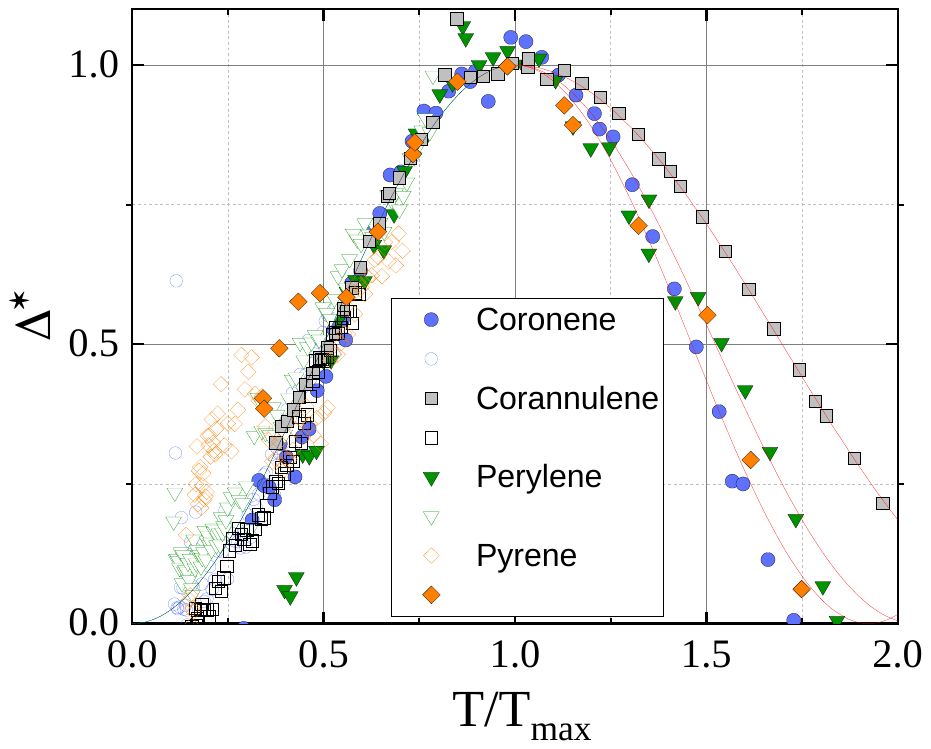}\end{minipage}

\medskip
\begin{minipage}{\linewidth}\small \textbf{Supplementary Figure 3 \textbar{}} Normalized heat-capacity excess. Δ*(x) versus x = T/Tₘₐₓ. a, Corannulene; black and red curves show the left and right GCC branches, respectively. Red dashed curves indicate the parabolic approximations near x = 1; dash-dotted curves show the two components of the two-oscillator Einstein fit. b, Coronene, corannulene, perylene and pyrene. The black curve is the raised-cosine function ½{[}1 − cos(πx){]}; red curves are the compound-specific right branches from Table 1. The rising branches are consistent with a common function within the resolution of the data, whereas the decays differ among compounds.\end{minipage}
\end{figure}

For corannulene, coronene and perylene, the automated refit reproduces the adopted parameters reasonably well. For tetracene and pentacene, whose excesses are weakest, the automated procedure does not recover the adopted maximum: it converges to Tₘₐₓ ≈ 1.45 K for tetracene and finds no stable maximum for pentacene. For pyrene, it returns appreciably different values of c₃ and d. These differences demonstrate sensitivity to the fitting window and initialization, especially for weak or broad features. The adopted parameters should therefore be treated as phenomenological descriptors, not as uniquely determined material constants.

\subsection*{1.6 Comparison with Einstein-oscillator descriptions}

A localized excess heat capacity is often described by a single Einstein term\textsuperscript{1,2},

\emph{C}\textsubscript{E}(\emph{T}) = 3\emph{nR}(θ\textsubscript{E}/\emph{T})\textsuperscript{2} exp(θ\textsubscript{E}/\emph{T})/{[}exp(θ\textsubscript{E}/\emph{T}) − 1{]}\textsuperscript{2}, (S11)

added to the Debye contribution. Two points distinguish it from the GCC description.

First, a single Einstein oscillator has a fixed temperature dependence once its characteristic temperature is chosen. Its reduced heat-capacity contribution can be asymmetric, but its rise and decay cannot be varied independently; it therefore does not generally reproduce both sides of the measured anomaly over a broad interval. A sum of Einstein terms\textsuperscript{3--6} can improve the fit, but introduces additional characteristic temperatures and amplitudes. The GCC form instead imposes a common normalized rising branch and describes the decay with one additional parameter d. The comparison is empirical and does not establish that either representation corresponds to a unique microscopic spectrum.

Second, for \emph{T} ≪ θ\textsubscript{E},

\emph{C}\textsubscript{E}(\emph{T}) → 3\emph{nR}(θ\textsubscript{E}/\emph{T})\textsuperscript{2} exp(−θ\textsubscript{E}/\emph{T}), (S12)

In contrast, the GCC excess approaches a T\textsuperscript{5} contribution at low temperature {[}equation (S7){]}. This is consistent with the adopted low-temperature expansion, but the available data do not by themselves establish that the excess follows this power law over an asymptotically broad temperature interval. As in any phenomenological decomposition, the acoustic background and the excess must be considered together.

\subsection*{\texorpdfstring{1.7 Candidate physical origins of the shape parameter \emph{d}}{1.7 Candidate physical origins of the shape parameter d}}

The GCC model is phenomenological. No microscopic derivation of the raised-cosine form is available, and none is claimed here. Several theoretical frameworks contain features that could be compatible with parts of the observed asymmetry, but these connections remain hypotheses until they are tested against the present data.

First, the leading ω\textsuperscript{4} onset of the excess density of states implied by equation (S7) is consistent with the low-frequency behaviour reported for quasi-localized modes in glasses, although it is not specific to them. The model of Gurevich, Parshin and Schober predicts a crossover from this steep ω\textsuperscript{4} regime to a much shallower, approximately linear regime above the boson-peak frequency\textsuperscript{7}, which gives an excess that is steep on the rise and shallow on the decay.

Second, in heterogeneous-elasticity theory the boson peak arises from spatial fluctuations of the shear modulus. These produce Rayleigh-like scattering of transverse phonons, Γ(ω) ∝ ω\textsuperscript{4}, and an Ioffe--Regel crossover at the peak frequency\textsuperscript{8}. The rise and the decay are then governed by different physics, with no requirement of a common scale on both sides.

Third, and most directly relevant to fully ordered molecular crystals, an avoided crossing between acoustic and low-lying optical (librational) branches produces a smoothed van Hove-like pile-up of states\textsuperscript{9,10}. Its shape is set by the generally unequal curvatures of the two branches on either side of the anticrossing and is therefore asymmetric by construction. This mechanism requires no positional or orientational disorder.

For the compounds studied here, the fitted d values are generally below π, indicating a slower decay than rise within the adopted parametrization. The mechanisms outlined above can produce asymmetric spectral features, but they do not all make a quantitative prediction for d or require d \textless{} π. Distinguishing them will require lattice-dynamics calculations and/or direct measurements of the low-energy vibrational density of states.

\section*{\texorpdfstring{Supplementary Note 2\textbf{. Materials}}{Supplementary Note 2. Materials}}

All compounds were commercial products used as received (Supplementary Table 2).

\needspace{8\baselineskip}
\textbf{Supplementary Table 2 \textbar{} Specifications of the PAH materials as declared by the suppliers.}\par\nopagebreak
\begin{longtable}[]{@{}
  >{\raggedright\arraybackslash}p{(\columnwidth - 10\tabcolsep) * \real{0.2562}}
  >{\raggedright\arraybackslash}p{(\columnwidth - 10\tabcolsep) * \real{0.1357}}
  >{\raggedright\arraybackslash}p{(\columnwidth - 10\tabcolsep) * \real{0.1506}}
  >{\raggedright\arraybackslash}p{(\columnwidth - 10\tabcolsep) * \real{0.1357}}
  >{\raggedright\arraybackslash}p{(\columnwidth - 10\tabcolsep) * \real{0.1479}}
  >{\raggedright\arraybackslash}p{(\columnwidth - 10\tabcolsep) * \real{0.1739}}@{}}
\toprule\noalign{}
\begin{minipage}[b]{\linewidth}\raggedright
Material
\end{minipage} & \begin{minipage}[b]{\linewidth}\raggedright
Formula
\end{minipage} & \begin{minipage}[b]{\linewidth}\raggedright
Molar mass (g mol⁻\textsuperscript{1})
\end{minipage} & \begin{minipage}[b]{\linewidth}\raggedright
Declared purity
\end{minipage} & \begin{minipage}[b]{\linewidth}\raggedright
Sigma-Aldrich no.
\end{minipage} & \begin{minipage}[b]{\linewidth}\raggedright
CAS RN
\end{minipage} \\
\midrule\noalign{}
\endhead
\bottomrule\noalign{}
\endlastfoot
Coronene & C\textsubscript{24}H\textsubscript{12} & 300.35 & 99.99\% & 335355 & 191-07-1 \\
Perylene & C\textsubscript{20}H\textsubscript{12} & 252.31 & ≥99.0\% & P11204 & 198-55-0 \\
Pyrene & C\textsubscript{16}H\textsubscript{10} & 202.25 & 99.0\% & 571245 & 129-00-0 \\
Pentacene & C\textsubscript{22}H\textsubscript{14} & 278.35 & 99.0\% & P1802 & 135-48-8 \\
Tetracene (benz{[}b{]}anthracene) & C\textsubscript{18}H\textsubscript{12} & 228.29 & 99.99\% & 698415 & 92-24-0 \\
Corannulene & C\textsubscript{20}H\textsubscript{10} & 250.30 & 95.0\% & FLUH99CDD621 & 5821-51-2 \\
\end{longtable}

\section*{\texorpdfstring{Supplementary Note 3. \textbf{Low-frequency Raman spectroscopy: methods and per-compound analysis}}{Supplementary Note 3. Low-frequency Raman spectroscopy: methods and per-compound analysis}}

\subsection*{3.1 Consistency check of the population-factor calculation}

The occupation columns in four exported data files (tetracene, corannulene, pentacene and perylene) correspond numerically to a Bose--Einstein factor evaluated at 300 K over the 341-point grid. This verifies the temperature parameter used in the data export, not an independent measurement of the sample temperature. The coronene file did not contain an occupation column; 300 K was therefore assumed for its Bose correction. The actual sample temperature and possible laser heating were not monitored independently.

\subsection*{3.2 Bose correction and reduced spectra}

The raw intensities were averaged over the available repeat scans (except for perylene; see Section 3.3) and divided by n(\ensuremath{\tilde{\omega}},T) + 1. This removes the Stokes thermal-population factor under the assumed sample temperature. The resulting reduced spectrum remains weighted by the frequency-dependent Raman coupling, optical response and any residual background; it is not a direct measurement of the vibrational density of states.

\subsection*{3.3 Exclusion of one perylene scan}

Perylene scan 2 contained a large, nearly frequency-independent additive offset (about 127,000--136,000 counts, four to five times the baseline of scans 1 and 3). Scans 1 and 3 agreed in spectral shape up to an approximately constant scale factor (about 1.3--1.5). The offset is consistent with a fluorescence or instrumental background, but its origin cannot be identified uniquely from these spectra. Scan 2 was excluded from the averaged spectrum; this exclusion and the resulting sensitivity of the centroid are noted in Supplementary Note 5.

\subsection*{3.4 Band fitting}

Bands were identified by a prominence-based search in the Bose-corrected spectra and fitted locally with a Lorentzian plus a linear baseline, I(\ensuremath{\tilde{\omega}}) = Aγ\textsuperscript{2}/{[}(\ensuremath{\tilde{\omega}} − \ensuremath{\tilde{\omega}}₀)\textsuperscript{2} + γ\textsuperscript{2}{]} + m\ensuremath{\tilde{\omega}} + b. The fitted centre is \ensuremath{\tilde{\omega}}₀ and the full width at half maximum is 2γ. For corannulene, only the two most clearly resolved features could be fitted reliably. Other shoulders are reported as approximate positions, not as established normal modes. The fitted values are summarized in Supplementary Table 3.

\needspace{8\baselineskip}
\textbf{Supplementary Table 3 \textbar{}} Fitted low-frequency Raman bands at nominal T = 300 K. Positions and FWHM are in cm⁻\textsuperscript{1}. Relative intensities are qualitative and refer only to bands within the same spectrum.\par\nopagebreak
\begin{longtable}[]{@{}
  >{\raggedright\arraybackslash}p{(\columnwidth - 6\tabcolsep) * \real{0.1824}}
  >{\raggedright\arraybackslash}p{(\columnwidth - 6\tabcolsep) * \real{0.2700}}
  >{\raggedright\arraybackslash}p{(\columnwidth - 6\tabcolsep) * \real{0.1217}}
  >{\raggedright\arraybackslash}p{(\columnwidth - 6\tabcolsep) * \real{0.4258}}@{}}
\toprule\noalign{}
\begin{minipage}[b]{\linewidth}\raggedright
Compound
\end{minipage} & \begin{minipage}[b]{\linewidth}\raggedright
Position
\end{minipage} & \begin{minipage}[b]{\linewidth}\raggedright
FWHM
\end{minipage} & \begin{minipage}[b]{\linewidth}\raggedright
Relative intensity
\end{minipage} \\
\midrule\noalign{}
\endhead
\bottomrule\noalign{}
\endlastfoot
Tetracene & 41.9 & 4.8 & medium \\
& 47.3 & 4.2 & medium \\
& 58.4 & 6.7 & medium \\
& 117.6 & 11.0 & strong \\
& 130.4 & 11.6 & strongest \\
& 169.1 & 6.7 & weak \\
Corannulene & 31--34 & \textasciitilde16 & best resolved, still broad \\
& 163 & \textasciitilde18 & best resolved, still broad \\
& 55--66, 97, 118, 179 & -- & shoulders only, not resolved \\
Coronene & 59.0 & 11.5 & strong \\
& 86.4 & 11.0 & strongest \\
Pentacene & 32.0 & 5.2 & strong, narrow \\
& 50.8 & 9.6 & medium \\
& 119.8 & 10.5 & strong \\
& 137.1 & 10.5 & strong \\
& 184.2 & 12.0 & medium \\
Perylene\textsuperscript{a} & 33.2 & 10.0 & strong \\
& 61.6 & 5.2 & medium \\
& 73.0 & 7.0 & strongest \\
& \textasciitilde99.2 & -- & weak shoulder \\
& \textasciitilde191 & -- & broad rise at the edge of the scan \\
Pyrene & 17.8 & 6.0 & medium \\
& 29.8 & 6.4 & medium \\
& 40.2 & 4.0 & medium \\
& 55.1 & 7.3 & strongest \\
& 76.6 & 5.4 & weak \\
& 92.8 & 13.4 & strong, broadest \\
& 125.9 & 6.3 & weak \\
\end{longtable}

\textsuperscript{a}Fitted to the average of scans 1 and 3 (Section 3.3).

\subsection*{\texorpdfstring{3.5 Tetracene (benz{[}b{]}anthracene, C\textsubscript{18}H\textsubscript{12})}{3.5 Tetracene (benz{[}b{]}anthracene, C18H12)}}

\begin{figure}[!htbp]
\centering
\includegraphics[width=0.6\linewidth]{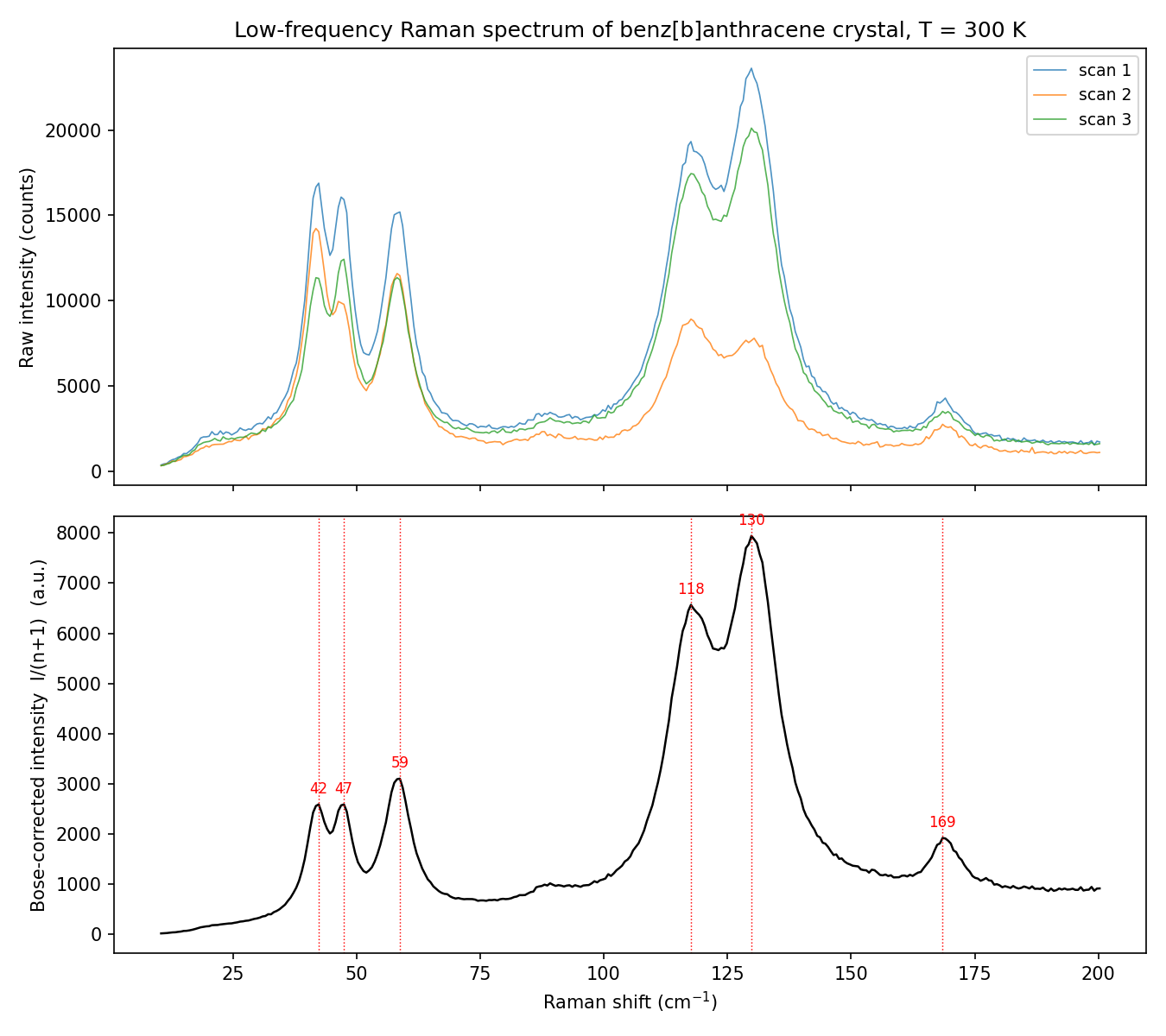}

\medskip
\begin{minipage}{\linewidth}\small \textbf{Supplementary Figure 4 \textbar{} Raman spectra of tetracene.} Top, raw spectra of the three repeat scans. Bottom, Bose-corrected spectrum.\end{minipage}
\end{figure}

Six sharp, well-separated bands (FWHM 4--12 cm⁻\textsuperscript{1}) appear on a background that returns nearly to zero above and below the band envelope. Tetracene crystallizes in the triclinic space group P-1 with \emph{Z} = 2 and molecules on inversion centres\textsuperscript{11}. The 6\emph{Z} = 12 external modes then comprise three acoustic and nine optical branches; by the mutual-exclusion rule, the six librational \emph{A}\textsubscript{g} modes are Raman active. The agreement of this count with the six resolved bands is indicative only: a factor-group and polarization analysis was not made, and weak or overlapping modes may be unresolved. The anthracene--pyromellitic dianhydride (PMDA) charge-transfer crystal, a related but not identical system, shows six low-frequency modes below 140 cm⁻\textsuperscript{1} with a strong mode at 130 cm⁻\textsuperscript{1} (ref. 12), numerically coincident with our strongest band.

\subsection*{\texorpdfstring{3.6 Corannulene (C\textsubscript{20}H\textsubscript{10})}{3.6 Corannulene (C20H10)}}

\begin{figure}[!htbp]
\centering
\includegraphics[width=0.6\linewidth]{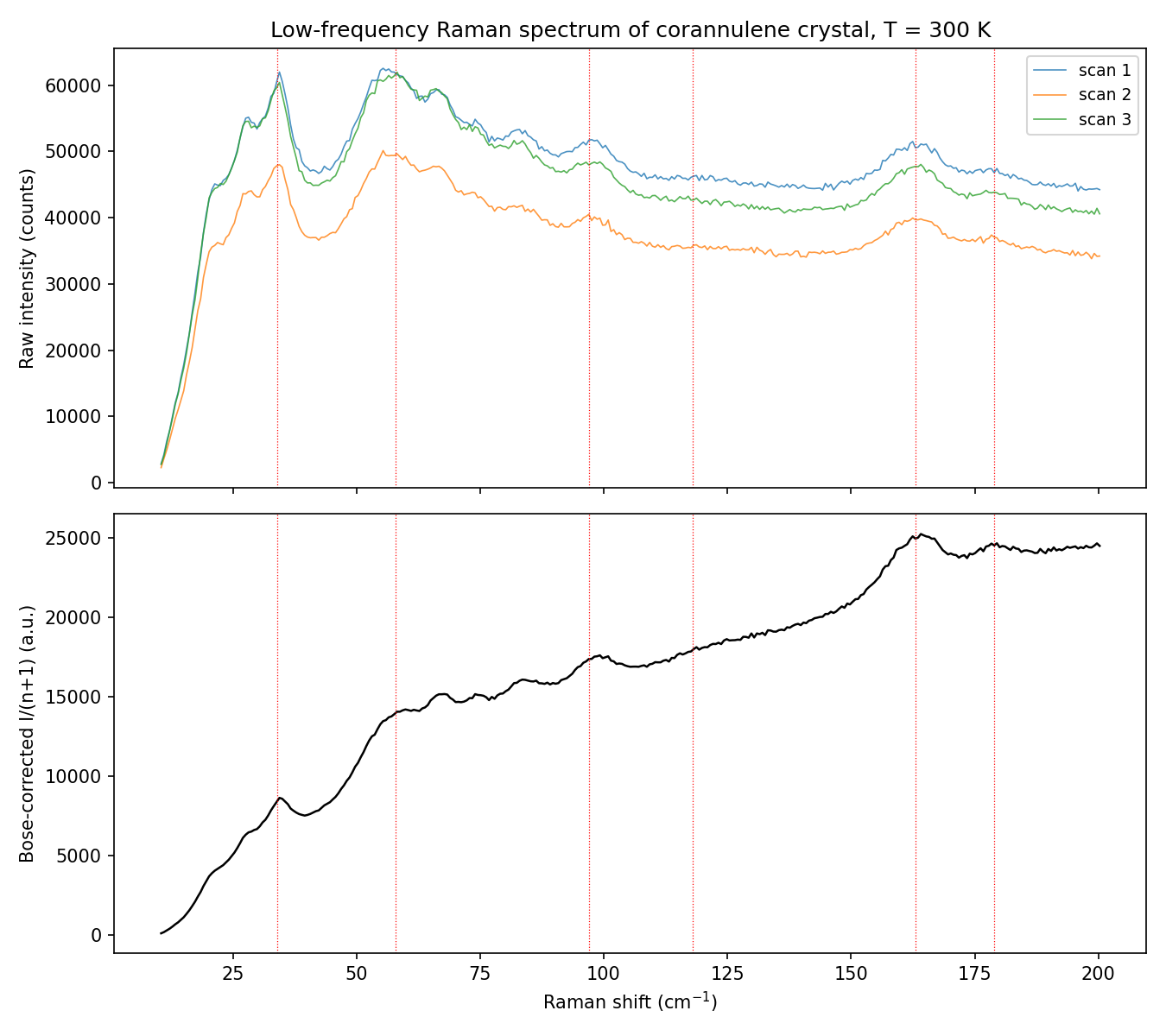}

\medskip
\begin{minipage}{\linewidth}\small \textbf{Supplementary Figure 5 \textbar{} Raman spectra of corannulene.} Top, raw spectra of the three repeat scans. Bottom, Bose-corrected spectrum.\end{minipage}
\end{figure}

The reduced spectrum contains a broad, rising background with only two clearly resolved features. Their widths (about 16--18 cm⁻\textsuperscript{1}) exceed those of the narrowest bands in the other samples. This broad response is compatible with several possibilities, including overlapping modes, inhomogeneous broadening, a fluorescence/instrumental background, or dynamical disorder. Because the raw spectrum is nearly flat over much of the measured interval, a frequency-independent background alone produces a rising contribution after Bose correction. The Raman data therefore do not, by themselves, establish disorder-driven relaxational scattering. Crystallographic reports of bowl reorientation and split site occupancy provide independent evidence for orientational disorder in corannulene; they do not prove that this disorder is the origin of the observed Raman continuum.

\subsection*{\texorpdfstring{3.7 Coronene (C\textsubscript{24}H\textsubscript{12})}{3.7 Coronene (C24H12)}}

\begin{figure}[!htbp]
\centering
\includegraphics[width=0.6\linewidth]{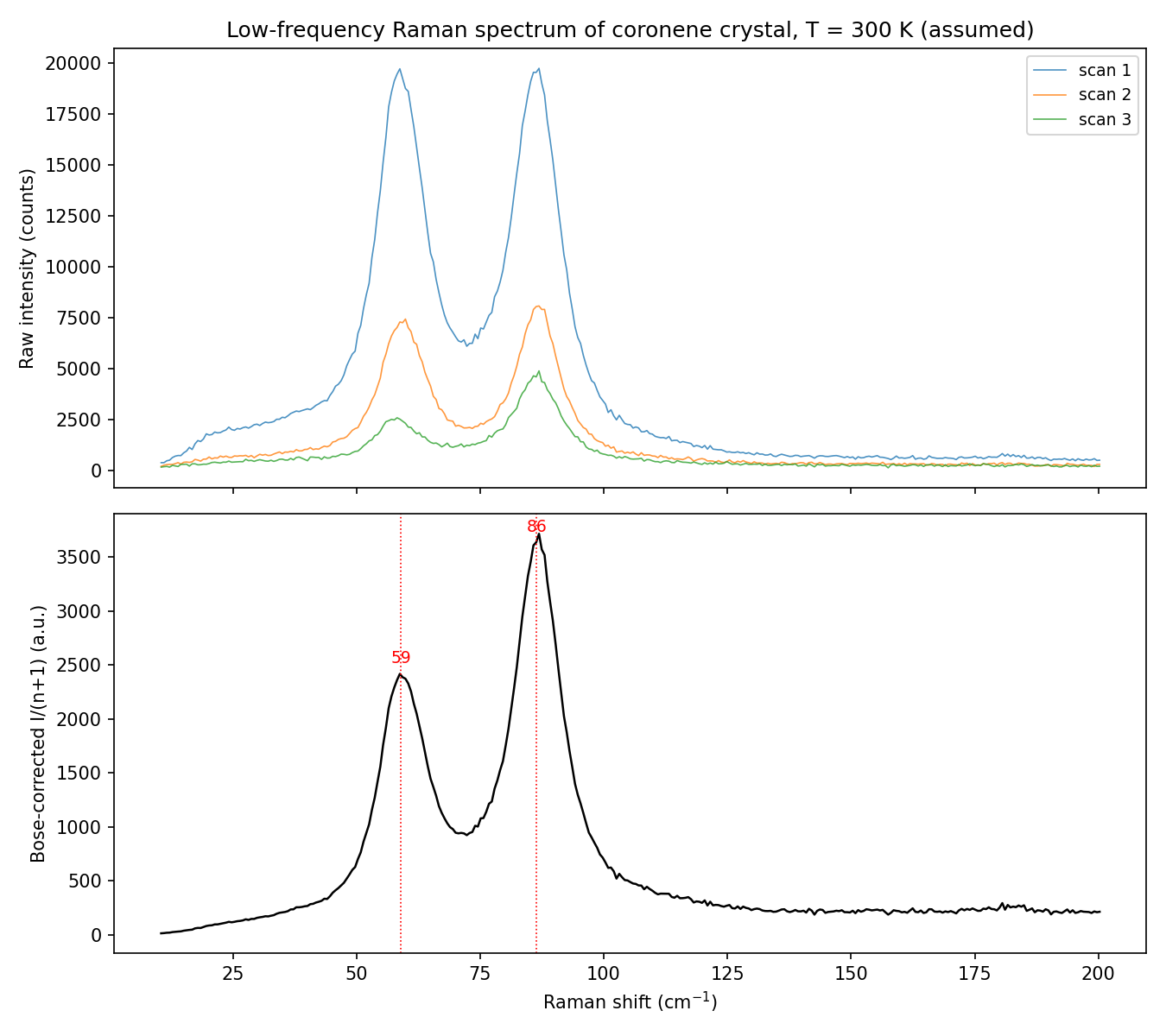}

\medskip
\begin{minipage}{\linewidth}\small \textbf{Supplementary Figure 6 \textbar{} Raman spectra of coronene.} Top, raw spectra of the three repeat scans. Bottom, Bose-corrected spectrum (\emph{T} = 300 K assumed; Section 3.1).\end{minipage}
\end{figure}

Only two strong external-mode bands are resolved; the spectrum is otherwise comparatively flat above about 100 cm⁻\textsuperscript{1}. Intensity ratios vary between repeat scans, which may reflect sample orientation, polarization selection, focus or other experimental factors; the present measurements do not isolate these effects. Dispersion-corrected DFT-D3 lattice-dynamics calculations for γ- and β-coronene predict dominant modes at 63.1, 65.6 and 92.85 cm⁻\textsuperscript{1} for the γ form and at 52.34 and 93.77 cm⁻\textsuperscript{1} for the β form (0 K values)\textsuperscript{15}. The observed 300 K bands (59.0 and 86.4 cm⁻\textsuperscript{1}) lie between these predictions, allowing for temperature and polymorph effects, and do not identify the polymorph present in the measured pellet.

\subsection*{\texorpdfstring{3.8 Pentacene (C\textsubscript{22}H\textsubscript{14})}{3.8 Pentacene (C22H14)}}

\begin{figure}[!htbp]
\centering
\includegraphics[width=0.6\linewidth]{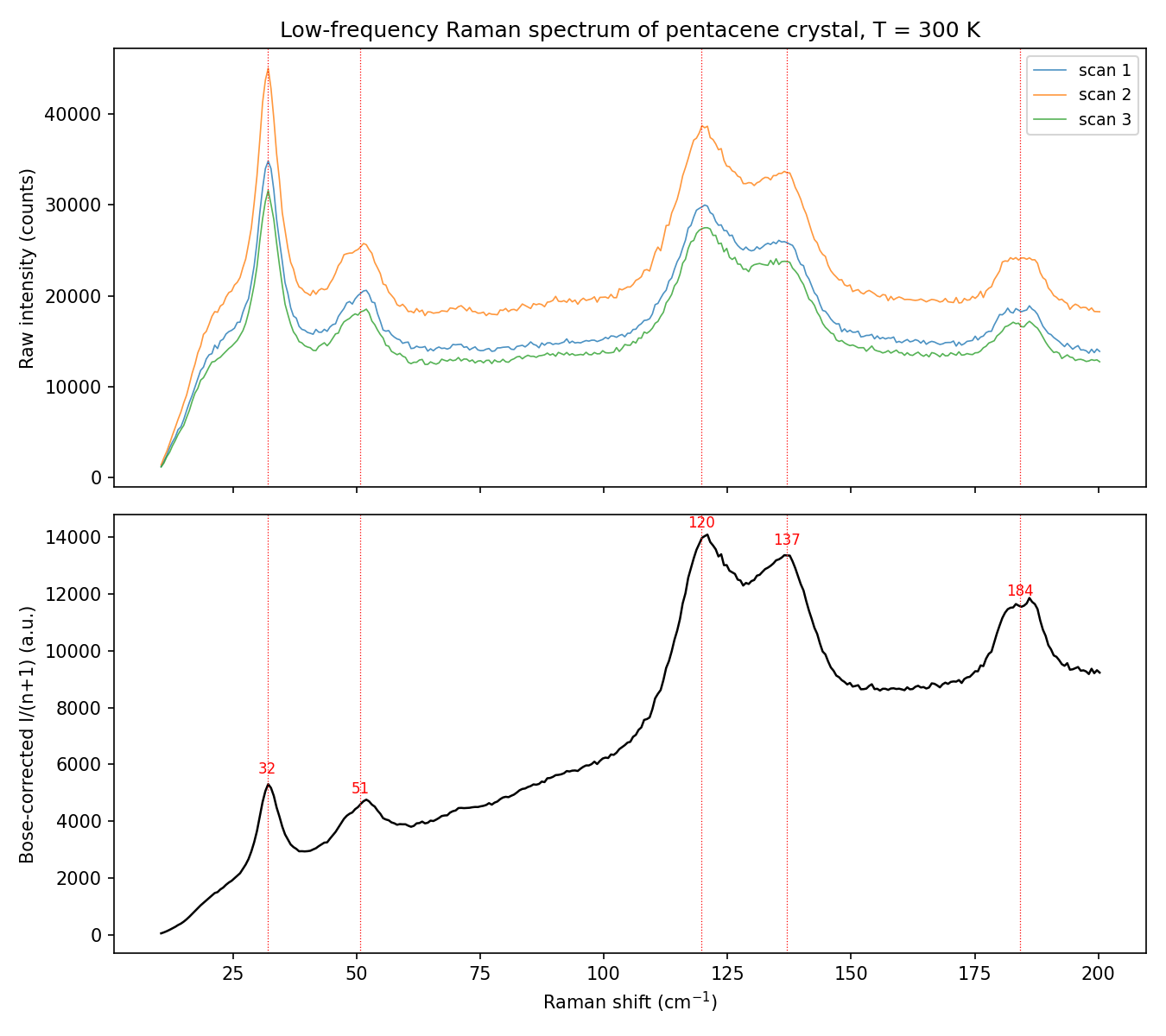}

\medskip
\begin{minipage}{\linewidth}\small \textbf{Supplementary Figure 7 \textbar{} Raman spectra of pentacene.} Top, raw spectra of the three repeat scans. Bottom, Bose-corrected spectrum.\end{minipage}
\end{figure}

Five narrow bands (FWHM about 5--12 cm⁻\textsuperscript{1}) are resolved on a weakly rising background. Bulk pentacene is triclinic (P-1, Z = 2), with molecules on inversion centres\textsuperscript{16}; the crystal factor group and selection rules therefore determine which external modes are Raman active\textsuperscript{17}. Five low-frequency modes have also been reported for single crystals\textsuperscript{18}, although their frequencies depend on polymorph and measurement conditions\textsuperscript{19}. The observed mode count is consistent with those reports, but does not establish a one-to-one assignment. The rising background may include intrinsic spectral weight from anharmonic lattice dynamics\textsuperscript{20}, fluorescence or instrumental contributions; the present spectra do not distinguish these possibilities.

\subsection*{\texorpdfstring{3.9 Perylene (C\textsubscript{20}H\textsubscript{12})}{3.9 Perylene (C20H12)}}

\begin{figure}[!htbp]
\centering
\includegraphics[width=0.6\linewidth]{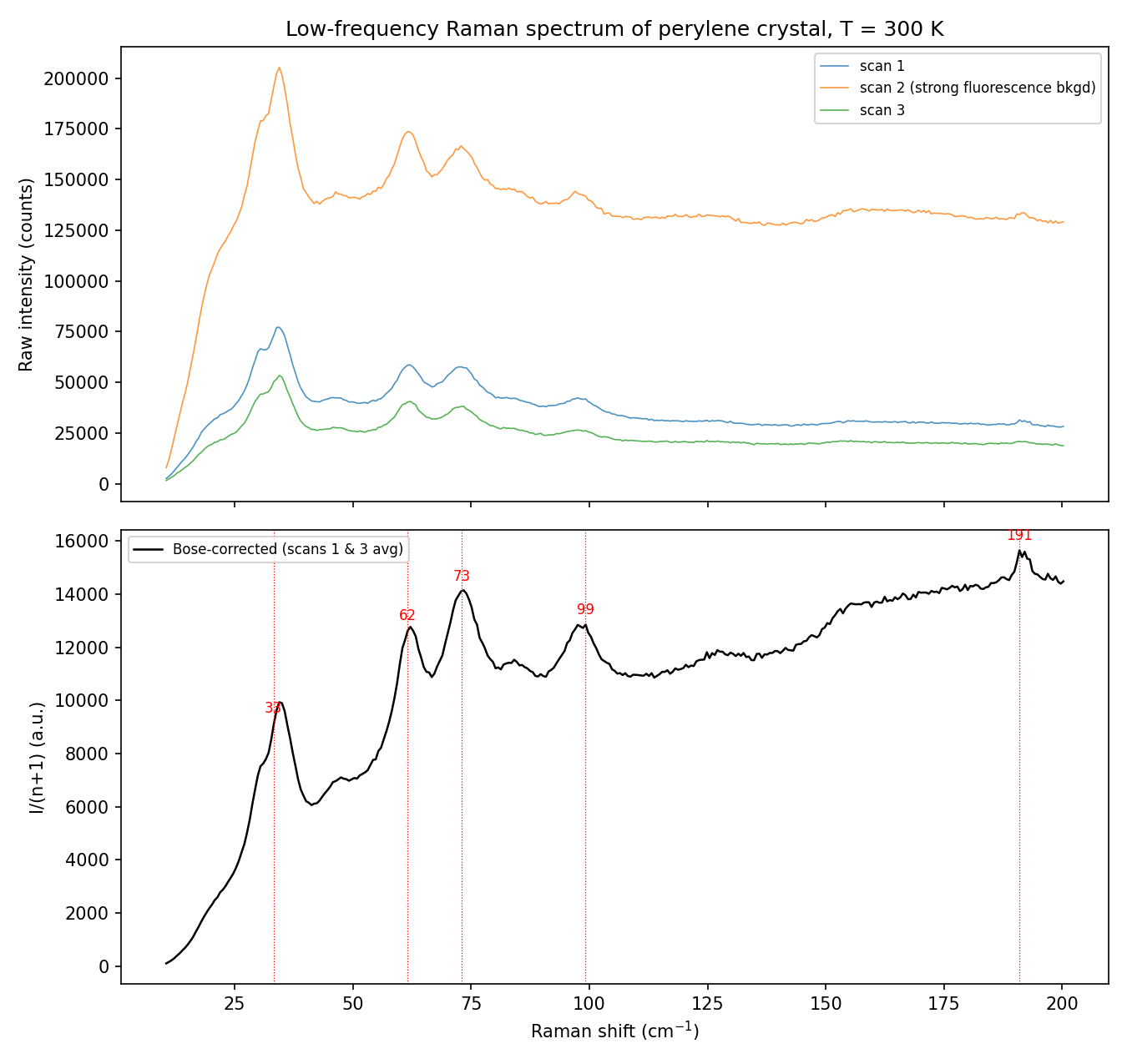}

\medskip
\begin{minipage}{\linewidth}\small \textbf{Supplementary Figure 8 \textbar{} Raman spectra of perylene.} Top, raw spectra of the three repeat scans. Bottom, Bose-corrected spectrum from scans 1 and 3 (Section 3.3).\end{minipage}
\end{figure}

Three relatively narrow bands below 80 cm⁻\textsuperscript{1} are resolved, with weaker structure at higher frequency and a rising background above about 150 cm⁻\textsuperscript{1}. The reported room-temperature structure is monoclinic P2₁/a with Z = 4 and has been described as related to that of pyrene\textsuperscript{21}. This structural similarity is consistent with, but does not prove, a larger number of external modes than in the Z = 2 crystals. Perylene is fluorescent, which is consistent with the additive offset observed in one scan; the spectral background should not be interpreted as a phonon continuum without independent background separation.

\subsection*{\texorpdfstring{3.10 Pyrene (C\textsubscript{16}H\textsubscript{10})}{3.10 Pyrene (C16H10)}}

\begin{figure}[!htbp]
\centering
\includegraphics[width=0.6\linewidth]{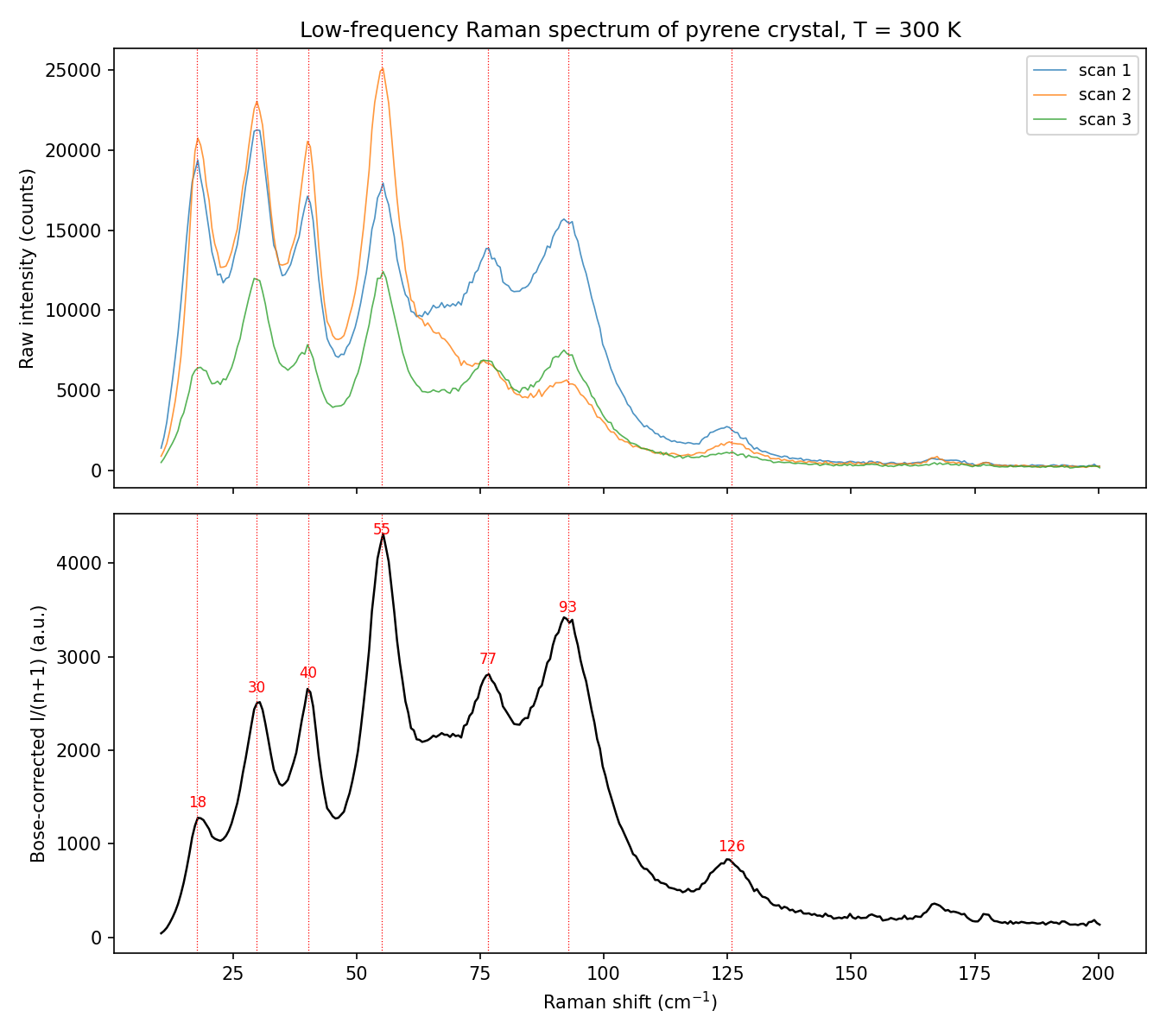}

\medskip
\begin{minipage}{\linewidth}\small \textbf{Supplementary Figure 9 \textbar{} Raman spectra of pyrene.} Top, raw spectra of the three repeat scans. Bottom, Bose-corrected spectrum.\end{minipage}
\end{figure}

Pyrene shows seven resolved bands, with FWHM values of about 4--13 cm⁻\textsuperscript{1}, on a background that returns close to zero above about 100 cm⁻\textsuperscript{1}. Its reported room-temperature structure is monoclinic P2₁/a with Z = 4. The larger number of resolved features is compatible with the larger number of external branches available in a Z = 4 unit cell, although Raman selection rules, overlap and mode intensities also determine how many bands are observed. Narrow fitted lines alone do not provide a quantitative phonon lifetime without accounting for instrumental resolution and other broadening mechanisms.

\subsection*{3.11 Comparison between compounds}

\needspace{8\baselineskip}
\textbf{Supplementary Table 4 \textbar{}} Molecular symmetry, reported room-temperature crystal structure, resolved Raman-band count and qualitative spectral background for the six PAH crystals.\par\nopagebreak
\begin{longtable}[]{@{}
  >{\raggedright\arraybackslash}p{(\columnwidth - 10\tabcolsep) * \real{0.1428}}
  >{\raggedright\arraybackslash}p{(\columnwidth - 10\tabcolsep) * \real{0.1904}}
  >{\raggedright\arraybackslash}p{(\columnwidth - 10\tabcolsep) * \real{0.1905}}
  >{\raggedright\arraybackslash}p{(\columnwidth - 10\tabcolsep) * \real{0.1905}}
  >{\raggedright\arraybackslash}p{(\columnwidth - 10\tabcolsep) * \real{0.1429}}
  >{\raggedright\arraybackslash}p{(\columnwidth - 10\tabcolsep) * \real{0.1429}}@{}}
\toprule\noalign{}
\begin{minipage}[b]{\linewidth}\raggedright
Compound
\end{minipage} & \begin{minipage}[b]{\linewidth}\raggedright
Molecular point group
\end{minipage} & \begin{minipage}[b]{\linewidth}\raggedright
Space group; Z
\end{minipage} & \begin{minipage}[b]{\linewidth}\raggedright
Resolved Raman bands
\end{minipage} & \begin{minipage}[b]{\linewidth}\raggedright
Background
\end{minipage} & \begin{minipage}[b]{\linewidth}\raggedright
Lattice/spectral description
\end{minipage} \\
\midrule\noalign{}
\endhead
\bottomrule\noalign{}
\endlastfoot
Tetracene & D₂h & P-1; Z = 2 (Ref. 11) & 6 & Flat, near zero & Six resolved narrow bands \\
Corannulene & C₅v (bowl-shaped) & P2₁/c; Z = 8 (Refs. 13, 14) & 2 (broad) & Rising continuum & Broad response; orientational disorder reported crystallographically \\
Coronene & D₆h & P2₁/a; Z = 2 (Ref. 23) & 2 & Flat, near zero & Two dominant resolved bands \\
Pentacene & D₂h & P-1; Z = 2 (Ref. 16) & 5 & Weakly rising & Five resolved narrow bands \\
Perylene & D₂h & P2₁/a; Z = 4 (Ref. 21) & 3 + shoulders & Rising above \textasciitilde150 cm⁻\textsuperscript{1} & Three fitted bands; weaker high-frequency structure \\
Pyrene & D₂h & P2₁/a; Z = 4 (Ref. 22) & 7 & Flat, near zero & Seven resolved bands \\
\end{longtable}

Point groups refer to the isolated molecules and are consistent with Table 2 of the main text. Space groups and \emph{Z} are literature values for the room-temperature polymorphs.

The space groups and Z values in Supplementary Table 4 are literature values for reported room-temperature structures. The actual polymorph and phase purity of the pressed pellets used in this study were not independently established by diffraction in the information supplied here. In particular, the coronene polymorph present in the pellets should not be assigned as γ or β without sample-specific structural data.

\section*{\texorpdfstring{Supplementary Note 4. \textbf{Thermal conductivity: transport model, single-compound analysis and fit parameters}}{Supplementary Note 4. Thermal conductivity: transport model, single-compound analysis and fit parameters}}

\subsection*{4.1 Background}

In well-ordered crystals, thermal conductivity often exhibits a pronounced phonon maximum that can be described using a Peierls--Boltzmann framework\textsuperscript{24}. In glasses and strongly disordered solids, thermal conductivity is typically lower and may show an approximately T\textsuperscript{2} regime at very low temperature, a broad plateau at intermediate temperature and a renewed rise at higher temperature. These are characteristic trends rather than universal criteria; their detailed form depends on the material and the temperature interval.

The temperature of the plateau correlates with the temperature of the boson peak in \emph{C}/\emph{T}\textsuperscript{3}\textsuperscript{25--35}. The standard tunnelling model\textsuperscript{36,37} accounts for the ∝ \emph{T}\textsuperscript{2} behaviour below 1 K. The soft-potential model\textsuperscript{38--40} accounts for the low-temperature behaviour, the approach to the plateau and the plateau itself. In these models, acoustic excitations in the ballistic regime are scattered resonantly by two-level systems and quasi-local vibrations.

The Allen--Feldman framework\textsuperscript{41--44} classifies harmonic vibrational modes according to their transport character. Propagons are extended, wave-like modes; diffusons are extended but non-propagating modes that transport energy diffusively\textsuperscript{45--47}; locons are localized modes and generally contribute little to macroscopic heat transport. This classification is useful for disordered solids, but it should not be identified term by term with the off-diagonal coherence contribution in the Wigner formulation.

\subsection*{4.2 Unified Peierls--coherence model}

A generalized quantum-mechanical theory of thermal conductivity\textsuperscript{48--51} describes crystal-like and glass-like behaviour as two limits and covers intermediate cases, although it has been applied only above the thermal conductivity plateau. It includes the particle-like propagation of phonon wave packets (Peierls term) and wave-like tunnelling between vibrational eigenstates (coherence term). The total thermal conductivity is the sum of the two contributions,

κ\textsubscript{tot}(\emph{T}) = κ\textsubscript{P}(\emph{T}) + κ\textsubscript{C}(\emph{T}). (S13)

Following our previous work on heat transport in disordered and complex solids\textsuperscript{52,53}, the low-temperature Peierls term is described by Matthiessen's rule\textsuperscript{54}, which is applicable in the quasiparticle picture of weakly interacting excitations in the ballistic regime:

{[}κ\textsubscript{P}(\emph{T}){]}⁻\textsuperscript{1} = (β\emph{T\textsuperscript{n}})⁻\textsuperscript{1} + {[}κ\textsubscript{ph}(\emph{T}){]}⁻\textsuperscript{1}. (S14)

The term β\emph{T\textsuperscript{n}} describes low-temperature scattering processes; β is their strength and \emph{n} ≈ 2 in the glass-like case of resonant scattering by two-level systems. The high-temperature phonon term is

κ\textsubscript{ph}(\emph{T}) = \emph{A}/\emph{T} + \emph{B}, (S15)

The coefficient A controls the 1/T part of κ\textsubscript{ph}(T) and therefore the temperature-dependent reduction of this phenomenological phonon channel. The constant B provides a finite high-temperature limit; in the parameterization used here it is identified with κ\textsubscript{pl}. The fitted parameters are descriptive and should not be interpreted as direct measurements of individual scattering rates.

\subsection*{4.3 Arrhenius approximation of the coherence term}

The coherence term κ\textsubscript{C}(\emph{T}) in equation (S13) is in general complex\textsuperscript{48}, and its temperature dependence is computationally demanding. Previous work\textsuperscript{53,55--57} showed, however, that it is well approximated by an Arrhenius-type expression,

κ\textsubscript{C}(\emph{T}) = κ\textsubscript{0} exp(−\emph{E}/\emph{T}), (S16)

In this work, κ\textsubscript{C}(T) is represented phenomenologically by an Arrhenius expression. Here κ₀ is a pre-exponential factor and E is a characteristic activation parameter expressed in kelvin. Although this term is discussed in relation to the wave-like coherence contribution in Wigner transport theory, the Arrhenius fit does not independently establish that microscopic assignment. Nor should κ\textsubscript{C}(T) be equated automatically with the Allen--Feldman diffuson contribution. The total thermal conductivity is therefore

κ\textsubscript{tot}(\emph{T}) = {[}(β\emph{T\textsuperscript{n}})⁻\textsuperscript{1} + (\emph{A}/\emph{T} + κ\textsubscript{pl})⁻\textsuperscript{1}{]}⁻\textsuperscript{1} + κ\textsubscript{0} exp(−\emph{E}/\emph{T}), (S17)

which is equation (3) of the main text.

\subsection*{4.4 Pyrene: single-compound analysis}

Supplementary Fig. 10 shows κ(\emph{T}) of pyrene measured from 4.5 to 280 K. The thermal conductivity is glass-like and small over the whole range. Three regions can be distinguished: a low-temperature region (\emph{T} \textless{} 30 K) where κ(T) increases, an intermediate region (30--150 K) where κ(T) approaches and follows a plateau, and a high-temperature region (\emph{T} \textgreater{} 150 K) where κ(T) increases again. Unlike amorphous polymers\textsuperscript{53}, the plateau of pyrene lies at higher temperature and extends over a temperature range several times wider.

\begin{figure}[!htbp]
\centering
\includegraphics[width=0.6\linewidth]{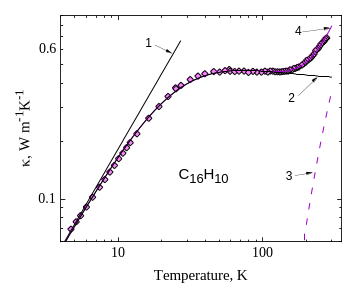}

\medskip
\begin{minipage}{\linewidth}\small \textbf{Supplementary Figure 10 \textbar{}} Thermal conductivity of pyrene. Symbols, experimental data. Curves: (1) low-temperature power-law guide κ(T) ∝ T\textsuperscript{1}·\textsuperscript{3}; (2) Peierls contribution from Matthiessen's rule, equation (S14); (3) phenomenological activated contribution κ\textsubscript{C}(T), equation (S16); (4) total fit, equation (S17). Parameters are listed in Supplementary Table 5.\end{minipage}
\end{figure}

The Peierls contribution describes the low- and intermediate-temperature data, while the additional activated term improves the description of the upward deviation at higher temperatures. The relative size of this term varies between compounds and does not exceed 46\% of the total thermal conductivity at 300 K in the five planar PAHs. It is therefore not appropriate to describe it as universally dominant. The parameter extraction by subtraction is sensitive to uncertainties in both the measured thermal conductivity and the fitted Peierls term.

For a descriptive estimate of the activated contribution, κ\textsubscript{P}(T) was subtracted from the measured κ(T), giving κ\textsubscript{C}(T) = κ(T) − κ\textsubscript{P}(T). Where this difference is positive and sufficiently above the experimental uncertainty, its logarithm can be plotted against 1/T:

ln κ\textsubscript{C}(\emph{T}) = ln{[}κ(\emph{T}) − κ\textsubscript{P}(\emph{T}){]} = ln κ\textsubscript{0} − \emph{E}/\emph{T}. (S18)

The slope and intercept of this representation provide estimates of E and κ₀. Because κ\textsubscript{C}(T) is obtained as the difference between measured and fitted quantities, these estimates inherit uncertainty and covariance from both terms; they should not be treated as independent observables.

\begin{figure}[!htbp]
\centering
\includegraphics[width=0.6\linewidth]{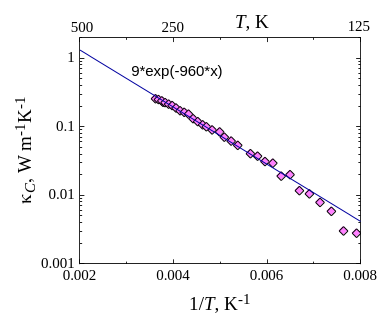}

\medskip
\begin{minipage}{\linewidth}\small \textbf{Supplementary Figure 11 \textbar{}} Residual activated contribution estimated for pyrene as κ\textsubscript{C}(T) = κ(T) − κ\textsubscript{P}(T), plotted against 1/T on a semi-logarithmic scale. The line is the Arrhenius representation κ\textsubscript{C}(T) = 9 exp(−960/T) W m⁻\textsuperscript{1} K⁻\textsuperscript{1}. The subtraction inherits uncertainty from both the measured κ(T) and fitted κ\textsubscript{P}(T).\end{minipage}
\end{figure}

\subsection*{4.5 All six compounds}

Figure 6a of the main text shows κ(T) for the five planar PAHs. The measured behaviour ranges from glass-like (pyrene and pentacene) to crystal-like, with a pronounced phonon maximum (tetracene, perylene and coronene). The Peierls term describes the low- and intermediate-temperature behaviour, while the activated term accounts for an additional high-temperature contribution in the fitted curves. Its contribution at 300 K ranges from 12\% to 46\% of the total thermal conductivity (main text), rather than dominating in every compound.

Corannulene differs from the five planar PAHs. Its thermal conductivity rises approximately as T\textsuperscript{1}·\textsuperscript{0}, reaches a plateau near 0.3 W m⁻\textsuperscript{1} K⁻\textsuperscript{1} above about 40 K and remains near this value up to 300 K. No activated contribution is resolved within the sensitivity of the present measurements (κ₀ \textless{} 0.01 W m⁻\textsuperscript{1} K⁻\textsuperscript{1}). The Peierls part of equation (S17) describes the measured curve with A = 9.1 W m⁻\textsuperscript{1} and κ\textsubscript{pl} = 0.31 W m⁻\textsuperscript{1} K⁻\textsuperscript{1}. The absence of a resolved activated term should not be taken as proof that the microscopic coherence contribution is exactly zero.

\needspace{8\baselineskip}
\textbf{Supplementary Table 5 \textbar{}} Best-fit parameters of the phenomenological thermal conductivity model, equation (S17), for the six PAH crystals. Best-fit values; formal parameter uncertainties were not determined.\par\nopagebreak
\begin{longtable}[]{@{}
  >{\raggedright\arraybackslash}p{(\columnwidth - 12\tabcolsep) * \real{0.1831}}
  >{\raggedright\arraybackslash}p{(\columnwidth - 12\tabcolsep) * \real{0.1971}}
  >{\raggedright\arraybackslash}p{(\columnwidth - 12\tabcolsep) * \real{0.0845}}
  >{\raggedright\arraybackslash}p{(\columnwidth - 12\tabcolsep) * \real{0.1409}}
  >{\raggedright\arraybackslash}p{(\columnwidth - 12\tabcolsep) * \real{0.1409}}
  >{\raggedright\arraybackslash}p{(\columnwidth - 12\tabcolsep) * \real{0.1409}}
  >{\raggedright\arraybackslash}p{(\columnwidth - 12\tabcolsep) * \real{0.1127}}@{}}
\toprule\noalign{}
\begin{minipage}[b]{\linewidth}\raggedright
Compound
\end{minipage} & \begin{minipage}[b]{\linewidth}\raggedright
β (W m⁻\textsuperscript{1} K⁻⁽\textsuperscript{1}⁺ⁿ⁾)
\end{minipage} & \begin{minipage}[b]{\linewidth}\raggedright
\emph{n}
\end{minipage} & \begin{minipage}[b]{\linewidth}\raggedright
\emph{A} (W m⁻\textsuperscript{1})
\end{minipage} & \begin{minipage}[b]{\linewidth}\raggedright
κ\textsubscript{pl} (W m⁻\textsuperscript{1} K⁻\textsuperscript{1})
\end{minipage} & \begin{minipage}[b]{\linewidth}\raggedright
κ\textsubscript{0} (W m⁻\textsuperscript{1} K⁻\textsuperscript{1})
\end{minipage} & \begin{minipage}[b]{\linewidth}\raggedright
\emph{E} (K)
\end{minipage} \\
\midrule\noalign{}
\endhead
\bottomrule\noalign{}
\endlastfoot
Pyrene & 0.0092 & 1.3 & 13 & 0.40 & 9.0 & 960 \\
Tetracene & 0.006 & 1.73 & 33 & 0.48 & 1.22 & 680 \\
Perylene & 0.0105 & 1.2 & 7.2 & 0.142 & 0.69 & 600 \\
Pentacene & 0.00325 & 1.6 & 9.2 & 0.451 & 0.5 & 600 \\
Coronene & 0.0098 & 1.2 & 13.8 & 0.221 & 2.8 & 820 \\
Corannulene & 0.0205 & 1.0 & 9.1 & 0.31 & \textless{} 0.01 & -- \\
\end{longtable}

For corannulene no coherence contribution is detected up to 300 K (κ\textsubscript{0} \textless{} 0.01 W m⁻\textsuperscript{1} K⁻\textsuperscript{1}), so \emph{E} cannot be determined.

The table reports best-fit values only; formal parameter uncertainties (covariance matrices or confidence intervals) were not determined for these fits.

\subsection*{4.6 Meyer--Neldel compensation}

The five fitted pairs (κ₀, E) show an approximately linear relation between ln κ₀ and E, reminiscent of the Meyer--Neldel compensation rule\textsuperscript{58,59}. With only five compounds, and because κ₀ and E are extracted from the same fit and may be strongly correlated, this trend is exploratory. The common crossing temperature and thermal conductivity should therefore be reported as descriptive fit features, not as an independently established compensation law.

\begin{figure}[!htbp]
\centering
\begin{minipage}[t]{0.49\linewidth}\raggedright\textbf{a}\\\centering\includegraphics[width=0.95\linewidth]{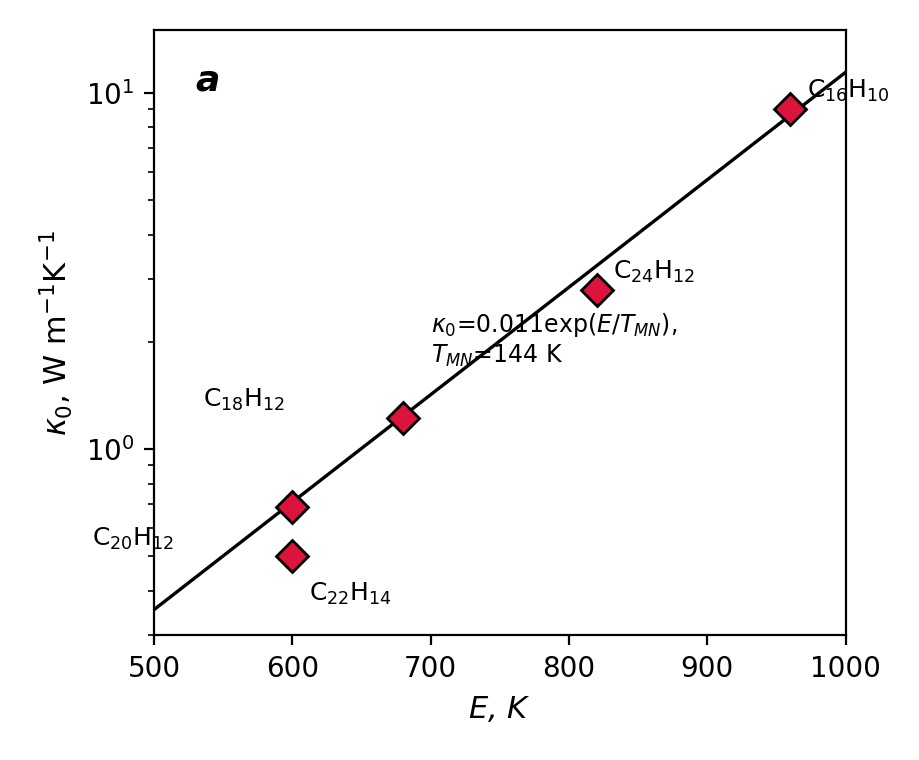}\end{minipage}\hfill\begin{minipage}[t]{0.49\linewidth}\raggedright\textbf{b}\\\centering\includegraphics[width=0.95\linewidth]{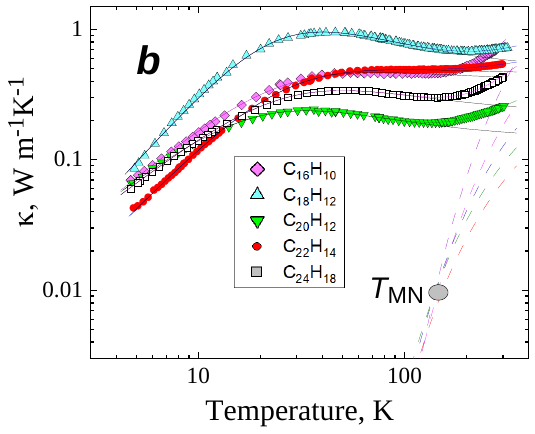}\end{minipage}

\medskip
\begin{minipage}{\linewidth}\small \textbf{Supplementary Figure 12 \textbar{}} Exploratory Meyer--Neldel-type compensation of the fitted activated term. a, ln κ₀ versus E for the five compounds with a resolved activated contribution. b, The corresponding κ\textsubscript{C}(T) curves. The approximate common crossing near 144 K is a descriptive feature of these five fits; parameter covariance and the small sample size prevent a strong claim of a universal compensation law.\end{minipage}
\end{figure}

\section*{\texorpdfstring{Supplementary Note 5. \textbf{Preliminary correlation between the activation energy and the Raman spectrum}}{Supplementary Note 5. Preliminary correlation between the activation energy and the Raman spectrum}}

Five compounds have both a fitted activation parameter E and a low-frequency Raman spectrum: pyrene, tetracene, perylene, pentacene and coronene. Corannulene is excluded because no activated term is resolved below 300 K. We examined several Raman descriptors against E, A and κ\textsubscript{pl} using Pearson correlation coefficients. Given the very small sample (n = 5) and the exploratory selection of descriptors, these analyses are hypothesis-generating rather than confirmatory.

\needspace{8\baselineskip}
\textbf{Supplementary Table 6 \textbar{} Thermal conductivity parameters and Raman descriptors used in the correlation analysis.} Wavenumbers in cm⁻\textsuperscript{1}.\par\nopagebreak
\begin{longtable}[]{@{}
  >{\raggedright\arraybackslash}p{(\columnwidth - 14\tabcolsep) * \real{0.1486}}
  >{\raggedright\arraybackslash}p{(\columnwidth - 14\tabcolsep) * \real{0.0946}}
  >{\raggedright\arraybackslash}p{(\columnwidth - 14\tabcolsep) * \real{0.1081}}
  >{\raggedright\arraybackslash}p{(\columnwidth - 14\tabcolsep) * \real{0.1351}}
  >{\raggedright\arraybackslash}p{(\columnwidth - 14\tabcolsep) * \real{0.1216}}
  >{\raggedright\arraybackslash}p{(\columnwidth - 14\tabcolsep) * \real{0.1351}}
  >{\raggedright\arraybackslash}p{(\columnwidth - 14\tabcolsep) * \real{0.1216}}
  >{\raggedright\arraybackslash}p{(\columnwidth - 14\tabcolsep) * \real{0.1351}}@{}}
\toprule\noalign{}
\begin{minipage}[b]{\linewidth}\raggedright
Compound
\end{minipage} & \begin{minipage}[b]{\linewidth}\raggedright
\emph{E} (K)
\end{minipage} & \begin{minipage}[b]{\linewidth}\raggedright
\emph{A} (W m⁻\textsuperscript{1})
\end{minipage} & \begin{minipage}[b]{\linewidth}\raggedright
κ\textsubscript{pl} (W m⁻\textsuperscript{1} K⁻\textsuperscript{1})
\end{minipage} & \begin{minipage}[b]{\linewidth}\raggedright
Highest band
\end{minipage} & \begin{minipage}[b]{\linewidth}\raggedright
Mean band position
\end{minipage} & \begin{minipage}[b]{\linewidth}\raggedright
Number of modes
\end{minipage} & \begin{minipage}[b]{\linewidth}\raggedright
Centroid ⟨\ensuremath{\tilde{\omega}}⟩
\end{minipage} \\
\midrule\noalign{}
\endhead
\bottomrule\noalign{}
\endlastfoot
Pyrene & 960 & 13 & 0.40 & 125.9 & 62.6 & 7 & 74.4 \\
Tetracene & 680 & 33 & 0.48 & 169.1 & 94.1 & 6 & 116.9 \\
Perylene & 600 & 7.2 & 0.142 & 73.0\textsuperscript{a} & 55.9\textsuperscript{a} & 3\textsuperscript{a} & 121.4 \\
Pentacene & 600 & 9.2 & 0.451 & 184.2 & 104.8 & 5 & 129.1 \\
Coronene & 820 & 13.8 & 0.221 & 86.4 & 72.7 & 2 & 89.0 \\
\end{longtable}

\textsuperscript{a}For perylene the band-based descriptors use only the three cleanly fitted bands (33.2, 61.6 and 73.0 cm⁻\textsuperscript{1}); the weak, unfitted shoulder near 191 cm⁻\textsuperscript{1} is excluded. The centroid uses the full measured spectrum and is insensitive to this choice. Band-based descriptors for the other compounds use the Lorentzian fits of Supplementary Table 3.

The band-based descriptors (highest resolved band, strongest-band position, mean band position and number of resolved modes) do not show robust correlations with E in this five-compound set. The apparent correlation between κ\textsubscript{pl} and the highest band is especially sensitive to how the weak high-frequency perylene shoulder is classified. Because the spectra do not establish whether that shoulder is a distinct band, this correlation is not interpreted physically and is not used in the main text.

The intensity-weighted centroid of the full Bose-corrected spectrum, ⟨\ensuremath{\tilde{\omega}}⟩ = ΣR(\ensuremath{\tilde{\omega}})\ensuremath{\tilde{\omega}}/ΣR(\ensuremath{\tilde{\omega}}), calculated over 10.5--200.3 cm⁻\textsuperscript{1}, shows a strong negative Pearson correlation with E in the uncorrected spectra (r = −0.986; nominal parametric p = 0.0020). The centroid uses the full measured spectral shape, but it is also sensitive to the baseline and Bose-correction procedure. The result is therefore a preliminary association, not evidence that the centroid measures the activation energy or identifies the microscopic transport channel.

\needspace{8\baselineskip}
\textbf{Supplementary Table 7 \textbar{}} Leave-one-out Pearson correlations between E and the Raman centroid. The full five-compound correlation is r = −0.986; the leave-one-out values are shown as a sensitivity check, not as independent validation.\par\nopagebreak
\begin{longtable}[]{@{}
  >{\raggedright\arraybackslash}p{(\columnwidth - 4\tabcolsep) * \real{0.5382}}
  >{\raggedright\arraybackslash}p{(\columnwidth - 4\tabcolsep) * \real{0.2658}}
  >{\raggedright\arraybackslash}p{(\columnwidth - 4\tabcolsep) * \real{0.1960}}@{}}
\toprule\noalign{}
\begin{minipage}[b]{\linewidth}\raggedright
Compound omitted
\end{minipage} & \begin{minipage}[b]{\linewidth}\raggedright
\emph{r} (\emph{n} = 4)
\end{minipage} & \begin{minipage}[b]{\linewidth}\raggedright
\emph{p}
\end{minipage} \\
\midrule\noalign{}
\endhead
\bottomrule\noalign{}
\endlastfoot
Pyrene & −0.974 & 0.026 \\
Tetracene & −0.988 & 0.012 \\
Perylene & −0.992 & 0.008 \\
Pentacene & −0.986 & 0.014 \\
Coronene & −0.990 & 0.010 \\
\end{longtable}

An exact permutation test over the 5! = 120 assignments of centroid values to E values gives two permutations with \textbar r\textbar{} at least as large as the observed value, corresponding to p = 2/120 = 0.017 under the stated two-sided permutation convention. This indicates an extreme rank association within this five-point set; it does not remove the limitations arising from the small sample, descriptor screening or systematic spectral background.

\textbf{Sensitivity to the spectral background, the frequency range and individual scans.} Because the reduced spectrum is obtained by dividing by \emph{n} + 1, a frequency-independent raw background becomes a rising function in \emph{R}(\ensuremath{\tilde{\omega}}) (Supplementary Note 3, Section 3.2). On the 341-point grid (10.5--200.3 cm⁻\textsuperscript{1}) at 300 K, such a background alone has a centroid of ≈129.1 cm⁻\textsuperscript{1}, essentially the value found for pentacene (129.1 cm⁻\textsuperscript{1} for the average spectrum; 128.9, 129.3 and 129.1 cm⁻\textsuperscript{1} for the three scans). This does not show that the pentacene spectrum is dominated by background, but it shows that such a value can arise without phonon-specific information: the centroid combines the phonon response with fluorescence and the instrumental background, and the larger the background, the closer the centroid is pulled towards about 129 cm⁻\textsuperscript{1}. The leave-one-out test above concerns the choice of compounds, not this systematic effect. We therefore recomputed the correlation for several treatments of the spectra (Supplementary Table 8).

\needspace{8\baselineskip}
\textbf{Supplementary Table 8 \textbar{} Sensitivity of the correlation between \emph{E} and the Raman centroid to the treatment of the spectra.} Pearson \emph{r} and \emph{p} for the five compounds.\par\nopagebreak
\begin{longtable}[]{@{}
  >{\raggedright\arraybackslash}p{(\columnwidth - 4\tabcolsep) * \real{0.7429}}
  >{\raggedright\arraybackslash}p{(\columnwidth - 4\tabcolsep) * \real{0.1286}}
  >{\raggedright\arraybackslash}p{(\columnwidth - 4\tabcolsep) * \real{0.1286}}@{}}
\toprule\noalign{}
\begin{minipage}[b]{\linewidth}\raggedright
Treatment of the spectra
\end{minipage} & \begin{minipage}[b]{\linewidth}\raggedright
\emph{r}
\end{minipage} & \begin{minipage}[b]{\linewidth}\raggedright
\emph{p}
\end{minipage} \\
\midrule\noalign{}
\endhead
\bottomrule\noalign{}
\endlastfoot
Full spectrum, no background subtraction (as in Fig. 7) & −0.99 & 0.002 \\
Constant subtracted (5th percentile of raw \emph{I}, \ensuremath{\tilde{\omega}} \textgreater{} 60 cm⁻\textsuperscript{1}) & −0.64 & 0.25 \\
Constant subtracted (10th percentile of raw \emph{I}, \ensuremath{\tilde{\omega}} \textgreater{} 60 cm⁻\textsuperscript{1}) & −0.62 & 0.26 \\
Upper limit of the range 175 cm⁻\textsuperscript{1} & −0.96 & 0.011 \\
Upper limit of the range 150 cm⁻\textsuperscript{1} & −0.87 & 0.054 \\
\end{longtable}

Subtracting a constant lowers the centroids, most strongly for the spectra with the largest background: perylene 121 → 77 cm⁻\textsuperscript{1}, coronene 89 → 78 cm⁻\textsuperscript{1}, pentacene 129 → 125 cm⁻\textsuperscript{1}, pyrene 74 → 70 cm⁻\textsuperscript{1} and tetracene 117 → 111 cm⁻\textsuperscript{1}. The centroids also scatter between repeat scans, from 68.6 to 77.6 cm⁻\textsuperscript{1} for pyrene and from 86.6 to 97.8 cm⁻\textsuperscript{1} for coronene. The constant subtraction used here is deliberately crude: it shows the sensitivity of the result but is not a justified background model, and it may also remove real broad phonon weight. The conclusion is nevertheless clear. The correlation of Fig. 7 depends on the treatment of the background and cannot be interpreted as a link between \emph{E} and the phonon spectrum until the background is fitted together with the bands (as a constant, linear or polynomial term), subtracted from the raw spectra before the Bose correction, and the centroid is recomputed for each scan and for several justified background models.

The analysis has important limitations. It includes only five compounds, and several descriptors were examined before the centroid was identified as the most strongly correlated. The Pearson and permutation p values do not account for this descriptor selection. In addition, E (600--960 K, equivalent to roughly 420--670 cm⁻\textsuperscript{1}) lies above the 10--200 cm⁻\textsuperscript{1} Raman window. The correlation therefore links a fitted thermal activation scale to the shape of the measured external-mode spectrum, not to a directly observed Raman mode. Any proposed physical explanation remains speculative until tested on a larger compound set and with background-resolved spectra.

\section*{Supplementary Note 6. Previous thermal studies of PAHs and boson-peak-like anomalies in ordered molecular crystals}

\subsection*{6.1 Classical thermodynamic characterization}

Most of the PAHs studied here were characterized thermodynamically long ago, mainly by the adiabatic calorimetry of Wong and Westrum. The first paper of their series reported heat capacities and enthalpies of fusion of pyrene, together with fluoranthene and triphenylene\textsuperscript{60}. The second extended this to perylene, coronene and naphthacene (tetracene), with heat capacities measured from 5 K and, for perylene, into the liquid at 578 K\textsuperscript{61}. The reported \emph{C}\textsubscript{p} values at 298.15 K were 274.9 J mol⁻\textsuperscript{1} K⁻\textsuperscript{1} (perylene), 313.8 J mol⁻\textsuperscript{1} K⁻\textsuperscript{1} (coronene) and 236.6 J mol⁻\textsuperscript{1} K⁻\textsuperscript{1} (tetracene). Wong and Westrum found a reproducible broad thermal anomaly in coronene near 225--230 K (crystal II → crystal I; Δ\emph{H} ≈ 0.44 kJ mol⁻\textsuperscript{1}, Δ\emph{S} ≈ 1.84 J mol⁻\textsuperscript{1} K⁻\textsuperscript{1}), but did not reproduce a transition previously reported for tetracene.

Fulem et al.~later measured the heat capacity of solid tetracene and pentacene from 258 to 600 K by differential scanning calorimetry, in good agreement with earlier compilations, which they also corrected in places\textsuperscript{62}. Chickos et al.~determined the enthalpies of sublimation of corannulene, perylene and coronene at 298.15 K by correlation gas chromatography combined with fusion enthalpies: 118.3 ± 3.3, 138.1 ± 6.2 and 155 ± 7.0 kJ mol⁻\textsuperscript{1}, respectively\textsuperscript{63}. A combustion-calorimetry study by Torres et al.~revised the thermochemistry of coronene, refining its enthalpies of formation and sublimation\textsuperscript{64}. On the structural side, Litasov et al.~measured the thermal expansion of coronene by X-ray diffraction from 185 to 416 K, finding that α increases from 4.8 × 10⁻\textsuperscript{5} K⁻\textsuperscript{1} at 185 K to 4.9 × 10⁻\textsuperscript{4} K⁻\textsuperscript{1} at 416 K (1.9 × 10⁻\textsuperscript{4} K⁻\textsuperscript{1} at 298 K), and noted that the thermal expansion of PAHs tends to decrease with the number of rings\textsuperscript{65}.

Earlier adiabatic calorimetry by Wong and Westrum includes measurements beginning near 5 K for several PAHs, so it would be inaccurate to state that all previous calorimetry was restricted to temperatures above 50 K. Those studies provide valuable thermodynamic data, but they do not offer the matched, systematic combination of low-temperature heat capacity, thermal conductivity and low-frequency Raman measurements across the structurally varied PAH series examined here.

\subsection*{6.2 Boson-peak-like anomalies in ordered molecular crystals}

Recent theoretical and experimental studies show that boson-peak-like vibrational and thermal anomalies can occur in translationally ordered molecular crystals as well as in glasses\textsuperscript{9,10,66,67}. Proposed mechanisms include hybridization between low-energy optical and acoustic branches, avoided crossings and clustering of low-frequency optical modes. These studies establish plausible routes to glass-like anomalies without requiring conventional amorphous disorder, but their microscopic relevance must be assessed separately for each crystal.

The existing literature spans chemically and structurally different molecular crystals. It therefore does not provide a controlled comparison in which molecular shape, symmetry and packing are varied within a closely related chemical family. PAHs offer a useful, though not perfectly controlled, test family because they share predominantly van der Waals intermolecular interactions while differing in molecular shape and crystal packing.

\subsection*{6.3 Two-level-system universality}

Whether the density of tunnelling two-level systems is an intrinsic, universal property of the amorphous state remains debated. A widely cited theoretical reassessment concluded that the low-temperature universality of glasses does not by itself prove a single tunnelling mechanism\textsuperscript{68}. Direct tests give different answers for different glass formers: ultrastable vapour-deposited indomethacin glasses show a suppressed linear-in-\emph{T} term that reappears on thermal rejuvenation\textsuperscript{69}, whereas 110-million-year-old amber shows an unchanged two-level-system density and boson peak compared with rejuvenated samples\textsuperscript{70}. Kinetic lattice Monte Carlo simulations have begun to reproduce a linear-in-\emph{T} heat capacity directly from localized two-level defects\textsuperscript{71}. These results do not apply directly to the fully ordered PAH crystals studied here, which lack the compositional and positional disorder that defines two-level systems in the classical sense, but they frame the broader question of disorder versus structural determinism that motivates this work.

\subsection*{6.4 The resulting gap}

The gap addressed here is not the absence of all earlier low-temperature calorimetry on PAHs. Rather, it is the lack of a systematic, side-by-side study combining low-temperature heat capacity, thermal conductivity and low-frequency Raman spectroscopy for a structurally diverse series of PAH crystals, together with a common analysis of their boson-peak-like anomalies.

\section*{Supplementary References}\small

\begin{enumerate}
\def\labelenumi{\arabic{enumi}.}
\tightlist
\item
  Zhou, Z. et al.~On the existence of Einstein oscillators and thermal conductivity in bulk metallic glass. \emph{Appl. Phys. Lett.} \textbf{89}, 031924 (2006).
\item
  Tian, Y., Li, Z. Q. \& Jiang, E. Y. Low temperature specific heat and thermal conductivity of bulk metallic glass (Cu50Zr50)94Al6. \emph{Solid State Commun.} \textbf{149}, 1527--1530 (2009).
\item
  Gamsjäger, E. \& Wiessner, M. Low temperature heat capacities and thermodynamic functions described by Debye--Einstein integrals. \emph{Monatsh. Chem.} \textbf{149}, 357--368 (2018).
\item
  van Hattem, A. et al.~Low-temperature heat capacity of CsPbI3, Cs4PbI6, and Cs3Bi2I9. \emph{J. Phys. Chem. C} \textbf{127}, 22808 (2023).
\item
  Li, K. et al.~High-entropy strategy to achieve electronic band convergence for high-performance thermoelectrics. \emph{J. Am. Chem. Soc.} \textbf{146}, 14318--14327 (2024).
\item
  Gamsjäger, E. \& Wiessner, M. Extended regression analysis for Debye--Einstein models describing low temperature heat capacity data of solids. \emph{Entropy} \textbf{26}, 452 (2024).
\item
  Gurevich, V. L., Parshin, D. A. \& Schober, H. R. Anharmonicity, vibrational instability, and the Boson peak in glasses. \emph{Phys. Rev.~B} \textbf{67}, 094203 (2003).
\item
  Schirmacher, W. \& Ruocco, G. in \emph{Low-Temperature Thermal and Vibrational Properties of Disordered Solids} (ed.~Ramos, M. A.) Ch. 9 (World Scientific, 2022). Preprint at https://arxiv.org/abs/2009.05970.
\item
  Baggioli, M. \& Zaccone, A. Low-energy optical phonons induce glassy-like vibrational and thermal anomalies in ordered crystals. \emph{J. Phys. Mater.} \textbf{3}, 015004 (2020).
\item
  Krivchikov, A. I. et al.~Role of optical phonons and anharmonicity in the appearance of the heat capacity boson peak-like anomaly in fully ordered molecular crystals. \emph{J. Phys. Chem. Lett.} \textbf{13}, 5061--5067 (2022).
\item
  Robertson, J. M., Sinclair, V. C. \& Trotter, J. The crystal and molecular structure of tetracene. \emph{Acta Crystallogr.} \textbf{14}, 697--704 (1961).
\item
  Raman scattering from the molecular charge-transfer crystal anthracene--PMDA. Solid State Commun. (1979). {[}Bibliographic details incomplete in the source file; authors, volume and page range must be verified before submission.{]}
\item
  Hanson, J. C. \& Nordman, C. E. The crystal and molecular structure of corannulene, C20H10. \emph{Acta Crystallogr. B} \textbf{32}, 1147--1153 (1976).
\item
  Gaboardi, M. et al.~Curvature-induced vitrification and polymorphism in corannulene. \emph{Commun. Chem.} \textbf{9}, 173 (2026).
\item
  Bannister, N. et al.~Lattice vibrations of γ- and β-coronene from Raman microscopy and theory. \emph{Phys. Rev.~Mater.} \textbf{3}, 125601 (2019).
\item
  Mattheus, C. C. et al.~Polymorphism in pentacene. \emph{Acta Crystallogr. C} \textbf{57}, 939--941 (2001).
\item
  Gnoli, L., Venuti, E., Salzillo, T., Masino, M. \& Graziosi, P. Electrons and phonons in pentacene: coupling patterns reveal the microscopic origin of the phonon-limited mobility. \emph{J. Phys. Chem. C} \textbf{129}, 21738 (2025).
\item
  Lynch, P. G., Baughman, S., Mihm, T. N., Sharifzadeh, S. \& Frontiera, R. R. Identifying driving and spectator phonon modes in pentacene exciton transport. \emph{J. Am. Chem. Soc.} \textbf{147}, 23705--23714 (2025).
\item
  Brillante, A., Bilotti, I., Della Valle, R. G., Venuti, E. \& Girlando, A. Probing polymorphs of organic semiconductors by lattice phonon Raman microscopy. \emph{CrystEngComm} \textbf{10}, 937--946 (2008).
\item
  Asher, M. et al.~Anharmonic lattice vibrations in small-molecule organic semiconductors. \emph{Adv. Mater.} \textbf{32}, 1908028 (2020).
\item
  Donaldson, D. M., Robertson, J. M. \& White, J. G. The crystal and molecular structure of perylene. \emph{Proc. R. Soc. Lond. A} \textbf{220}, 311--321 (1953).
\item
  Camerman, A. \& Trotter, J. The crystal and molecular structure of pyrene. \emph{Acta Crystallogr.} \textbf{18}, 636--643 (1965).
\item
  Potticary, J. et al.~An unforeseen polymorph of coronene by the application of magnetic fields during crystal growth. \emph{Nat. Commun.} \textbf{7}, 11555 (2016).
\item
  Peierls, R. Zur kinetischen Theorie der Wärmeleitung in Kristallen. \emph{Ann. Phys.} \textbf{395}, 1055--1101 (1929).
\item
  Zeller, R. C. \& Pohl, R. O. Thermal conductivity and specific heat of noncrystalline solids. \emph{Phys. Rev.~B} \textbf{4}, 2029--2041 (1971).
\item
  Nataj, Z. E. et al.~Cryogenic characteristics of graphene composites---evolution from thermal conductors to thermal insulators. \emph{Nat. Commun.} \textbf{14}, https://doi.org/10.1038/s41467-023-38508-3 (2023).
\item
  Jäckel, M., v. Schoenebeck, F., Escher, U. \& Gladun, A. Low-temperature thermal properties of amorphous polycarbonate. \emph{Czech. J. Phys.} \textbf{46}, 2249--2250 (1996).
\item
  Duval, E. et al.~Low-energy excitations in polyvinyl chloride: Raman scattering and thermal properties. \emph{J. Non-Cryst. Solids} \textbf{190}, 258--263 (1995).
\item
  Nakamura, S., Fujii, T., Matsukawa, S., Katagiri, M. \& Fukuyama, H. Specific heat, thermal conductivity, and magnetic susceptibility of cyanate ester resins -- an alternative to commonly used epoxy resins. \emph{Cryogenics} \textbf{95}, 76--81 (2018).
\item
  Kelham, S. \& Rosenberg, H. M. The thermal conductivity and specific heat of epoxy-resin from 0.1--80 K. \emph{J. Phys. C} \textbf{14}, 1737--1749 (1981).
\item
  Nicholls, C. I. \& Rosenberg, H. M. The effects of cross-link length on the thermal properties of epoxy-resins from 1.5 to 80 K. \emph{Physica B+C} \textbf{108}, 1015--1016 (1981).
\item
  Nicholls, C. I. \& Rosenberg, H. M. The excitation spectrum of epoxy resins; neutron diffraction, specific heat and thermal conductivity at low temperatures. \emph{J. Phys. C} \textbf{17}, 1165--1178 (1984).
\item
  Choy, C. L., Hunt, R. G. \& Salinger, G. L. Specific heat of amorphous polymethyl methacrylate and polystyrene below 4 K. \emph{J. Chem. Phys.} \textbf{52}, 3629--3633 (1970).
\item
  Jäckel, M., Geilenkeuser, R. \& Gladun, A. Influence of high pressure on the specific heat of amorphous polymers at low temperatures. \emph{Physica B} \textbf{316--317}, 535--538 (2002).
\item
  Ramos, M. A. (ed.) \emph{Low-Temperature Thermal and Vibrational Properties of Disordered Solids: A Half-Century of Universal ``Anomalies'' of Glasses} (World Scientific, 2022).
\item
  Anderson, P. W., Halperin, B. I. \& Varma, C. M. Anomalous low-temperature thermal properties of glasses and spin glasses. \emph{Phil. Mag.} \textbf{25}, 1--9 (1972).
\item
  Phillips, W. A. Tunneling states in amorphous solids. \emph{J. Low Temp. Phys.} \textbf{7}, 351--360 (1972).
\item
  Karpov, V. G., Klinger, M. I. \& Ignatiev, P. N. Atomic tunneling states and low-temperature anomalies of thermal properties in amorphous materials. \emph{Solid State Commun.} \textbf{44}, 333--337 (1982).
\item
  Buchenau, U. et al.~Interaction of soft modes and sound waves in glasses. \emph{Phys. Rev.~B} \textbf{46}, 2798--2808 (1992).
\item
  Ramos, M. A. \& Buchenau, U. Low-temperature thermal conductivity of glasses within the soft-potential model. \emph{Phys. Rev.~B} \textbf{55}, 5749--5754 (1997).
\item
  Allen, P. B. \& Feldman, J. L. Thermal conductivity of glasses: theory and application to amorphous Si. \emph{Phys. Rev.~Lett.} \textbf{62}, 645 (1989).
\item
  Allen, P. B. \& Feldman, J. L. Thermal conductivity of disordered harmonic solids. \emph{Phys. Rev.~B} \textbf{48}, 12581 (1993).
\item
  Allen, P. B., Feldman, J. L., Fabian, J. \& Wooten, F. Diffusons, locons and propagons: character of atomic vibrations in amorphous Si. \emph{Phil. Mag. B} \textbf{79}, 1715--1731 (1999).
\item
  Feldman, J. L., Kluge, M. D., Allen, P. B. \& Wooten, F. Thermal conductivity and localization in glasses: numerical study of a model of amorphous silicon. \emph{Phys. Rev.~B} \textbf{48}, 12589 (1993).
\item
  Tlili, A. et al.~Enhancement and anticipation of the Ioffe--Regel crossover in amorphous/nanocrystalline composites. \emph{Nanoscale} \textbf{11}, 21502 (2019).
\item
  Desmarchelier, P., Carré, A., Termentzidis, K. \& Tanguy, A. Ballistic heat transport in nanocomposite: the role of the shape and interconnection of nanoinclusions. \emph{Nanomaterials} \textbf{11}, 1982 (2021).
\item
  Tanguy, A. Vibrations and heat transfer in glasses: the role played by disorder. \emph{C. R. Phys.} \textbf{24}, 73 (2023).
\item
  Simoncelli, M., Marzari, N. \& Mauri, F. Unified theory of thermal transport in crystals and glasses. \emph{Nat. Phys.} \textbf{15}, 809--813 (2019).
\item
  Simoncelli, M., Marzari, N. \& Mauri, F. Wigner formulation of thermal transport in solids. \emph{Phys. Rev.~X} \textbf{12}, 041011 (2022).
\item
  Simoncelli, M., Mauri, F. \& Marzari, N. Thermal conductivity of glasses: first-principles theory and applications. \emph{npj Comput. Mater.} \textbf{9}, 106 (2023).
\item
  Simoncelli, M. et al.~Temperature-invariant heat conductivity from compensating crystalline and glassy transport: from the Steinbach meteorite to furnace bricks. Preprint at https://arxiv.org/abs/2405.13161 (2024).
\item
  Korolyuk, O. A. et al.~Bridging glass-like and crystal-like thermal transport: a unified model for disordered and complex solids. \emph{J. Non-Cryst. Solids} \textbf{690}, 124273 (2026).
\item
  Krivchikov, A. I. \& Korolyuk, O. A. Empirical universal approach to describing the thermal conductivity of amorphous polymers: effects of pressure, radiation and the Meyer--Neldel rule. \emph{Low Temp. Phys.} \textbf{50}, 328--335 (2024).
\item
  Berman, R. \emph{Thermal Conduction in Solids} (Clarendon, 1976).
\item
  Krivchikov, A., Andersson, O., Korolyuk, O. \& Kryvchikov, O. Thermal conductivity of solid triphenyl phosphite. \emph{Molecules} \textbf{27}, 8399 (2022).
\item
  Krivchikov, A. I. et al.~Exponential approximation of the coherence contribution to the thermal conductivity of complex clathrate-type crystals. \emph{Materialia} \textbf{32}, 101944 (2023).
\item
  Korolyuk, O. A., Sagan, V. V., Krivchikov, A. I., Konstantinov, V. A. \& Horbatenko, Yu. V. Multi-channel heat transfer in molecular crystals and their solutions. \emph{Low Temp. Phys.} \textbf{51}, 1048 (2025).
\item
  Meyer, W. \& Neldel, H. Über die Beziehungen zwischen der Energiekonstanten ε und der Mengenkonstanten a in der Leitwerts-Temperaturformel bei oxydischen Halbleitern. \emph{Z. Tech. Phys.} \textbf{18}, 588--593 (1937).
\item
  Bernges, T. et al.~Considering the role of ion transport in diffuson-dominated thermal conductivity. \emph{Adv. Energy Mater.} \textbf{12}, 2200717 (2022).
\item
  Wong, W.-K. \& Westrum, E. F. Jr.~Thermodynamics of polynuclear aromatic molecules. I. Heat capacities and enthalpies of fusion of pyrene, fluoranthene, and triphenylene. \emph{J. Chem. Thermodyn.} \textbf{3}, 105--124 (1971).
\item
  Wong, W.-K. \& Westrum, E. F. Jr.~Thermodynamics of polynuclear aromatic molecules. II. Low temperature thermal properties of perylene, coronene, and naphthacene. \emph{Mol. Cryst. Liq. Cryst.} \textbf{61}, 207--228 (1980).
\item
  Fulem, M., Laštovka, V., Straka, M., Růžička, K. \& Shaw, J. M. Heat capacities of tetracene and pentacene. \emph{J. Chem. Eng. Data} \textbf{53}, 2175--2181 (2008).
\item
  Chickos, J. S. et al.~The enthalpy of vaporization and sublimation of corannulene, coronene, and perylene at T = 298.15 K. \emph{J. Chem. Thermodyn.} \textbf{34}, 1195--1206 (2002).
\item
  Torres, L. A., Campos, M., Martínez, M. \& Rojas, A. The thermochemistry of coronene revisited. \emph{J. Chem. Thermodyn.} \textbf{41}, 957--965 (2009).
\item
  Litasov, K. D. et al.~Thermal expansion of coronene C24H12 at 185--416 K. \emph{J. Therm. Anal. Calorim.} \textbf{119}, 1183--1189 (2015).
\item
  Gebbia, J. F. et al.~Glassy anomalies in the low-temperature thermal properties of a minimally disordered crystalline solid. \emph{Phys. Rev.~Lett.} \textbf{119}, 215506 (2017).
\item
  Szewczyk, D. et al.~Specific heat at low temperatures in quasiplanar molecular crystals: origin of glassy anomalies in minimally disordered crystals. \emph{Phys. Rev.~B} \textbf{110}, 174204 (2024).
\item
  Leggett, A. J. \& Vural, D. C. ``Tunneling two-level systems'' model of the low-temperature properties of glasses: are ``smoking-gun'' tests possible? \emph{J. Phys. Chem. B} \textbf{117}, 12966--12971 (2013).
\item
  Pérez-Castañeda, T., Rodríguez-Tinoco, C., Rodríguez-Viejo, J. \& Ramos, M. A. Suppression of tunneling two-level systems in ultrastable glasses of indomethacin. \emph{Proc. Natl Acad. Sci. USA} \textbf{111}, 11275--11280 (2014).
\item
  Pérez-Castañeda, T., Jiménez-Riobóo, R. J. \& Ramos, M. A. Two-level systems and boson peak remain stable in 110-million-year-old amber glass. \emph{Phys. Rev.~Lett.} \textbf{112}, 165901 (2014).
\item
  Gao, X.-Y., Deng, H.-Y., Lee, C.-S., You, J. Q. \& Lam, C.-H. Emergence of two-level systems in glass formers: a kinetic Monte Carlo study. \emph{Soft Matter} \textbf{18}, 2211--2221 (2022).
\end{enumerate}

\end{document}